\documentclass[sigconf, nonacm]{acmart}

\usepackage{modifiedfrompvldbforarxiv}

\renewcommand\vldbdoi{XX.XX/XXX.XX}
\renewcommand\vldbpages{XXX-XXX}
\renewcommand\vldbavailabilityurl{} 

\usepackage{mathtools,algorithm,enumitem,bm,listings,subcaption,tikz}
\usepackage[indLines=true,noEnd=true]{algpseudocodex}

\theoremstyle{acmplain}

\newtheorem{claim}{Claim}[section]
\newtheorem{lemma}{Lemma}[section]
\newtheorem{corollary}{Corollary}[section]

\theoremstyle{acmdefinition}
\newtheorem{definition}{Definition}[section]

\theoremstyle{remark}

\newtheorem{remark}{Remark}

\newcommand{\updown}[2]{\genfrac{}{}{0pt}{}{#1}{#2}}
\newcommand{\eg}{e.$\,$g.\@}
\newcommand{\ie}{i.$\,$e.\@}
\newcommand{\aka}{a.$\,$k.$\,$a.\@}
\newcommand{\eps}{\varepsilon}
\newcommand{\Cx}[2]{\mathrlap{C_{#2}}\phantom{C}^{\hspace{0.5pt}\raisebox{1.5pt}{$\scriptstyle{#1}$}}_{\hphantom{#2}}}

\newcommand{\bprime}[1]{\mathrlap{b_{\hspace{0.5pt}#1}}\phantom{b}^{\raisebox{1pt}{$\scriptstyle\prime$}}_{\hphantom{\hspace{0.5pt}#1}}}
\newcommand{\bpprime}[1]{\mathrlap{b_{\hspace{0.5pt}#1}}\phantom{b}^{\raisebox{1pt}{$\scriptstyle\prime\prime$}}_{\hphantom{\hspace{0.5pt}#1}}}

\DeclareMathOperator*{\argmin}{arg\,min}
\DeclareMathOperator*{\diag}{diag}
\DeclareMathOperator*{\E}{\mathbf{E}}
\DeclareMathOperator*{\Var}{\mathbf{Var}}
\DeclareMathOperator*{\Varh}{\mathbf{V\mkern-3mu\widehat{\mkern3mu \vphantom{V}a\mkern3mu}\mkern-3mu r}}
\DeclareMathOperator*{\Cov}{\mathbf{Cov}}
\DeclareMathOperator*{\sd}{\mathbf{St.D.}}
\DeclareMathOperator{\rank}{rank}
\DeclareMathOperator{\yes}{yes}
\DeclareMathOperator{\expmone}{expm1}
\DeclareMathOperator{\logonep}{log1p}

\renewcommand{\emptyset}{\varnothing}

\newlength\templength

\algnewcommand{\IfThen}[2]{\State \algorithmicif\ #1\ \algorithmicthen\ #2}
\algnewcommand{\ForAllDone}[2]{\State \algorithmicforall\ #1\ \algorithmicdo\ #2}

\usepackage[normalem]{ulem}
\newif\ifshowedits \showeditstrue

\DeclareRobustCommand{\del}[1]{\ifshowedits{\color{red}\sout{#1}}\fi}
\DeclareRobustCommand{\delm}[1]{\ifshowedits{\color{red}#1}\fi}

\AtBeginDocument{\pdfstringdefDisableCommands{\def\del#1{}\def\delm#1{}}}

\begin{document}
\title{Blocked Bloom Filters for Overlap Estimation}

\author{Alexandre Evfimievski}
\affiliation{%
  \institution{Adobe Inc.}
  \streetaddress{345 Park Ave.}
  \city{San Jose}
  \state{CA}
  \country{USA}
}
\email{aevfimievski@adobe.com}

\author{Anup Rao}
\affiliation{%
  \institution{Adobe Inc.}
  \streetaddress{345 Park Ave.}
  \city{San Jose}
  \state{CA}
  \country{USA}
}
\email{anuprao@adobe.com}

\author{Yeshwanth Vijayakumar}
\affiliation{%
  \institution{Adobe Inc.}
  \streetaddress{345 Park Ave.}
  \city{San Jose}
  \state{CA}
  \country{USA}
}
\email{yvijayak@adobe.com}

\author{Tung Mai}
\affiliation{%
  \institution{Adobe Inc.}
  \streetaddress{345 Park Ave.}
  \city{San Jose}
  \state{CA}
  \country{USA}
}
\email{tumai@adobe.com}

\begin{abstract}
Estimating the cardinality of set overlaps and set-algebra expressions over sketched sets is a common operation in marketing research and audience analytics. We develop efficient methods for these estimates over large-capacity single and blocked Bloom filters (BFs). We study cardinality and overlap estimation from truncated BFs and derive first-order variance approximations to guide partial-read budgeting. We extend these estimators to multi-block Bloom filters (MBFs) that partition a sketch into hash-routed blocks for efficient parallelization in relational databases. Truncation in MBFs becomes a choice of how many blocks we must read. Building on this, we give a three-phase method for accuracy-limited cardinality estimation of set-algebra expressions and implement it for pairwise overlaps, where reading only enough blocks to meet a user-specified precision yields a 10–100× speedup. We scale the method to the cross-overlap problem: estimating all pairwise overlaps between two collections of MBFs to the same target accuracy. Our SQL cross-overlap implementation controls I/O and storage costs through careful shared-block management, adaptive batching, and compressed block representation. We adapt a known rANS codec to compress run-length encoded sparse bit-vectors to within a few percent of the Shannon bound. We evaluate the framework on Snowflake and analyze how MBF parameters shape the accuracy vs. efficiency vs. storage trade-off.

\end{abstract}

\maketitle

\vldbtopmatter

\section{Introduction}
\label{sec:intro}

Marketing platforms maintain thousands of audiences---segments of anonymized customer identifiers---with sizes from millions to billions of items.
A catalog of a few hundred such audiences already stores tens of billions of identifiers.
The core operations are (i)~overlap and set-expression cardinality (``how many customers are in $A$ and $B$ but not~$C$''), (ii)~\emph{cross-overlap} search of one or many audiences against a catalog, ranked by overlap, and (iii)~membership (``is this customer in this audience'').
Many search queries only test whether a pair's overlap clears a minimum commercially viable threshold.
These audiences live in a relational cloud warehouse and are queried with SQL.

\paragraph{Why Bloom filters.}
Storing identifiers, or even plain $64$-bit hashes, costs $64$~bits per item. 
Catalog-wide overlap queries are dominated by I/O, not arithmetic, so the representation must shrink both what is stored and what is read per query.
Standard cardinality sketches (HyperLogLog, MinHash/Theta) reduce storage but give up per-item membership. %
Bloom filters (BFs) keep one-sided membership (no false negatives), control false-positive rate, are union-friendly,
and their bit occupancy yields cardinality and, via inclusion--exclusion, overlaps and $\{\cup,\cap,\setminus\}$-expressions (Section~\ref{sec:overlap}).

\paragraph{Why blocked Bloom filters.}
A single filter for a billion-item audience exceeds $1$~GiB uncompressed (Section~\ref{sec:block} gives a $1.116$~GiB example); it does not fit a table cell, cannot be parallelized by the SQL engine, and must be read whole.
We use the item-routed blocked layout of Putze et~al., with database-scale blocks: each item is hashed to one of $b$ blocks (Definition~\ref{def:block:mbf}).
One block per row lets the engine parallelize and prune, and makes ``read only $b'$ of $b$ blocks'' a truncation whose accuracy is predictable (Section~\ref{sec:block}) and budgetable (Section~\ref{sec:accur}).
On a $200\times 200$ catalog, a full-read cross-overlap took $1$h~$54$m; the accuracy-limited run took $8$m~$22$s at $\sigma=5{,}000$ and $3$m~$24$s at $\sigma=10{,}000$ (Table~\ref{tab:exper:comp-speed}).
The corresponding reduction in block-pair \emph{reads} is $14$--$53\times$ (Section~\ref{sec:exper:tradeoff}).

\paragraph{Main Challenge.}
Given compatible blocked BF sketches of sets $S_1,\ldots,S_L$ stored in a relational database, estimate $|F(S_1,\ldots,S_L)|$ for a $\{\cup,\cap,\setminus\}$-formula $F$---in particular, all pairwise overlaps over collections $L\times R$---to a user-specified standard deviation $\sigma$, reading as few blocks as possible, while keeping the sketches compact and membership-queryable.

\subsubsection*{Contributions}
\begin{enumerate}[leftmargin=*]
\item \textit{Joint occupancy analysis for partial observations.}
We develop %
first-order variance approximations for BF overlap and set-expression estimates (Sec{.}~\ref{sec:overlap}).
Under the i.i.d.-bit model, the occupancy-inversion estimator attains the Cram\'er--Rao bound to first order; %
Wald intervals follow (Appendix~\ref{sec:app:stt}).
\item \textit{Set-expression evaluation via inclusion--exclusion.}
Using union-compatibility, we %
apply a M\"obius inversion to obtain IE-partition sizes; any $\{\cup,\cap,\setminus\}$-formula then becomes a disjoint sum over these partitions (Sec{.}~\ref{sec:overlap})
\item \textit{Blocked BF estimator, robust to block saturation.}
For $b$ blocks we invert the sum of bits across blocks, which remains finite if a few blocks saturate, and the same inversion on the first $b'$ blocks, with numerically stable evaluation (Sec{.}~\ref{sec:block}).
\item \textit{Accuracy-limited estimation.}
We turn the overlap variance into an accuracy dial for the number of blocks to read, give a closed-form budget expression,
develop and implement a three-phase Seed~/~Main~/~Top-up estimation method (Sec{.}~\ref{sec:accur:limit}).
\item \textit{Cross-overlaps at scale in SQL.}
Algorithms~\ref{alg:accur:cross} and~\ref{alg:accur:sweep} estimate all pairwise overlaps over collections $L$ and $R$ to a common $\sigma$: engine-side aggregation of per-block bit counts, adaptive batching, 
join optimization, %
and a reverse-order Top-up sweep for block sharing (Sec{.}~\ref{sec:accur:scale}).
\item \textit{Compressed block representation.}
We adapt range ANS to run-length encoded sparse bit-vectors, reaching within a few percent of the Shannon bound, with a decode-optimized implementation (Alg{.}~\ref{alg:compress:rans-decode-faster}) and Roaring bitmaps in memory (Sec{.}~\ref{sec:compress} and~\ref{sec:accur:scale}).
\item \textit{Evaluation.}
We analyze the accuracy--I/O--storage tradeoff for truncated BFs, Theta sketches, and MBFs, finding that $k=1$ dominates under compression, and report Snowflake measurements on a $200\times 200$ synthetic catalog (Sec{.}~\ref{sec:exper}).
\end{enumerate}

\section{Related Work}
\label{sec:related}

\paragraph{Bloom-filter cardinality and overlap estimation.}
Bloom filters support approximate membership testing and a range of applications reviewed by Broder and Mitzenmacher~\cite{Broder2004:bloom-survey}.
Their bit occupancy also provides information about cardinality and similarity.
Mullin~\cite{Mullin1993:join} uses Bloom filters or filter segments to estimate natural-join size, providing an early database precedent for estimation from partial filter information.
Swamidass and Baldi~\cite{Swamidass2007:fingerprint} develop collision corrections for similarity measures computed from folded fingerprints.
Papapetrou et al.~\cite{Papapetrou2010:cardinality} study Bloom-filter cardinality estimation, unions and intersections, probabilistic bounds, and dynamic length adaptation.
For two ordinary, untruncated filters, their intersection estimator is algebraically equivalent to inclusion--exclusion using the exact occupancy inversion; the $-m/k$ form used here applies a large-$m$ approximation to its denominator.
Scholler et al.~\cite{Scholler2022:instream} improve cardinality estimation using additional information collected during insertion.
Grandi~\cite{Grandi2018:analysis} derives the distribution, expectation, and exact variance of single-filter occupancy, accounting for dependence between bit positions.
Our analysis develops joint zero-count statistics across overlapping sets and partially observed filters, with first-order variance propagation to overlaps and set expressions.

\paragraph{Blocked layouts and length adaptation.}
Putze et al.~\cite{Putze2007:cache} develop cache-efficient Bloom-filter layouts; their blocked construction places all probes for an item within one selected block.
Lang et al.~\cite{Lang2019:filtering} further study register-blocked and cache-sectorized variants for database filtering workloads.
Our construction uses the same item-to-block routing principle, with blocks serving as separately stored units of overlap-query processing.
Michael et al.~\cite{Michael2007:joins} develop extended Bloom-filter operations for distributed joins and use \emph{block-partitioned Bloom filters}, subsequently studied by Papapetrou et al.~\cite{Papapetrou2010:cardinality}.
In that construction, every item is represented in every component filter, and components can be discarded to reduce length.
Here, each item is assigned to a single block; we use \emph{blocked Bloom filter} for this routing topology.
For a fixed number of accessed blocks, we derive covariances that account for both within-block and cross-block dependence, and use them to model the accuracy obtained from partial reads.

\paragraph{Truncation and space allocation.}
Mersy et al.~\cite{Mersy2024:collections} define truncated Bloom filters by retaining a prefix and checking only hashes that fall within it.
This is the truncation operation used here.
They allocate space across filter collections to minimize weighted false-positive cost and also discuss faster bitwise set operations on shortened filters.
We study cardinality and overlap estimation from partial observations and use estimated variance to guide read budgeting.

\paragraph{Coordinated sampling and set expressions.}
Beyer et al.~\cite{Beyer2007:synopses} develop distinct-value synopses that can be combined under unions, intersections, and differences, together with unbiased estimators and error analyses.
The Theta-sketch framework of Dasgupta et al.~\cite{Dasgupta2016:theta} provides a general treatment of stream-expression cardinalities and sufficient conditions for unbiased estimation under adaptive thresholds.
These methods retain coordinated item hashes, allowing set operations on the retained samples.
Our representation retains bit occupancy and estimates expression cardinalities through partial unions and their joint statistics.

\paragraph{Register sketches and joint estimation.}
HyperLogLog~\cite{Flajolet2007:HyperLogLog} and its engineering refinements~\cite{Heule2013:HyperLogLog} provide compact cardinality summaries.
For overlap estimation, relevant alternatives include joint estimators that exploit the relationship between two sketches.
Ertl~\cite{Ertl2017:jointHLL} develops maximum-likelihood estimators for joint cardinalities from HyperLogLog sketches, improving on inclusion--exclu\-sion approaches.
SetSketch~\cite{Ertl2021:setsketch} combines features of MinHash and HyperLogLog and supports cardinality and joint estimation.
These methods are important comparators for overlap accuracy per byte.

\paragraph{Compression.}
Mitzenmacher~\cite{Mitzenmacher2001:compressed} studies compressed Bloom filters and the resulting tradeoffs between representation size and membership error.
Our encoding builds on asymmetric numeral systems~\cite{Duda2013:ans}, adapting range ANS to the zero runs in sparse Bloom-filter bitvectors.
Compression is lossless: the estimators operate on the decoded occupancy statistics.
Independently encoded blocks allow query execution to decode selected portions of a stored filter, coupling the representation with the partial-read design.

\paragraph{Progressive query processing.}
Online aggregation~\cite{Hellerstein1997:online} develops the database perspective of progressively refining estimates and reporting uncertainty as processing proceeds.
Here, refinement operates over precomputed sketch blocks.
Our execution scheme uses initial blocks to estimate a read budget, performs a main computation, and adds a top-up phase when the revised budget requires more blocks.
It shares block reads across pairwise overlaps and exposes block-level statistics to SQL aggregation.

\section{Truncated Bloom Filters}
\label{sec:basics}

\begin{definition}
\label{def:basics:bloom}
A \emph{truncated Bloom filter} over a universe $\Omega$ is defined by its hash range $\mathcal{H} \subset \mathbb{R}$ of size $|\mathcal{H}| = m$ and its $k$ hash functions $h := (h_1,\ldots, h_k)$ that map $\Omega\to\mathcal{H}$. Its instance for a set $S\subset\Omega$ and a threshold~$\theta\in\mathbb{R}$ is a pair $\left(\theta, h(S)|_{<\theta}\right)$ where $h(S)|_{<\theta} = \cup_{i=1}^k h_i(S)\cap [0, \theta)$.
\end{definition}

Thus, truncated Bloom filters have parameters $(m, k, \theta)$ where $m$ is the number of "bits" (possible hash values), $k$ is the number of hash functions = hashes per item, and $\theta$ = threshold such that we keep only the hashes $<\,\theta$ and discard all hashes $\geq\,\theta$. For convenience we pick $\mathcal{H} = \{0, 1/m, 2/m, \ldots, (m\,{-}\,1)/m\}$ and $0 \leq \theta \leq 1$. If $m = 2^{64}$ and $k = 1$ we get the usual Theta sketches, and if $\theta = 1$, we get the usual Bloom filters. We assume that all hash functions are chosen uniformly at random from $\mathcal{H}^{\Omega}$, although in reality a pseudorandom algorithm is used. In order to perform set operations over sketches, we must ensure that these sketches are compatible, \eg{} that they use the same hash functions. This prefix-truncation construction is studied by Mersy et al.~\cite{Mersy2024:collections}.

Let us review some well-known properties of Bloom filters (BFs) and see how they apply to truncated BFs.

\paragraph{Membership query.} Truncated BFs allow us to test membership of any item $v\in\Omega$ in the sketched set~$S$ with one-sided guarantee: no false negatives, \ie{} every response ``$v\notin S$'' always implies that $v$ is not in~$S$. See the details in Appendix~\ref{app:basic-props}. 

\paragraph{Size estimation.} Following~\cite{Papapetrou2010:cardinality}, we estimate the size~$n$ of set~$S$ from its truncated BF by inverting the expected value formula for its sum-of-bits statistic $\varsigma := \big|h(S)|_{<\theta}\big|$:
\begin{equation}
\E\varsigma \,=\, m'\Big[1 - \Big(1\hspace{0.5pt}{-}\hspace{0.5pt}\frac{1}{m}\Big)^{kn}\Big] \;\;\implies\;\;
\hat{n} \,:=\, \frac{1}{k}\,{\cdot}\,\frac{\log(1 - \varsigma/m')}{\log(1 - 1/m)},
\label{eq:basics:set-size-predictor}
\end{equation}
often approximated as $\hat{n} \approx {-}(m/k)\log(1\,{-}\,\varsigma/m')$. Following the full filter analysis by Grandi~\cite{Grandi2018:analysis}, we approximate the variance of the truncated BF cardinality estimator $\hat{n}(\varsigma)$ from $\E\varsigma$ and $\Var\varsigma$ using the Delta method applied to~(\ref{eq:basics:set-size-predictor}):
\begin{equation}
\Var\hat{n} \,\,\approx\,\, \Big(\frac{m}{k}\Big)^2 \!{\cdot}\, \frac{\Var\varsigma}{(m' - \E\varsigma)^2} \,\,\approx\,\, 
\frac{m^2}{k^2}\Big(\frac{1}{m' - \E\varsigma} - \frac{1}{m'}\Big) \,-\, \frac{n}{k} \label{eq:basics:set-size-predictor-variance}
\end{equation}
Replace $\E\varsigma$ with just $\varsigma$ and replace $n$ with~$\hat{n}$ in~(\ref{eq:basics:set-size-predictor-variance}) to obtain a practical variance estimator. See more in Appendix~\ref{app:basic-props}.

\paragraph{Unions of BFs.} If we have several sets $S_1, \ldots, S_L$ where each $S_i$ is represented by a truncated BF $\left(\theta, h(S_i)|_{<\theta}\right)$ that uses the same $k$ hash functions and the same~$\theta$, then the union of the sketches is the sketch of the union:
\begin{equation}
\textstyle
\bigcup_{l=1}^L \, h(S_l)|_{<\theta} \;=\; \Big(\bigcup_{l=1}^L \, h(S_l)\Big)\,\cap\, [0, \theta) \;=\; h\Big(\bigcup_{l=1}^L \, S_l\Big)\Big|_{<\theta}
\label{eq:basics:union-bf}
\end{equation}
If the truncated BFs have different thresholds $\theta_1, \ldots, \theta_L$ for different sets, we take the minimum $\theta = \min\{\theta_1, \ldots, \theta_L\}$ and make it a single threshold for all sketches before performing the union. Any hashes in $h(S_l) \cap [\theta, \theta_l)$ will be ignored.

Note that the intersection of several BF sketches is usually \emph{not} the sketch of the set intersection, but a superset of that sketch:
\begin{equation*}
h(S_1) \cap \ldots \cap h(S_L) \;\supseteq\; h(S_1 \cap \ldots \cap S_L)
\end{equation*}
This is because the same hash may appear in each $h(S_i)$, but come from different items $v_i\in S_i$ via hash collisions.

\section{Overlaps and Expressions}
\label{sec:overlap}

\subsection{Inclusion-Exclusion Principle}
\label{sec:overlap:incl-excl}

Let $S_1, \ldots, S_L$ be $L$ subsets of universe~$\Omega$. For each $L$-bit Boolean vector $x\in\{0, 1\}^L$ where $x\neq\bar{0}$ let $U_x := \bigcup_{l:\,x_l=1} S_l$ be the union of all sets marked with ``1'' in~$x$. We call such $U_x$ the \emph{partial unions} of $S_1, \ldots, S_L$. On the other hand, for each $\bar{0}\neq x\in\{0, 1\}^L$ define
\begin{equation}
\label{eq:overlap:ie-parts}
P_x \;:=\; \bigcap_{l=1}^L \;\begin{cases} S_l, & \text{if $x_l=1$}\\ \Omega \setminus S_l, & \text{if $x_l=0$} \end{cases}
\end{equation}
We call such $P_x$ the \emph{IE-partitions} of $S_1, \ldots, S_L$, where IE means ``inclusion-exclusion''. We drop $P_{\bar{0}}$ because the ``non-$S$'' part of~$\Omega$ is irrelevant. Note that the IE-partitions are disjoint and each $U_x$ is the union of all $P_y$ that contain items from \emph{any} of the $S_l$'s where $x_l=1$: $U_x = \bigcup_{y:\, x\land y\,>\,\bar{0}} P_y$.

Some notation: for $x, y\in\{0, 1\}^L$ let $x\lor y$ and $x\land y$ denote bitwise OR and AND, let $x\leq y$ denote $\forall l\in\{1,\ldots, L\}\!:$ $x_l \leq y_l$, let $x<y$ denote ``$x\leq y$ and $x\neq y$,'' and let $|x| := |\{l: x_l = 1\}|$.

\begin{claim}[Inclusion-exclusion principle]
\label{th:overlap:incl-excl}
Given the sizes $|U_x|$ of the partial unions of sets $S_1, \ldots, S_L$, we can compute the sizes $|P_y|$ of their IE-partitions as follows:
\begin{equation}
|P_y| \;= \sum_{x:\, \updown{x\lor y \,=\, \bar{1}}{x \,>\, \bar{0}}} (-1)^{|x \land y| \,+\, 1} \cdot |U_x|
\label{eq:overlap:incl-excl}
\end{equation}
\end{claim}
\begin{proof} See Appendix~\ref{app:ie-transform}. \end{proof}

The IE principle allows us to estimate all IE-partition sizes of $S_1, \ldots, S_L$ if we can estimate the sizes of their partial unions using sketches. We do so by replacing $|U_x|$ with their estimators $\hat{n}[U_x]$ in Eq.~(\ref{eq:overlap:incl-excl}). The IE-mapping is linear, which simplifies accuracy analysis in Section~\ref{sec:accur}. Note that, if the partial unions are estimated with some error, (\ref{eq:overlap:incl-excl})~can return a negative estimate; a real-life implementation must handle such estimates.

\subsection{Expressions over Sketches}
\label{sec:overlap:formula}

Estimating all IE-partition sizes allows us to predict the size of sets defined by expressions over $S_1, \ldots, S_L$ written with set-algebra operations $\cup\,, \cap\,, \setminus$. Any such formula represents a disjoint union of ``atomic'' IE-partitions, because each set $S_l = \bigcup_{x:\, x_l=1} P_x$ is a disjoint union of IE-partitions and operations $\cup\,, \cap\,, \setminus$ preserve this property. In fact, there is a simple way to find exactly which IE-partitions make up the result of a set expression. Given a $\{\cup\,, \cap\,, \setminus\}$-formula $F(S_1, \ldots, S_L)$, convert it into a Boolean formula $F^B(x_1, \ldots, x_L)$ where each occurrence of $S_l$ converts to $x_l$ and each instance of\/ $\cup\,, \cap\,, \setminus$ converts to Boolean operation $\lor\,, \land\,, \land\neg$ respectively. Then, we have:

\begin{claim}
$F(S_1, \ldots, S_L) \,=\, \bigcup_{x:\,F^B(x)=1} P_x$. Since the union is disjoint, its size estimator is:$\;\hat{n}[F] \,=\, \sum_{x:\,F^B(x)=1} \hat{n}[P_x]$.
\label{th:overlap:formula}
\end{claim}
\begin{proof} See Appendix~\ref{app:overlap:formula}. \end{proof}

We start by estimating the partial union sizes $|U_x|$ for all $x \in \{0, 1\}^L$, $x\neq \bar{0}$. Given truncated BFs for each set~$S_l$, we use~(\ref{eq:basics:union-bf}) to form the sketches for all~$U_x$, then use the size estimator~(\ref{eq:basics:set-size-predictor}) to compute~$\hat{n}[U_x]$:
\begin{equation}
\hat{n}[U_x] \,=\, \frac{\log(1 - \varsigma_x/m')}{k\,{\cdot}\,\log(1 - 1/m)} \,\approx\, -\frac{m}{k}\big(\log Z_x\,{-}\,\log m'\big),
\label{eq:overlap:union-from-Z_x}
\end{equation}
where $Z_x = m' - \varsigma_x$ are \emph{zero-bit} counts in the truncated BFs for~$U_x$ and $m' = \big|\mathcal{H}|_{<\theta}\big|$. Then, we replace $|U_x|$ with $\hat{n}[U_x]$ in~(\ref{eq:overlap:incl-excl}) to obtain a linear (affine) mapping from vector $\langle \log Z_x \rangle$ to vector $\langle \hat{n}[P_y] \rangle$. Finally, combining Claim~\ref{th:overlap:formula} with Eqs{.}~(\ref{eq:overlap:incl-excl}) and~(\ref{eq:overlap:union-from-Z_x}) gives an affine expression for $\hat{n}[F]$ applied to vector~$\langle \log Z_x \rangle$. %

The $L$-ary overlap $\bigcap_{l=1}^L S_l$ is one of the IE-partitions, namely the $P_y$ for $y = \bar{1}$, for which the mapping simplifies to
\begin{equation}
\hat{n}\left[\textstyle{\bigcap_{l=1}^L S_l}\right] \,\,\approx\,\, \frac{m}{k} \sum_{x \,>\, \bar{0}}\, (-1)^{|x|} \log Z_x \,+\, \frac{m}{k}\log m'.
\label{eq:overlap:overlap-from-Z_x}
\end{equation}
For $L=2$ and ordinary untruncated filters, the exact occupancy-inversion form is algebraically equivalent to the intersection estimator of Papapetrou et al.~\cite{Papapetrou2010:cardinality}. The expression~(\ref{eq:overlap:overlap-from-Z_x}) with coefficient $m/k$ uses the corresponding large-$m$ approximation.

Algorithm~\ref{alg:overlap:ie-predict} summarizes this approach. As long as $L$ is small, we can estimate the size of multiple expressions over $S_1, \ldots, S_L$ ``in one go'' by estimating the sizes of all IE-partitions and then adding up those estimates that correspond to the satisfying assignments for the converted Boolean formulas. %

\begin{algorithm}[!t]
\caption{Predicts size of the set defined by a $\{\cup\,, \cap\,, \setminus\}$-for\-mula $F(S_1, \ldots, S_L)$ over BF sketches from all their hashed unions}
\label{alg:overlap:ie-predict}
\begin{algorithmic}[1]%
\Require%
\begin{itemize}\setlength{\itemindent}{-44pt}
\item $F(S_1, \ldots, S_L)$ given by all IE-partitions it has: $F^B \subset \{0, 1\}^L$
\item All hashed union sizes: $\varsigma_x = \big|h(\bigcup_{l:\,x_l=1} S_l)|_{<\theta}\big|$ for $x \,{\in}\,\{0, 1\}^L$
\item BF hash range:~$m = |\mathcal{H}|$, $m'\,{=}\:|\mathcal{H}\,{\cap}\,[0, \theta)|$; hash functions:~$k$
\end{itemize}
\Ensure $\hat{n}[F]$ approximates the predicted size of $F(S_1, \ldots, S_L)$
\Procedure{IE-Predict}{$F, \langle \varsigma_x \rangle_{x \in \{0, 1\}^L\,}, m, m', k$}
    \State $\forall y \in\{0, 1\}^L \setminus \{\bar{0}\}: \,\hat{n}[P_y] \gets 0$ \Comment{$\hat{n}[P_y]$ will predict $|P_y|$}
    \ForAll{$x\in \{0, 1\}^L \setminus \{\bar{0}\}$}
        \If{$\varsigma_x < m'$}
            \State $\hat{n}[U_x] \gets 1/k\cdot\logonep(- \varsigma_x/m')\mathop{/}\logonep(- 1/m)$
        \Else
            \State $\hat{n}[U_x] \gets +\infty$ \Comment{No upper limit for $|U_x|$}
        \EndIf
        \ForAll{$y\in \{0, 1\}^{\mathstrut L} \setminus \{\bar{0}\}$ such that $x\,{\lor}\,y = \bar{1}$}
            \State $\hat{n}[P_y] \gets \hat{n}[P_y] + (-1)^{\mathstrut|x\land y| \,+\, 1}\!\cdot \hat{n}[U_x]$
        \EndFor
    \EndFor
    \State \textbf{return} $\hat{n}[F] \gets \sum_{y\in F^B} \hat{n}[P_y]$
\EndProcedure
\end{algorithmic}
\end{algorithm}

\subsection{Variance Estimation}
\label{sec:overlap:variance}

Here we derive a variance formula for estimator $\hat{n}[F]$ that predicts the size of $\{\cup\,, \cap\,, \setminus\}$-expres\-sion $F(S_1, \ldots, S_L)$ over $L$ sets represented by their Bloom filters. Recall from Section~\ref{sec:overlap:formula} that $\hat{n}[F]$ is an affine function applied to vector $\langle \log Z_x \rangle$ where $x\in\{0, 1\}^L$ excluding $\bar{0}$ and where $Z_x = m' - \varsigma_x$ are zero-bit counts for the BFs of partial unions $U_x := \bigcup_{l:\,x_l=1} S_l$:
\begin{align}
\hat{n}[F] \,= \sum_{y\in F^B} \hat{n}[P_y] \,&= 
\sum_{y\in F^B} \sum_{\updown{x:\, x\hspace{0.2pt} > \bar{0}\hspace{0.5pt},}{x\lor y \,=\, \bar{1}}} \frac{(-1)^{|x \land y| \,+\, 1}}{k\log\big(1\hspace{0.5pt}{-}\hspace{0.5pt}\frac{1}{m}\big)} \big(\log Z_x - \log m'\big)
\nonumber\\
{}&=\,\, \bm{a}^T \langle \log Z_x \rangle \,+\, \bm{b}.\label{eq:overlap:affine}
\end{align}
The same Delta method applied in the variance formula~(\ref{eq:basics:set-size-predictor-variance}) for the single-set size estimator can be used here to approximate the variance of $\hat{n}[F]$ as follows:
\begin{align}
\Var \hat{n}[F] \,&=\, \bm{a}^T \Cov \langle \log Z_x \rangle \,\bm{a}  \,\approx{}\nonumber\\
{}&\approx\, \bm{a}^T \diag\langle 1\,/\,\E Z_x \rangle \,\Cov \langle Z_x \rangle \,\diag\langle 1\,/\,\E Z_x \rangle \,\bm{a}\label{eq:overlap:delta}
\end{align}
The key step is computing the covariance matrix $\Cov \langle \log Z_x \rangle$, \ie{} computing each matrix entry $\Cov(Z_x, Z_{x'}\!)$ for any two partial unions \hbox{$S = U_x$} and \hbox{$S' = U_{x'}$} including the case \hbox{$x = x'$}:

\begin{claim}
\label{th:overlap:cov-z}
For sets $S, S' \subset \Omega$ let $Z(S)$ and $Z(S')$ be zero-bit counts observed as we sketch $S$ and $S'$ in a truncated Bloom filter with the same parameters $(m, k, \theta)$ and hash functions $h = (h_1, \ldots, h_k)$. In other words, let
\begin{equation*}
Z(S) \,:=\, m' - \big|h(S)|_{<\theta}\big| \quad\textrm{and}\quad Z(S') \,:=\, m' - \big|h(S')|_{<\theta}\big|\,,
\end{equation*}
where $m' = \big|\mathcal{H}|_{<\theta}\big|$ is the truncated hash range. Assuming that all hashes are random, independent, and uniformly distributed in~$\mathcal{H}$, the covariance between $Z(S)$ and $Z(S')$ is as follows:
\begin{align}
&\Cov\big(Z(S), Z(S')\big) \,\,=\,\, m' \Big[\Big(1\,{-}\,\frac{1}{m}\Big)^{k|S\hspace{0.5pt}\cup\hspace{0.5pt}S'|}  \!\!{-}\,\Big(1\,{-}\,\frac{1}{m}\Big)^{k(|S|\hspace{0.5pt}+\hspace{0.5pt}|S'|)} \Big] \,\,{+}\nonumber\\
&{+}\,\,m'(m'\,{-}\,1) \Big(1\,{-}\,\frac{1}{m}\Big)^{k(|S|\hspace{0.5pt}+\hspace{0.5pt}|S'|)}   \Big[ \Big(1\,{-}\,\frac{1}{(m\,{-}\,1)^2}\Big)^{k|S\hspace{0.5pt}\cap\hspace{0.5pt}S'|} \!\!{-}\, 1\Big].\label{eq:overlap:cov-z}
\end{align}
Note that if\/ $S = S'$ the covariance gives $\Var Z(S)$.
\end{claim}
\begin{proof} See Appendix~\ref{app:overlap:cov-z}. \end{proof}

\begin{corollary}
\label{th:overlap:cov-z-est}
When both $m$ and $m'$ are large, we can approximate~(\ref{eq:overlap:cov-z}) as follows: $\,\Cov\hspace{-0.5pt}\big(Z(S), Z(S')\big) \,\approx{}$
\begin{equation}
{\approx}\,\,\, \E Z(S\,{\cup}\,S') \,-\, \left(\frac{1}{m'} \,+\, \frac{k}{m^2}\,|S\,{\cap}\,S'|\right) \cdot \E Z(S) \,\E Z(S'). \label{eq:overlap:cov-z-appr}
\end{equation}
Here we used (\ref{eq:basics:set-size-predictor}) to replace $m'(1\,{-}\,1/m)^{k|S|}$ with $\E Z(S)$. If we do not know these expectations, we can estimate~(\ref{eq:overlap:cov-z-appr}) from the observed zero-counts: $\,\Cov\hspace{-0.5pt}\big(Z(S), Z(S')\big) \,\approx{}$
\begin{equation}
{\approx}\,\,\, Z(S\,{\cup}\,S') \,-\, \big(1/m' \,+\, k/m^2 \cdot \hat{n}[S\,{\cap}\,S']\big) \cdot Z(S) Z(S'),\label{eq:overlap:cov-z-est}
\end{equation}
where $\hat{n}[S\,{\cap}\,S']$ is estimated from $Z(S)$, $Z(S')$, and $Z(S\,{\cup}\,S')$ using inclusion-exclusion:
\begin{equation*}
\hat{n}[S\,{\cap}\,S'] \,=\, -\frac{m}{k}\Big(\log\frac{Z(S)}{m'} + \log\frac{Z(S')}{m'} - \log\frac{Z(S\,{\cup}\,S')}{m'}\Big).
\end{equation*}
\end{corollary}

If we use~(\ref{eq:overlap:cov-z-est}) to approximate all entries $\Cov(Z_x, Z_{x'}\!)$ in matrix $\Cov \langle \log Z_x \rangle$, as well as replace $1\,/\,\E Z_x$ with $1/Z_x$ in the diagonal matrices, then~(\ref{eq:overlap:delta}) becomes a practically useful estimator of $\Var \hat{n}[F]$ for any $\{\cup\,, \cap\,, \setminus\}$-expres\-sion $F$. Let us write it out for the important special case of pairwise overlap $\Var \hat{n}[S_1\,{\cap}\,S_2]$:

\begin{corollary}
\label{th:overlap:var-overlap}
For the overlap estimator $\hat{n}_\cap = \hat{n}[S_1\,{\cap}\,S_2]$ given by formula $\,- m/k\hspace{-0.5pt}\cdot\hspace{-0.5pt}\big(\log(Z_1/m') + \log(Z_2/m') - \log(Z_\cup/m')\big)$ where $Z_1$, $Z_2$, and $Z_\cup$ are zero-bit counts observed in truncated BFs for sets $S_1$, $S_2$, and $S_1\,{\cup}\,S_2$ respectively, the Delta method gives the following variance approximation:
\begin{equation}
\Var \hat{n}_\cap \,\approx\, \frac{m^2}{k^2} \left(\frac{2\hspace{0.5pt}Z_\cup}{Z_1 Z_2} + \frac{1}{Z_\cup} - \frac{1}{Z_1} - \frac{1}{Z_2} - \frac{1}{m'}\right) \,-\, \frac{\hat{n}_\cap}{k}.
\label{eq:overlap:var-overlap}
\end{equation}
\end{corollary}
\begin{proof} See Appendix~\ref{app:var-overlap}. \end{proof}

A Wald interval $\hat{n}_\cap \pm z_{1-\delta/2}\,\widehat{\sd}(\hat{n}_\cap)$ follows from~(\ref{eq:overlap:var-overlap}) under the i.i.d.-bit model (Appendix~\ref{app:bf-ci-two}).
The occupancy-inversion estimator attains the Cram\'er--Rao bound for the single-set size to first order (Proposition~\ref{prop:crlb:single-bf}); we do not claim that identification for the overlap functional itself.

\begin{remark}
\label{th:overlap:var-overlap-bound}
For Theta sketches, since we assume no collisions, their intersection is also a Theta sketch. If we denote by $\varsigma_{\cap}$ the number of hashes below~$\theta$ in that intersection, from the properties of the Binomial distribution we have:
\begin{equation*}
\hat{n}^{\mathrm{theta}}_\cap \,=\, \varsigma_{\cap} / \theta, \quad \Var \hat{n}^{\mathrm{theta}}_\cap \,=\, |S_1\,{\cap}\,S_2|\hspace{0.5pt}(1\,{-}\,\theta)/\theta \,\approx\, \varsigma_{\cap}(1\,{-}\,\theta)/\theta^2.
\end{equation*}
Theta-sketch variance provides the following lower bound on (\ref{eq:overlap:var-overlap}) if written with expectations, where $n_\cap = |S_1\,{\cap}\,S_2|$:
\begin{equation*}
\frac{m^2}{k^2} \left(\frac{2\,{\cdot}\hspace{0.5pt}\E Z_\cup}{\E Z_1 \E Z_2} + \frac{1}{\E Z_\cup} - \frac{1}{\E Z_1} - \frac{1}{\E Z_2} - \frac{1}{\theta m}\right) \,-\, \frac{n_\cap}{k} \,\geq\,
\frac{n_\cap\hspace{0.5pt}(1\,{-}\,\theta)}{k\hspace{0.5pt}{\cdot}\hspace{0.5pt}\theta}.
\end{equation*}
\begin{proof} See Appendix~\ref{app:overlap:var-overlap-bound}. \end{proof}
\end{remark}

\section{Compression}
\label{sec:compress}

The bit-vector of a Bloom filter can be viewed as a random Boolean vector of size~$m$ with a fixed number $C = |h(S)|$ of ones; for truncated BFs replace $m$ with~$m'$ and $h(S)$ with $h(S)|_{<\theta}$. The number of such equiprobable vectors is $\binom{m}{C}$, hence the number of bits to encode such vector without loss needs to be at least
\begin{gather*}
\big\lceil\log_2 {\textstyle\binom{m}{C}}\big\rceil \,\,=\,\, m\cdot H(C\,{/}\,m) \,+\, O(\log m), \;\;\textrm{where}\\
H(p) \, := {} - p \log_2 p - (1-p) \log_2 (1 - p) \;\,\textrm{is Shannon entropy.}
\end{gather*}
We can view $m\hspace{0.5pt}H(C/m)$ as the approximate lower bound on lossless compression size (``the Shannon bound''). A simple and efficient method to compress random vectors within a few percent of the Shannon bound is called Range Asymmetric Numeral Systems (rANS)~\cite{Duda2013:ans}. See Appendix~\ref{app:rans-bf} for a brief introduction.

An ordinary rANS algorithm that encodes characters one-by-one is less efficient for sparse bit-vectors with long stretches of zeros. Such bit-vectors are common in the context of Bloom filters and are often represented in ``run-length'' format as a sequence of integers $\langle r_1, r_2, \ldots, r_C, r_{C+1} \rangle$ where each $r_i$ is the number of consecutive zeros followed by the $i$-th one (or, in case of $r_{C+1}$, followed by the vector's end). Here we propose a slightly modified version of rANS that encodes an entire stretch of zeros ``in one go.'' It runs in $O\big(C \log(m/C)\big)$ steps on average, where $C$ is the number of ones. Our proposed rANS encoder is Algorithm~\ref{alg:compress:rans-encode} and the decoder is Algorithm~\ref{alg:compress:rans-decode-faster}. See Appendix~\ref{app:rans-bf} for more details.

\begin{algorithm}[!t]
\caption{Encodes a sparse bit vector represented in run-length format using rANS method.}
\label{alg:compress:rans-encode}
\begin{algorithmic}[1]%
\Require $\langle r_1, r_2, \ldots, r_C, r_{C+1} \rangle$ represents $m$-bit vector with $C$ ones;
         $r_i = {}$number of consecutive zeros followed by the $i$-th one (or, for $r_{C+1}$, by the vector's end). Assumes $2\hspace{0.5pt}C \leq m$.
\Ensure $v$ is the rANS-compressed bit vector.
\Procedure{RANS-Encode}{$r_1, r_2, \ldots, r_C, r_{C+1}$; $m$}
    \State $\xi \gets 2m - 1$; $\;v \gets {}$[$\,$]
    \ForAll{$i = 1, 2, \ldots, C, \,C+1$}
        \State $r \gets r_i$ \Comment{Encode the $i$-th run of $r_i$ zeros}
        \While{$r > 0$}
            \State $\displaystyle r_{\max} \gets \Big\lfloor \frac{m + \big\lfloor (2m - 1 - \xi) \,{/}\, 2 \big\rfloor}{C} \Big\rfloor - 1$
            \State \Call{Prepend-Bit}{$\xi \bmod 2^{\mathstrut} ;\; v$}
            \State $\xi \gets \lfloor\xi \,{/}\, 2\rfloor$
            \If{$r \geq r_{\max}$}
                \State $\xi \gets \xi + r_{\max}\hspace{0.5pt}C$; $\;r \gets r - r_{\max}$
            \Else
                \State $\xi \gets \xi + r\hspace{0.5pt}C$; $\;r \gets 0$
            \EndIf
        \EndWhile
        \If{$i \leq C$} \Comment{This run ends with a one}
            \While{$\xi \geq 2\hspace{0.5pt}C$}
                \State \Call{Prepend-Bit}{\hspace{0.5pt}$\xi \bmod 2 ;\; v$}
                \State $\xi \gets \lfloor \xi \,{/}\, 2 \rfloor$
            \EndWhile
            \State $\xi \gets \xi + 2(m - C)$
        \EndIf
    \EndFor
    \State \Call{Prepend-Integer}{\hspace{0.5pt}$\xi - m ;\; v$}; $\;$\Call{Prepend-Integer}{\hspace{0.5pt}$C ;\; v$}
    \State \textbf{return} $v$
\EndProcedure
\end{algorithmic}
\end{algorithm}

\begin{algorithm}[!t]
\caption{Faster decoder from rANS-compressed encoding to run-length representation.}
\label{alg:compress:rans-decode-faster}
\begin{algorithmic}[1]%
\Require $v = {}$the rANS-compressed form of sparse $m$-bit vector
\Ensure $\langle r_1, r_2, \ldots, r_C, r_{C+1} \rangle = {}$the bit vector in run-length format
\Procedure{RANS-Decode}{$v, m$}
    \State $C \gets {}$\Call{Read-Integer}{$v$}; $\;\xi \gets m + {}$\Call{Read-Integer}{$v$}
    \ForAll{$i = C+1,\, C, \ldots, 2, 1$}
        \State $r_i \gets 0$ \Comment{Decode the $i$-th run of zeros}
        \If{$\xi \leq 2m - 1 - C$}
            \State $q \gets \big\lfloor(\xi - m) \,{/}\, C\big\rfloor$; $\;q\hspace{0.5pt}C \gets q\cdot C$ \Comment{Once per~$i$}
            \Repeat
                \State $r_i  \gets r_i + q + 1$
                \State $\psi \gets (m \bmod C) + 2(\xi - m - q\hspace{0.5pt}C) + {}$\Call{Read-Bit}{$v$}
                \State $\xi  \gets \psi + \big(2(m - C) - (m \bmod C)\big)$ \Comment{$\psi + {}$const}
                \State $q \gets \lfloor m/C \rfloor - 2$; $\;q\hspace{0.5pt}C \gets q\cdot C$ \Comment{constants}
                \IfThen{$\psi \geq  C$}{$\,\langle q,\, q\hspace{0.5pt}C\rangle \gets \langle q + 1,\, q\hspace{0.5pt}C + C\rangle$}
                \IfThen{$\psi \geq 2C$}{$\,\langle q,\, q\hspace{0.5pt}C\rangle \gets \langle q + 1,\, q\hspace{0.5pt}C + C\rangle$}
            \Until{$\xi > 2m - 1 - C$}
        \EndIf
        \State $\xi \gets \xi - 2(m - C)$
        \State $\nu \gets \lfloor\log_2(m + 1)\rfloor - \lfloor\log_2(\xi + 1)\rfloor$ \Comment{Use bit counts}
        \IfThen{$\nu > 0$}{$\xi \gets (\xi \mathbin{\texttt{<\hspace{-0.5pt}<}} \nu) + {}$\Call{Read-Bits}{$\nu\hspace{0.5pt}; v$}}
        \While{$\xi < m$}
            \State $\xi \gets 2\xi + {}$\Call{Read-Bit}{$v$}
        \EndWhile
    \EndFor
    \State \textbf{return} $r_1, r_2, \ldots, r_C, r_{C+1}$
\EndProcedure
\end{algorithmic}
\end{algorithm}

\section{Blocked Bloom Filters}
\label{sec:block}

A large BF implemented as one block is hard to store and parallelize in the context of database query execution. For example, sketching a set of size $10^9$ items with FPP${} = 1\%$ calls for $k = 7$ hash functions and $m = 9.6 \cdot 10^9$ bits (1.116~GiB) of uncompressed memory. The same $k$ and~$m$ applied to smaller sets result in a sparse bit-vector that can be compressed up to its Shannon entropy size lower bound, which for sets of 10M and 50M items gives 71~MiB and 255~MiB respectively. %
Even 255~MiB size is too large to keep in a single cell of a relational table or to process efficiently with a user-defined function in a SQL query. 

A natural solution is to partition the BF bit-vector into blocks and store each block in its own table cell. How we do it, though, significantly influences the efficiency of our implementation. If each BF block can store hashes from any set item, a membership query requires inspecting up to $k$ blocks per item, performing a join computation of size up to $nk$ where $n = {}$query size and $k = {}$the number of BF hash functions. The same $nk$-size join appears when creating a BF for a set of size~$n$. A better approach is to ensure that all hashes of one item are stored in a single, easily identifiable BF block. This motivates the following definition:

\begin{definition}
\label{def:block:mbf}
A \emph{multi-block} Bloom filter (MBF) over a universe $\Omega$ with parameters $(b, m, k)$ is a collection of $b$ blocks where each block $B_j$ is a Bloom filter with the same hash range $\mathcal{H}$ of size $m$ and with $k$ hash functions $h|_j := (h_1|_j,\ldots, h_k|_j)$ that map $\Omega$ to~$\mathcal{H}$. Addi\-ti\-onally, there is a hash function $h_0:\Omega\to\{1, \ldots, b\}$ that determines which block stores the hashes of a given item. To simplify notation, we denote by $h_i$ functions such that $h_i(v) := h_i|_{h_0(v)}(v)$ and use the same $(h_1,\ldots, h_k)$ for all blocks.

The MBF instance for a given set $S\subset\Omega$ is a tuple $(H|_1, \ldots, H|_b)$ where $H|_j := \cup_{i=1}^k h_i(S|_j)$ and $S|_j := \{v \in S: h_0(v)=j\} = S \cap \Omega|_j$. In other words, $S|_j$ are the items of $S$ hashed into block~$B_j$ and $H|_j$ are their hashes stored in~$B_j$. Notation ``$|_j$'' constrains the set items and their hashes to block~$B_j$.
\end{definition}

This is the item-routed blocked layout of Putze et al.~\cite{Putze2007:cache}, which we use with database-scale blocks instead of cache-sized ones. We keep the abbreviation MBF for this layout.

An MBF membership query for a given set $S'\subset\Omega$ proceeds in two steps. First, for each item $v\in S'$ we evaluate $j = h_0(v)$ and match with its block $B_j$ in a join operation. Second, given a pair $(v, B_j)$, we compute the $k$ hashes $h_i(v)$ and return ``positive'' iff $\forall i=1, \ldots, k$: $h_i(v)\in H|_j$. The one-sided guarantee is the same as for a single-block BF, and the total false positive probability equals the average one-block FPP across all blocks. For efficiency, we can group together all items matching a given block and check their membership in a separate process per block, making it easy to parallelize.

As with ordinary BFs, we do not have to read and aggregate all $b$ blocks before computing a set size prediction. Suppose we have only seen the first $b'<b$ blocks $B_1, \ldots, B_{b'}$ of the sketch for set~$S$. Let $\varsigma|_{\leq b'}$ denote the sum-of-bits statistic aggregated over the first $b'$ blocks. Inverting the formula for $\E\varsigma|_{\leq b'}$ gives us the following set size predictor:
\begin{multline}
\hat{n}[S] \,\,=\,\, \log\Big(1 - \frac{\varsigma|_{\leq b'}}{b'm}\Big) \,\,\Big/\,\, \log\Big[\,\frac{b-1}{b} \;+ \; \frac{1}{b}\cdot \Big(1 - \frac{1}{m}\Big)^k\,\Big]\,\,={}\\
{}=\,\, \logonep\Big({-}\,\frac{\varsigma|_{\leq b'}}{b'm}\Big)\,\,\Big/\,\, \logonep \Big[\,\frac{1}{b} \,\expmone\Big(k\, \logonep\Big({-}\,\frac{1}{m}\Big)\Big)\Big]
\label{eq:block:mbf-size-prime}
\end{multline}
where functions $\expmone(x) := e^x - 1$ and $\logonep(x) := \ln(1 + x)$ help reduce roundoff errors. See Appendix~\ref{app:block:cardinal} for the derivation.

Algorithm~\ref{alg:block:mbf-predict} summarizes the above discussion and provides the steps to predict set size for an $\{\cup\,, \cap\,, \setminus\}$-formula over MBF sketches after reading only the first $b'$ blocks with estimator~(\ref{eq:block:mbf-size-prime}).

\begin{algorithm}[!t]
\caption{Predicts size of the set defined by a $\{\cup\,, \cap\,, \setminus\}$-for\-mula $F(S_1, \ldots, S_L)$ over multi-block BF sketches from hashed union sizes}
\label{alg:block:mbf-predict}
\begin{algorithmic}[1]%
\Require%
\begin{itemize}\setlength{\itemindent}{-44pt}    %
\item $F(S_1, \ldots, S_L)$ given by all IE-partitions it has: $F^B \subset \{0, 1\}^L$
\item Seen hashed union sizes: $\varsigma_x = \sum_{j=1}^{b'} \big|h(U_x|_j)\big|$ for $x \,{\in}\,\{0, 1\}^L$
\item MBF blocks: $b$ all, $b'$ seen; $m$ bits per block, $k$ hash functions
\end{itemize}
\Ensure $\hat{n}[F]$ approximates the predicted size of $F(S_1, \ldots, S_L)$
\Procedure{MBF-Predict}{$F, \langle \varsigma_x \rangle_{x \in \{0, 1\}^L\,}, b, b', m, k$}
    \State $\alpha \gets \logonep \big[1/b \cdot \expmone\big(k\, \logonep(-1/m)\big)\big]$
    \State $\forall y \in\{0, 1\}^L \setminus \{\bar{0}\}: \,\hat{n}[P_y] \gets 0$ \Comment{$\hat{n}[P_y]$ will predict $|P_y|$}
    \ForAll{$x\in \{0, 1\}^L \setminus \{\bar{0}\}$}
        \If{$\varsigma_x < b'm$}
            \State $\hat{n}[U_x] \gets \logonep\big({-}\,\varsigma_x\mathop{/}(b'm)\big)\mathop{/}\alpha$
        \Else
            \State $\hat{n}[U_x] \gets +\infty$ \Comment{No upper limit for $|U_x|$}
        \EndIf
        \ForAll{$y\in \{0, 1\}^{\mathstrut L} \setminus \{\bar{0}\}$ such that $x\,{\lor}\,y = \bar{1}$}
            \State $\hat{n}[P_y] \gets \hat{n}[P_y] + (-1)^{\mathstrut|x\land y| \,+\, 1}\!\cdot \hat{n}[U_x]$
        \EndFor
    \EndFor
    \State \textbf{return} $\hat{n}[F] \gets \sum_{y\in F^B} \hat{n}[P_y]$
\EndProcedure
\end{algorithmic}
\end{algorithm}

\subsection{Variance Estimation for MBFs}
\label{sec:block:variance}

The Delta method from Section~\ref{sec:overlap:variance} extends naturally to multi-block Bloom filters. For any $\{\cup\,, \cap\,, \setminus\}$-expres\-sion $F(S_1, \ldots, S_L)$ over $L$ sets represented by their MBFs, we have the following affine cardinality estimator:
\begin{align}
\hat{n}[F] \,= \sum_{y\in F^B} \hat{n}[P_y] \,&= 
\sum_{y\in F^B} \sum_{\updown{x:\, x\hspace{0.2pt} > \bar{0}\hspace{0.5pt},}{x\lor y \,=\, \bar{1}}} \frac{(-1)^{|x \land y| \,+\, 1}}{\log p_0} \big(\log Z_x - \log m'\big)
\nonumber\\
{}&=\,\, \bm{a}^T \langle \log Z_x \rangle \,+\, \bm{b}.\label{eq:block:affine}
\end{align}
Here $Z_x := m' - \varsigma_x|_{\leq b'}$ is the zero-bit count observed within the first $b'\leq b$ blocks of the MBF for partial union $U_x := \bigcup_{l:\,x_l=1} S_l$, where $x\in\{0, 1\}^L$ excluding $\bar{0}$. The variance formula~(\ref{eq:overlap:delta}) from Section~\ref{sec:overlap:variance} applies without change. However, the covariance matrix $\Cov \langle \log Z_x \rangle$ in the MBF case has somewhat different entries. Let us give a formula for their estimation.

\begin{claim}
\label{th:block:cov-z}
Suppose we hash two sets $S, S' \subset \Omega$ into multi-block BFs with the same parameters $(b, m, k)$ and the same hash functions. Let $Z(S)$ and $Z(S')$ be zero-bit counts observed within the first $b'\leq b$ blocks of these MBFs. Assuming that all hashes are random, independent, and uniformly distributed, the covariance between $Z(S)$ and $Z(S')$ is as follows:
\begin{align}
&\Cov\big(Z(S), Z(S')\big) \,\,=\,\,\, b'm \, p_0^{|S \,\cup\, S'|} \,+\, \big[(b'\,{-}\,1)m\,(p_d/p_0^2)^{|S\,\cap\,S'|} \,+{}\nonumber\\
&\qquad{} +\, (m - 1)\, (p_s/p_0^2)^{|S\,\cap\,S'|} \,-\, b'm \big] \cdot b'm \,p_0^{|S| + |S'|}. \,
\label{eq:block:cov-z}
\end{align}
Here $p_0$, $p_d$, and $p_s$ denote the following expressions:
\begin{align*}
p_0 \,\,&:=\,\, (b-1) / b \,+\, 1/b \cdot (1-1/m)^k \\[-3pt]
p_d \,\,&:=\,\, (b-2) / b \,+\, 2/b \cdot (1-1/m)^k \\[-3pt]
p_s \,\,&:=\,\, (b-1) / b \,+\, 1/b \cdot (1-2/m)^k
\end{align*}
Note that if\/ $S = S'$ the covariance gives $\Var Z(S)$.
\end{claim}
\begin{proof} See Appendix~\ref{app:block:cov-z}. \end{proof}

\begin{corollary}
\label{th:block:cov-z-est}
When $m \gg k^2$, we can approximate~(\ref{eq:block:cov-z}) as follows:
\begin{align}
&\Cov\hspace{-0.5pt}\big(Z(S), Z(S')\big) \,\,\approx\,\, \E Z(S\,{\cup}\,S') \,+{}\nonumber\\
&\quad{}+\, \frac{1}{b'm}\Big(\frac{(b - b')k^2 - bk}{b^2 m}\,|S\,{\cap}\,S'| \,-\, 1\Big) \cdot \E Z(S) \,\E Z(S').
\label{eq:block:cov-z-appr}
\end{align}
If we do not know these expectations, we can estimate~(\ref{eq:block:cov-z-appr}) from the observed zero-counts: 
\begin{align}
&\Cov\hspace{-0.5pt}\big(Z(S), Z(S')\big) \,\,\approx\,\, Z(S\,{\cup}\,S') \,+{}\nonumber\\
&\quad{}+\, \frac{1}{b'm}\Big(\frac{(b - b')k^2 - bk}{b^2 m}\,\hat{n}[S\,{\cap}\,S'] \,-\, 1\Big) \cdot Z(S) \,Z(S').
\label{eq:block:cov-z-est}
\end{align}
where $\hat{n}[S\,{\cap}\,S']$ is estimated from $Z(S)$, $Z(S')$, and $Z(S\,{\cup}\,S')$ using inclusion-exclusion:
\begin{equation*}
\hat{n}[S\,{\cap}\,S'] \,=\, \frac{1}{\log p_0}\Big(\log\frac{Z(S)}{b'm} + \log\frac{Z(S')}{b'm} - \log\frac{Z(S\,{\cup}\,S')}{b'm}\Big).
\end{equation*}
\end{corollary}
\begin{proof} See Appendix~\ref{app:block:cov-z-est}. \end{proof}

Let us write out the variance of pairwise overlap $\Var \hat{n}[S_1\,{\cap}\,S_2]$ for MBFs with parameters $(b, m, k)$:

\begin{corollary}
\label{th:block:var-overlap}
For the overlap estimator $\hat{n}_\cap = \hat{n}[S_1\,{\cap}\,S_2]$ given by formula $\,1/\log p_0\hspace{-0.5pt}\cdot\hspace{-0.5pt}\big(\log Z_1$ $+$ $\log Z_2$ $-$ $\log Z_\cup$ $-$ $\log(b'm)\big)\,$ where $Z_1$, $Z_2$, and $Z_\cup$ are zero-bit counts observed within the first $b'$ blocks of MBFs for sets $S_1$, $S_2$, and $S_1\,{\cup}\,S_2$ respectively, and if $m \gg k^2$, the Delta method gives the following variance approximation: $\,\Var \hat{n}_\cap \,\,{\approx}$
\begin{equation}
\left(\frac{bm}{k}\right)^2 \!\left(\frac{2\hspace{0.5pt}Z_\cup}{Z_1 Z_2} \,{+}\, \frac{1}{Z_\cup} \,{-}\, \frac{1}{Z_1} \,{-}\, \frac{1}{Z_2} \,{-}\, \frac{1}{b'm}\right) \,+\, \left(\frac{b(k\,{-}\,1)}{b'k} - 1\right)\hat{n}_\cap.
\label{eq:block:var-overlap}
\end{equation}
\end{corollary}
\begin{proof} See Appendix~\ref{app:block:var-overlap}. \end{proof}

Under the i.i.d.-bit model the corresponding Cram\'er--Rao bound on set-size estimation scales as $1/b'$ and diverges as blocks saturate (Proposition~\ref{prop:crlb:mbf}), which is why the accuracy dial~(\ref{eq:accur:b-prime-overlap}) scales as $1/\sigma^2$.
A Wald interval for $\hat{n}_\cap$ uses~(\ref{eq:block:var-overlap}) (Appendix~\ref{app:bf-ci-two}).
The second-order bias of $\hat{n}_\cap$ is of the order of a few items, negligible relative to $\sigma$; clamping negative estimates adds bias only when $n_\cap$ is within about $2\sigma$ of zero (Appendix~\ref{app:stt:bias}).

\section{Accuracy vs.\ Efficiency Tradeoff}
\label{sec:accur}

At production scale, we compute overlaps over thousands of sketched sets, each containing up to billions of items. To enable membership queries with a reasonable FPP, each sketch consumes a lot of memory. Cross-overlapping thousands of MBFs to answer, \eg, an audience search query could take hours.

Fortunately, most audience search queries do not need maximum overlap precision. If we can meet our business goal by estimating overlaps at a lower accuracy, we can achieve $\sim$10${\times}$-100${\times}$ speedup by processing only enough MBF blocks to reach the required precision. The block partitioning framework and Algorithm~\ref{alg:block:mbf-predict} already let us estimate cardinality from fewer blocks; in this section we turn that capability into a principled, accuracy-driven policy.

\subsection{Accuracy Limited Overlaps}
\label{sec:accur:limit}

Suppose we want to estimate the cardinality of a $\{\cup\,, \cap\,, \setminus\}$-expres\-sion $F(S_1, \ldots, S_L)$ with the standard deviation of~$\sigma$ from the sets' MBF sketches. We plan on using Algorithm~\ref{alg:block:mbf-predict} that computes estimator~$\hat{n}[F]$. Applying the Delta method~(\ref{eq:overlap:delta}) to its affine form~(\ref{eq:block:affine}) gives the estimated variance $\Varh\hat{n}[F]$:
\begin{equation}
\Varh\hat{n}[F] \,:=\, \bm{a}^T \diag\langle 1/Z_x \rangle \,\Cov \langle Z_x \rangle \,\diag\langle 1/Z_x \rangle \,\bm{a},\label{eq:accur:delta}
\end{equation}
with matrix $\Cov \langle Z_x \rangle$ instantiated as in~(\ref{eq:block:cov-z-est}) from Corollary~\ref{th:block:cov-z-est}. $\Varh\hat{n}[F]$ is a function of observed zero-bit counts and of the MBF parameters including~$b'$, the number of processed BF-blocks. Given our target variance $\sigma^2$, we can solve equation $\Varh\hat{n}[F] = \sigma^2$ for $b'$ and estimate the minimum number of BF-blocks required to meet the target variance, providing us with an \emph{accuracy dial}.

Solving for $b'$ in general form can be done numerically, \eg{} by substituting all $b' = 1\ldots b$ and picking the smallest $b'$ such that $\Varh\hat{n}[F] \leq \sigma^2$. Since we have an explicit formula~(\ref{eq:block:var-overlap}) for the variance of pairwise overlaps, let's solve it for~$b'$ explicitly as well:
\begin{align}
&b'  \,=\,\, \min\left\{b,\,\left\lceil\frac{g \cdot b^2 m / k^2 \,+\, \hat{n}_\cap \cdot b(k\,{-}\,1)/k}{\sigma^2 \,+\, \hat{n}_\cap}\right\rceil\,\right\}, \;\; \textrm{where}\label{eq:accur:b-prime-overlap}\\
&g  \,:=\, \frac{2\hspace{0.5pt}z_\cup}{z_1 z_2} \,{+}\, \frac{1}{z_\cup} \,{-}\, \frac{1}{z_1} \,{-}\, \frac{1}{z_2} \,{-}\, 1,
\quad z_i := \frac{Z_i}{b'm},\quad z_\cup := \frac{Z_\cup}{b'm}\nonumber
\end{align}
Here $b'$ must be clamped to $1\le b'\le b$, because some target accuracies are simply out of reach: they need more blocks than exist. The key to solving~(\ref{eq:accur:b-prime-overlap}) is that the fractions $z_x\,{:=}\,Z_x/(b'm)$ estimate the probability of a bit being zero in their corresponding sketches, so their expectations do not depend on~$b'$.

We still need to compute the $z_x$ values for $\Varh\hat{n}[F]$ before we know how many blocks to read.
Our solution is to seed zero-bit counts by reading a small fixed number of blocks needed anyway for our $\hat{n}[F]$ estimation. We use them to estimate $b'$, and later as part of the final counts. But with so few blocks, the $Z_x$ are very noisy, and some can take extreme values.
To improve stability, we regularize the $z_x$ estimates toward a baseline set configuration that makes sense in our business application.
We revisit the $b'$ estimate once we have read more blocks and obtained more accurate $Z_x$ counts, then read further blocks if the revised $b'$ calls for them. This leads us to the following three-phase approach to accuracy limited estimation:
\begin{enumerate}[leftmargin=*]
\item \textbf{Seed.} Read one or a few leading blocks per MBF and use to initialize the $Z_x$ counts; map them to bit probabilities via regularized estimation; produce a first estimate of~$b'$.
\item \textbf{Main.} Read the remaining blocks up to $b'$; recompute the counts from the full read; re-estimate $b'$ from these more accurate counts, now without regularization.
\item \textbf{Top-up.} If the re-estimated $b'$ grew beyond what was read, process the extra required blocks; then estimate $\hat{n}[F]$ at the target variance level.
\end{enumerate}
This approach could be refined by estimating the variance of $b'$ and running the last phase in a loop until a safety criterion is reached. 
However, in an RDBMS additional phases may hamper batching and scaling to many parallel block-sharing operations.

Because \textsc{Top-up} reads fresh blocks and the budget depends on zero-fractions that concentrate, the first-order variance remains asymptotically valid for the data-dependent number of blocks (Appendix~\ref{app:stt:three-phase}). Finite-sample coverage under per-block adaptive stopping would require a confidence-sequence construction~\cite{Howard2021:confidence}.

While the three-phase approach is feasible for any $\{\cup,\cap,\setminus\}$-expression with few enough sets, this paper implements only the practically important special case of {\bf pairwise cross-overlaps}.

Our Phase~1 regularized estimate for overlaps assumes that the two sets are disjoint. This is useful for three reasons. First, most audience pairs have low overlap compared to their sizes. Second, many search queries only test if a pair's overlap clears a minimum threshold, placing low-overlap pairs under greater scrutiny. Finally, under disjointness the algebra collapses to a remarkably simple form that enables extra optimization.

Recall from Section~\ref{sec:block} that $\E Z(S) = b'm\hspace{0.5pt}p_0^{|S|}$, so for the sketch's zero-bit fraction $z_S\,{:=}\,Z(S)/(b'm)$ we have $\E z_S = p_0^{|S|}$. 
If~$S_1$ and $S_2$ are disjoint, then $|S_1 \cup S_2|$ $=$ $|S_1| + |S_2|$, and hence
\begin{equation*}
\E z_\cup \,=\, p_0^{|S_1| + |S_2|} \,=\, p_0^{|S_1|} p_0^{|S_2|} \,=\, \E z_1 \, \E z_2.
\end{equation*}
When performing regularized estimation in Phase~1, we disregard the zero-bit count $Z_\cup$ and instead set $z_\cup$ to satisfy the above exactly: $z_\cup^{\mathrm{reg}} := z_1\hspace{0.5pt}{\cdot}\hspace{0.5pt}z_2$. Define the \emph{odds} of a bit being one vs{.} zero as
\begin{equation}
o_i \,:=\, \frac{1 - z_i}{z_i} \,=\, \frac{m - Z(S_i)}{Z(S_i)}, \qquad i \in \{1, 2\},\label{eq:accur:odds}
\end{equation}
computed from the single first block of each sketch. Substituting $z_\cup = z_1 z_2$ and $\hat{n}_\cap = 0$ into~(\ref{eq:accur:b-prime-overlap}) makes the variance factor~$g$ collapse to $g = o_1 o_2$, giving the closed-form Phase~1 estimate:
\begin{equation}
b' \,=\, \min\bigl\{\,b,\ \lceil b_{\mathrm{ref}}\cdot o_1 o_2 \rceil\,\bigr\}, \qquad b_{\mathrm{ref}} \,:=\, (b^2 m)\,/\,(k^2 \sigma^2).\label{eq:accur:b-prime-disjoint}
\end{equation}
Crucially, $o_1$ and $o_2$ are \emph{per-audience} quantities read from a single block each, so every pair's Phase~1 budget is known without ever forming the $|L| \times |R|$ product of cross-overlap pairs~--- the fact that facilitates SQL scaling in Section~\ref{sec:accur:scale}.

Once we read the first $b'$ blocks during Phase~2, we update zero-bit counts $\langle Z_1, Z_2, Z_\cup \rangle$ and estimate $z_\cup$ as $Z_\cup/(b'm)$ without regularization. As discussed in Sec{.}~\ref{sec:exper}, the disjoint baseline usually \emph{under}-estimates the required $b'$, which is good because it prevents Phase~2 from doing unnecessary work. Then $b'$ is re-estimated from the observed counts using the full formula~(\ref{eq:accur:b-prime-overlap}), and Phase~3 reads the extra blocks it calls for.

To summarize, accuracy limited pairwise overlap estimation proceeds in three phases with the disjoint baseline of~(\ref{eq:accur:b-prime-disjoint}):
\begin{enumerate}[leftmargin=*]
\item \textbf{Seed.} Read one block per MBF, initialize the running $Z_1, Z_2, Z_\cup$ counts; compute the odds $o_1, o_2$; set $b' \!= \min\{b, \lceil b_{\mathrm{ref}}\, o_1 o_2\rceil\}$. %
\item \textbf{Main.} Read $b'\,{-}\,1$ block pairs; update the $Z_1, Z_2, Z_\cup$ counts; estimate $\hat{n}_\cap$ and use Eq{.}~(\ref{eq:accur:b-prime-overlap}) to get a refined $b'$~--- let's call it~$b''\!$.
\item \textbf{Top-up.} If $b'' > b'$, read the extra $b'' - b'$ block-pairs; finally, compute $\hat{n}_\cap$ from all blocks read.
\end{enumerate}
Note that the $b'$ blocks read for a pair of MBFs need not be blocks $1, \ldots, b'$: any subset of $b'$ distinct block-IDs is distributionally equivalent.
Section~\ref{sec:accur:scale} exploits this for block reuse across pairs.

\subsection{Scaling Overlaps in SQL}
\label{sec:accur:scale}

In marketing, a single similarity search query may need to estimate a large number of audience overlaps. Our focus here is the \emph{cross-overlap}: given a left collection $L$ and a right collection $R$ of MBF sketches over the same parameters $(b, m, k)$, estimate $\hat{n}_\cap[l, r]$ for every pair $(l, r) \in L \times R$. Given the target variance $\sigma^2$, each pair has its own number $\bprime{l,\,r}$ of BF-blocks required for $\hat{n}_\cap[l, r]$ to meet the variance constraint. With hundreds of audiences per side, 
the challenge shifts to limiting and reusing block reads across them.

We target a columnar cloud RDBMS (Snowflake in our deployment) where each MBF is identified by its $\mathit{bloom\_id}$ and stored one block per row, keyed by $\mathit{block\_id}$ $=$ the hash $h_0(v)$. Per-block arithmetic is expressed in SQL with a few Java UDFs that deserialize block binaries into RAM and perform operations on them. In particular, \textsc{Bit-Counts} UDF returns the sums of bits for a pair of same-ID BF-blocks and their union. Clustering the block rows by $(\mathit{bloom\_id}, \mathit{block\_id})$ lets the engine prune micro-partitions for sketches and blocks that a batch does not touch, translating the reductions below into real I/O savings.

Two costs dominate a cross-overlap: reading the BF-blocks and performing joins over large tables. The highest cost is reading block-pairs, deserializing, and counting their bits; the surrounding arithmetic is negligible by comparison. 
The other %
is incurred by the RDBMS itself: the cost of BF-block retrieval from large tables during join execution. We keep it in check by processing the sweep in batches (Alg{.}~\ref{alg:accur:sweep}) and materializing smaller per-batch intermediates. 
The rest of this subsection discusses optimizations against these costs, stated as general principles and then tied to the concrete realization mechanism.

Algorithms~\ref{alg:accur:cross} and~\ref{alg:accur:sweep} give the full cross-overlap procedure\footnote{For clarity we index the $b$ blocks of each MBF from $1$ to $b$; the implementation indexes them from $0$ to $b - 1$.}. Algorithm~\ref{alg:accur:cross} is the driver and follows the three phases of Section~\ref{sec:accur:limit} over all pairs at once.

\begin{algorithm}[!t]
\caption{Accuracy-limited cross-overlaps: per MBF pair, process only enough block-pairs to reach $\Var\hat{n}_\cap \leq \sigma^2$}
\label{alg:accur:cross}
\begin{algorithmic}[1]%
\Require%
\begin{itemize}\setlength{\itemindent}{-44pt}
\item Left and right MBF collections $L$, $R$ with parameters $(b, m, k)$
\item Target st.\ dev.\ $\sigma \geq 1$;\; $\Call{clamp}{x} := \min\{b, \max\{1, \lceil x \rceil\}\}$
\end{itemize}
\Ensure $\hat{n}_\cap[l, r]$ approximates $|S_l \cap S_r|$ for every pair $(l, r) \in L \times R$
\Procedure{Cross-Overlaps}{$L,\, R,\, \sigma;\,\, b, \,m, \,k$}
    \State $b_{\mathrm{ref}} \gets b^2 m \,/\, (k^2 \sigma^2)$
    \LComment{\textsc{Seed}\hspace{0.5pt}: read one block per MBF, get odds needed for $b'$}
    \ForAll{$(l, r) \in L \times R$}
        \State $\!\!\langle \Cx{L}{}\!, \Cx{R}{}\!, \Cx{\cup}{}\rangle_{l,\,r} \gets \Call{Bit-Counts}{l[1], r[1]}$ \Comment{read block $1$}
    \EndFor
    \ForAllDone{$l \in L$}{$o_l \gets \Cx{L}{l,\,\forall} / (m - \Cx{L}{l,\,\forall})$} \Comment{1-vs-0-bit odds}
    \ForAllDone{$r \in R$}{$o_r \gets \Cx{R}{\forall,\,r} / (m - \Cx{R}{\forall,\,r})$} \Comment{of block 1 only}
    \ForAllDone{$(l, r) \in L \times R$}{$\bprime{l,\, r} := \Call{clamp}{b_{\mathrm{ref}}\cdot o_l\, o_r}$}
    \LComment{\textsc{Main}\hspace{0.5pt}: forward sweep to each pair's disjoint budget $b'$}
    \State \Call{Batched-Sweep}{$\pi \gets \langle 2, \ldots, b\rangle;\; b'\,{-}\,1,\, o,\, \textsc{use-}\Gamma\hspace{-0.5pt}\gets{}$\textbf{true}}
    \LComment{Re-estimate each pair's budget from observed counts by solving~(\ref{eq:accur:b-prime-overlap}); schedule the shortfall}
    \ForAll{$(l, r) \in L \times R$}
        \State $\bpprime{l,\, r} \gets \Call{Required-Blocks}{\langle \Cx{L}{}\!, \Cx{R}{}\!, \Cx{\cup}{}\rangle_{l,\,r}\hspace{0.5pt};\, \bprime{l,\, r}, \sigma}$
        \State $\varDelta b_{l,\, r} \gets \max\{0,\; \bpprime{l,\, r} - \bprime{l,\, r}\}$ \Comment{missing blocks}
    \EndFor
    \LComment{\textsc{Top-up}\hspace{0.5pt}: reverse sweep over trailing blocks for the shortfall}
    \State \Call{Batched-Sweep}{$\pi \gets \langle b, \ldots, 1\rangle;\; \varDelta b,\, o,\, \textsc{use-}\Gamma\hspace{-0.5pt}\gets{}$\textbf{false}}
    \State \textbf{return} $\hat{n}_\cap[l, r] \gets \Call{Overlap}{\langle \Cx{L}{}\!, \Cx{R}{}\!, \Cx{\cup}{}\rangle_{l,\,r}\hspace{0.5pt}; \, \bprime{l,\, r}\,{+}\,\varDelta b_{l,\, r}}$  %
\EndProcedure
\end{algorithmic}
\end{algorithm}

\begin{algorithm}[!t]
\caption{Batched sweep with adaptive batch sizing and an optional $\Gamma$-shape filter; block cache budget $W$}
\label{alg:accur:sweep}
\begin{algorithmic}[1]%
\Require Block order $\pi$; remaining block budgets $\varDelta b_{l,\, r}$; flag \textsc{use-$\Gamma$}
\Ensure Each pair's next $\varDelta b_{l,\, r}$ blocks along $\pi$ are added to the aggregated bit-counts $\langle \Cx{L}{}\!, \Cx{R}{}\!, \Cx{\cup}{}\rangle_{l,\,r}$
\Procedure{Batched-Sweep}{$\pi;\, \varDelta b,\, \textrm{odds}, \, \textsc{use-$\Gamma$}$}
    \ForAllDone{$\mathrlap{l}\hphantom{r} \in L$}{$\beta_l \gets \max_{r\in R} \,\varDelta b_{l,\, r}$} \label{alg:accur:sweep:beta-l}
    \ForAllDone{$r \in R$}{$\beta_r \gets \max_{l\in L} \,\varDelta b_{l,\, r}$} \label{alg:accur:sweep:beta-r}
    \State $t \gets 0$
    \While{$\exists\, a$ with $\beta_a > t$} \Comment{some MBF still needs blocks}
        \State $L^{\mathrm{act}} \gets \{l : \beta_l > t\}$;\;\; $R^{\mathrm{act}} \gets \{r : \beta_r > t\}$ \Comment{active MBFs} \label{alg:accur:sweep:active-mbfs}
        \LComment{Adaptive width (number of blocks per pair in batch):}
        \State $w \gets \max\big\{1,\, \min\{\,|\pi| - t,\; \lfloor W / \max(|L^{\mathrm{act}}|, |R^{\mathrm{act}}|)\rfloor\,\}\big\}$ \label{alg:accur:sweep:adaptive-width}
        \If{\textsc{use-$\Gamma$}}
            \State rank $L^{\mathrm{act}}$ and $R^{\mathrm{act}}$ by odds descending
            \State $(i^{\ast}\!, j^{\ast}) \gets \Call{$\Gamma$-Shape-Split}{L^{\mathrm{act}}, R^{\mathrm{act}}, \rank, \varDelta b, t}$
        \Else
            \State $(i^{\ast}\!, j^{\ast}) \gets (0, |R^{\mathrm{act}}| + 1)$ \Comment{full active grid, any rank}
        \EndIf
        \LComment{Process $\Gamma$-cover pairs made of two disjoint rectangles:}
        \ForAll{$(l, r): \genfrac{[}{]}{0pt}{0}{l \in L^{\mathrm{act}}\;\textrm{and}\;\rank(r) < j^{\ast},\;\textrm{or}}{\rank(l) \leq i^{\ast}\;\textrm{and}\;\rank(r) \geq j^{\ast}}$}
            \ForAll{positions $t' \in (t,\, t\,{+}\,w]$ with $t' \leq \varDelta b_{l,\, r}$} \label{alg:accur:sweep:interval}
                \State $\langle \Cx{L}{}\!, \Cx{R}{}\!, \Cx{\cup}{}\rangle_{l,\,r} \;\,{+}{=}\;\, \Call{Bit-Counts}{l[\pi_{t'}], r[\pi_{t'}]}$ \label{alg:accur:sweep:bit-counts}
            \EndFor
        \EndFor
        \State $t \gets t + w$
    \EndWhile
\EndProcedure

\LComment{Finds minimal two-rectangle cover of the active staircase:}
\Procedure{$\Gamma$-Shape-Split}{$L^{\mathrm{act}},\, R^{\mathrm{act}},\, \rank,\, \varDelta b,\, t$}
    \State $\rho    \gets (\rank\,\textrm{on}\;R^{\mathrm{act}})^{-1}$ so that $\forall r \in R^{\mathrm{act}}\!: \, \rho(\rank(r)) = r$
    \ForAll{ranked positions $j = 1, \ldots, |R^{\mathrm{act}}|$}
        \State $\yes(j) \gets \big|\{\,l \in L^{\mathrm{act}} : \varDelta b_{l,\, \rho(j)} > t\,\}\big|$ \Comment{staircase height, non-increasing in $j$}
    \EndFor
    \LComment{Minimize area of two rectangles holding all active pairs:} %
    \State $j^{\ast} \gets \argmin_{j}\big[\,|L^{\mathrm{act}}|\cdot(j - 1) + \yes(j)\cdot(|R^{\mathrm{act}}| - (j - 1))\,\big]$ \label{alg:accur:sweep:gamma-area}
    \State \textbf{return} $(i^{\ast}\!, j^{\ast}) \gets \big(\yes(j^{\ast}),\; j^{\ast}\big)$
\EndProcedure
\end{algorithmic}
\end{algorithm}

\paragraph{Aggregating bit-counts in SQL} 
Rather than using a table UDTF to estimate~$\hat{n}_\cap$, we keep the Java side minimal: run a ``scalar'' UDF, \textsc{Bit-Counts}, over a pair of same-ID blocks (line~\ref{alg:accur:sweep:bit-counts} of Alg{.}~\ref{alg:accur:sweep}), letting the SQL engine handle the bit-count aggregation. \textsc{Bit-Counts} deserializes the two blocks and returns the bit sums for each block and their union as a small fixed-length vector. The SQL unpacks these per-block vectors %
and aggregates with a grouped \textsc{sum}, keyed by sketch IDs and the in-vector position:

\begin{lstlisting}[language=SQL,basicstyle=\small\ttfamily,columns=fullflexible]
WITH per_block AS (
    SELECT l_id, r_id, Bit_Counts(l.block, r.block) AS counts
    FROM   fl AS l JOIN fr AS r USING (block_id)
    WHERE  <active-pair filter for this batch>
)
SELECT   l_id, r_id, f.index AS pos, SUM(f.value) AS partial
FROM     per_block, LATERAL FLATTEN(input => counts) AS f
GROUP BY l_id, r_id, pos;
\end{lstlisting}
The DBMS parallelizes UDF calls across workers, chooses join and aggregation strategies, and spills to disk under memory pressure.

\paragraph{Adaptive batching} 
To reduce join costs, we group block-pair reads into batches that join smaller materialized tables. If we batched by MBF pairs,
the same blocks would be replicated across many batches. Moreover, a single dense $R$-sketch in a batch would prevent us from ``caching'' a small subset of $L$-side blocks.
We therefore partition by disjoint $\mathit{block\_id}$ intervals: a batch is responsible for $w$ consecutive blocks of every sketch it touches. %
Two ideas keep each batch small: restricting it to the still-active sketches, and sizing $w$ to a fixed row budget.

The batch width~$w$ adapts to how many MBFs remain \emph{active}, \ie{} have to process blocks in $(t,\, t\,{+}\,w]$
assuming that all blocks in $[1,\, t]$ are processed.
Given a cache budget $W$ of blocks per side, we set $w$ in line~\ref{alg:accur:sweep:adaptive-width} of Alg{.}~\ref{alg:accur:sweep} so the batch caches at most $W$ left rows and right rows. 
The batch drops all MBFs whose blocks are not needed by any pair (line~\ref{alg:accur:sweep:active-mbfs} of Alg{.}~\ref{alg:accur:sweep}), causing its active sets to shrink and its $\mathit{block\_id}$ interval to widen. 
Now cached $L$- and $R$-blocks are disjoint from other batches and size-bounded to prevent OOM errors.

\paragraph{Shared-block management} Our algorithm runs a forward sweep in phase \textsc{Main} and a reverse sweep in phase \textsc{Top-up}. Phase \textsc{Main} limits the number of blocks per MBF-pair by the odds factorization formula~(\ref{eq:accur:b-prime-disjoint}) that behaves like a bi-monotone multiplication table. 
Ranking both sides by decreasing odds makes pairs with $\bprime{l,\,r}\hspace{0.5pt}{>}\,t$ a \emph{monotone staircase} in the $(\rank l, \rank r)$ grid. 
We cover that staircase with disjoint rectangles, a wide short one and a narrow tall one (as~``$\Gamma$'') and restrict the batch join to these rectangles.
We choose the pivot $(i^{\ast}\!,\,j^{\ast})$ to minimize the $\Gamma$-shape area (line~\ref{alg:accur:sweep:gamma-area} of Alg{.}~\ref{alg:accur:sweep}).

Phase \textsc{Top-up} reads extra blocks for MBF-pairs with positive shortfall $\bpprime{l,\,r} - \bprime{l,\,r}$. 
To continue forward, we must start at different $\mathit{block\_id}$ positions per pair. 
The extra blocks for sketch $l\in L$ would be different for each partner $r\in R$, providing no opportunity for caching or sharing. 
Our solution is to sweep the blocks in \emph{reverse order} instead, starting with $\mathit{block\_id} = b$. 
This draws every pair's extra blocks from a common sequence $\langle b$, $b\,{-}\,1$, $\ldots\rangle$, allowing each sketch to cache the same few blocks for all its partners. 

\paragraph{Block representation} Our I/O-dominated runtime and storage cost is further reduced by lossless compression. %
Snowflake already uses general-purpose data compression; %
however, rANS compression (see Sec{.}~\ref{sec:compress}) is much better at compressing random vectors, getting to within 4\% of the Shannon bound if $C/m \notin (1/3,\, 2/3)$.
The compute cost of rANS decompression (Alg{.}~\ref{alg:compress:rans-decode-faster}) becomes noticeable for dense blocks, where rANS is not useful anyway, so we apply rANS only if $C/m \notin (1/3,\, 2/3)$.
In RAM, for deserialized blocks, we use Roaring bitmaps~\cite{Chambi2016:roaring} that
support set operations directly on compressed data and can be initialized from run-length output of our rANS decoder. 

\paragraph{Edge cases} Implementing MBF cross-overlaps (Alg{.}~\ref{alg:accur:cross}) requires addressing common edge-cases, including extremely sparse blocks, very small variance limit, negative predicted overlaps, and roundoff errors. %
We have a step that generates the missing all-zero blocks if no set-item hashed into them.
Very small (or zero) accuracy limit $\sigma$ can cause the integer $b'$ to overflow, so we clamp at $\sigma = 1$. The overlap estimator $\hat{n}_\cap$ and other formulas with $\log Z_x$ must guard against roundoff errors by using $\logonep\big({-}C/(b'm)\big)$ and $\expmone$ where appropriate. Negative $\hat{n}_\cap$ must be clamped at~0. Finally, if after phase \textsc{Main} we get a low overlap estimate (\eg{}~$\hat{n}_\cap < \sigma$), we recommend using a larger overlap in the $\bpprime{l,\,r}$ estimator~(\ref{eq:accur:b-prime-overlap}) with $z_\cup$ adjusted to match it; otherwise its effect on $\bpprime{l,\,r}$ may remain underestimated.

\section{Experimental Results}
\label{sec:exper}

\subsection{Tradeoff Analysis}
\label{sec:exper:tradeoff}

\typeout{textwidth=\the\textwidth columnwidth=\the\columnwidth}

We analyze the accuracy--I/O tradeoff of single- and multi-block BFs for large-scale overlap estimation. Filters support up to $\mathtt{MAX\_SIZE} = {}$3 billion items with full-filter membership FPP at $\mathtt{MAX\_FPP} = {}$1\%. Compressed storage is modeled as 2\% above the Shannon bound, capped at the uncompressed size; $\sigma$ denotes the standard deviation predicted by our variance approximations.

\subsubsection*{Compressed single-block BFs and Theta sketches.}

\begin{figure*}[!t]                 %
  \centering
  \begin{subfigure}[t]{0.32\textwidth}
    \phantomsubcaption
    \label{fig:sbf-props:storage}
    \begin{tikzpicture}
      \node[inner sep=0pt] (plot) {
        \includegraphics[width=\linewidth]{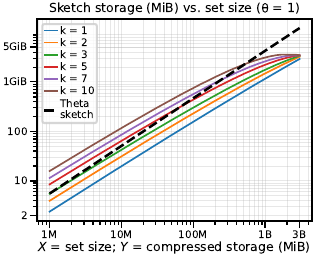}  %
      };
      \node[
        anchor=south east,
        xshift=-3mm,
        yshift=10mm,
        fill=white,
        inner sep=1pt,
        font=\small\bfseries
      ] at (plot.south east) {(\thesubfigure)};
    \end{tikzpicture}
  \end{subfigure}\hfill%
  \begin{subfigure}[t]{0.32\textwidth}
    \phantomsubcaption
    \label{fig:sbf-props:accuracy}
    \begin{tikzpicture}
      \node[inner sep=0pt] (plot) {
        \includegraphics[width=\linewidth]{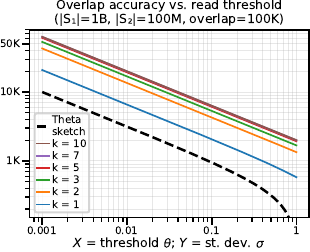}  %
      };
      \node[
        anchor=north east,
        xshift=-3mm,
        yshift=-10mm,
        fill=white,
        inner sep=1pt,
        font=\small\bfseries
      ] at (plot.north east) {(\thesubfigure)};
    \end{tikzpicture}
  \end{subfigure}\hfill%
  \begin{subfigure}[t]{0.32\textwidth}
    \phantomsubcaption
    \label{fig:sbf-props:theta-ol}
    \begin{tikzpicture}
      \node[inner sep=0pt] (plot) {
        \includegraphics[width=\linewidth]{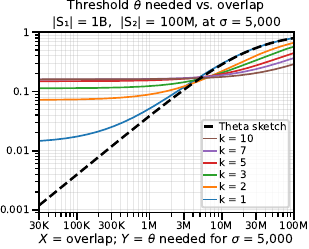}  %
      };
      \node[
        anchor=south east,
        xshift=-5mm,
        yshift=10mm,
        fill=white,
        inner sep=1pt,
        font=\small\bfseries
      ] at (plot.south east) {(\thesubfigure)};
    \end{tikzpicture}
  \end{subfigure}

  \medskip                         %
  \centering
  \begin{subfigure}[t]{0.32\textwidth}
    \phantomsubcaption
    \label{fig:sbf-props:bytes-ol}
    \begin{tikzpicture}
      \node[inner sep=0pt] (plot) {
        \includegraphics[width=\linewidth]{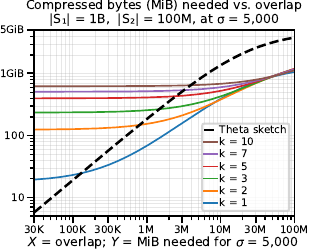}  %
      };
      \node[
        anchor=south east,
        xshift=-5mm,
        yshift=10mm,
        fill=white,
        inner sep=1pt,
        font=\small\bfseries
      ] at (plot.south east) {(\thesubfigure)};
    \end{tikzpicture}
  \end{subfigure}\hfill%
  \begin{subfigure}[t]{0.32\textwidth}
    \phantomsubcaption
    \label{fig:sbf-props:theta-sz}
    \begin{tikzpicture}
      \node[inner sep=0pt] (plot) {
        \includegraphics[width=\linewidth]{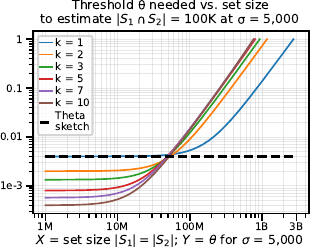}  %
      };
      \node[
        anchor=south east,
        xshift=-3mm,
        yshift=10mm,
        fill=white,
        inner sep=1pt,
        font=\small\bfseries
      ] at (plot.south east) {(\thesubfigure)};
    \end{tikzpicture}
  \end{subfigure}\hfill%
  \begin{subfigure}[t]{0.32\textwidth}
    \phantomsubcaption
    \label{fig:sbf-props:bytes-sz}
    \begin{tikzpicture}
      \node[inner sep=0pt] (plot) {
        \includegraphics[width=\linewidth]{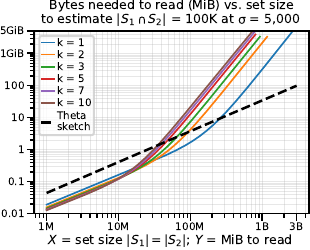}  %
      };
      \node[
        anchor=south east,
        xshift=-3mm,
        yshift=10mm,
        fill=white,
        inner sep=1pt,
        font=\small\bfseries
      ] at (plot.south east) {(\thesubfigure)};
    \end{tikzpicture}
  \end{subfigure}
  
  \captionsetup{skip=4pt}
  \caption{Truncated single-block Bloom filter properties for overlap estimation.}\label{fig:sbf-props}
\end{figure*}

For each $k$, we choose the smallest integer number of bits $m$ satisfying the full-filter FPP constraint:
\begin{equation*}
1 \,-\, (1 - 1/m)^{k\hspace{0.5pt}{\cdot}\hspace{0.5pt}\mathtt{MAX\_SIZE}}\,\,\leq\,\,\mathtt{MAX\_FPP}^{1/k}.
\end{equation*}

Truncated BFs reach a useful overlap accuracy $\sigma = {}$5,000 while reading only a few percent of the filter for billion-item sets paired with smaller sets (Fig{.}~\ref{fig:sbf-props:accuracy}). Among the evaluated configurations, $k = 1$ substantially outperforms $k > 1$ in accuracy per byte read.

The $k = 1$ BF also minimizes storage, using less than a Theta sketch at the same $\theta$ (Fig{.}~\ref{fig:sbf-props:storage}), making it attractive when approximate membership with FPP${} \leq 1\%$ suffices. Truncation sharply raises membership FPP and limits membership usefulness, although larger $k$ offers some resilience.

For Theta sketches, the target-accuracy threshold $\theta = 1\hspace{0.5pt}/\hspace{0.5pt}\big(1 + \sigma^2\hspace{0.5pt}/\hspace{0.5pt}|S_1\,{\cap}\,S_2|\big)$ depends only on overlap (Remark~\ref{th:overlap:var-overlap-bound} in Sec{.}~\ref{sec:overlap}; Fig{.}~\ref{fig:sbf-props:theta-sz}), but the expected retained-hash count grows linearly with set size (Fig{.}~\ref{fig:sbf-props:bytes-sz}). This threshold is lower than for $k = 1$ BFs; the two nearly match when collision-induced variance is small relative to overlap sampling variance, as for small sets or large overlaps in the plotted regime. For small sets, multiple hash placements at $k > 1$ can lower the required threshold (Fig{.}~\ref{fig:sbf-props:theta-ol}), but not necessarily the bytes read.

When thresholds are similar, the smaller $k = 1$ BF footprint means fewer bytes read for the same $\sigma$. BFs therefore favor small-to-medium sets or medium-to-large overlaps, whereas collision-free Theta sketches favor large sets with small overlaps (Figs{.}~\ref{fig:sbf-props:bytes-ol} and~\ref{fig:sbf-props:bytes-sz}). Theta sketches can also materialize intersections as sketches; $k = 1$ BFs instead offer better compression and ``plausible deniability'' in positive membership answers, possibly useful in privacy-sensitive applications.

\subsubsection*{Compressed multi-block BFs.}

\begin{figure*}[!t]
  \centering
  \begin{subfigure}[t]{0.32\textwidth}
    \phantomsubcaption
    \label{fig:mbf-props:accuracy-bfrac}
    \begin{tikzpicture}
      \node[inner sep=0pt] (plot) {
        \includegraphics[width=\linewidth]{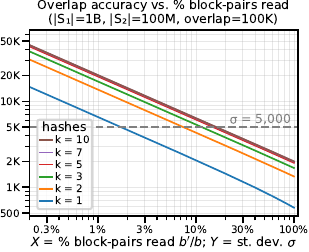}  %
      };
      \node[
        anchor=north east,
        xshift=-5mm,
        yshift=-10mm,
        fill=white,
        inner sep=1pt,
        font=\small\bfseries
      ] at (plot.north east) {(\thesubfigure)};
    \end{tikzpicture}
  \end{subfigure}\hfill%
  \begin{subfigure}[t]{0.32\textwidth}
    \phantomsubcaption
    \label{fig:mbf-props:bfrac-ol}
    \begin{tikzpicture}
      \node[inner sep=0pt] (plot) {
        \includegraphics[width=\linewidth]{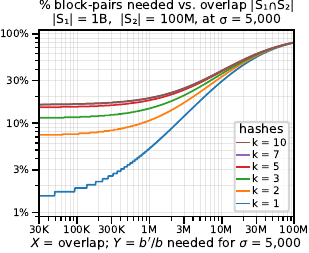}  %
      };    
      \node[
        anchor=north west,
        xshift=10mm,
        yshift=-10mm,
        fill=white,
        inner sep=1pt,
        font=\small\bfseries
      ] at (plot.north west) {(\thesubfigure)};
    \end{tikzpicture}
  \end{subfigure}\hfill%
  \begin{subfigure}[t]{0.32\textwidth}
    \phantomsubcaption
    \label{fig:mbf-props:stdev-ol}
    \begin{tikzpicture}
      \node[inner sep=0pt] (plot) {
        \includegraphics[width=\linewidth]{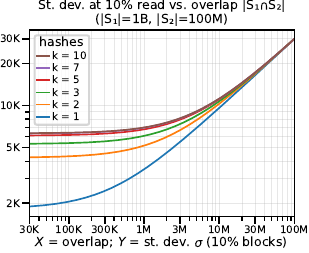}  %
      };
      \node[
        anchor=south east,
        xshift=-5mm,
        yshift=10mm,
        fill=white,
        inner sep=1pt,
        font=\small\bfseries
      ] at (plot.south east) {(\thesubfigure)};
    \end{tikzpicture}
  \end{subfigure}

  \nointerlineskip\vspace{5pt}

  \centering
  \begin{subfigure}[t]{0.32\textwidth}
    \phantomsubcaption
    \label{fig:mbf-props:bpairs-sz}
    \begin{tikzpicture}
      \node[inner sep=0pt] (plot) {
        \includegraphics[width=\linewidth]{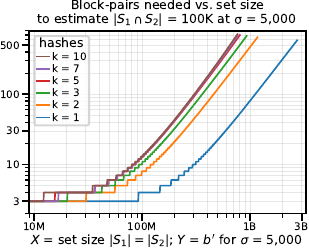}  %
      };
      \node[
        anchor=south east,
        xshift=-5mm,
        yshift=10mm,
        fill=white,
        inner sep=1pt,
        font=\small\bfseries
      ] at (plot.south east) {(\thesubfigure)};
    \end{tikzpicture}
  \end{subfigure}\hfill%
  \begin{subfigure}[t]{0.32\textwidth}
    \phantomsubcaption
    \label{fig:mbf-props:bytes-sz}
    \begin{tikzpicture}
      \node[inner sep=0pt] (plot) {
        \includegraphics[width=\linewidth]{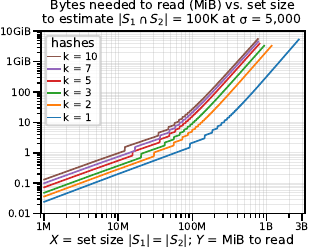}  %
      };
      \node[
        anchor=south east,
        xshift=-5mm,
        yshift=10mm,
        fill=white,
        inner sep=1pt,
        font=\small\bfseries
      ] at (plot.south east) {(\thesubfigure)};
    \end{tikzpicture}
  \end{subfigure}\hfill%
  \begin{subfigure}[t]{0.32\textwidth}
    \phantomsubcaption
    \label{fig:mbf-props:stdev-geom-skew}
    \begin{tikzpicture}
      \node[inner sep=0pt] (plot) {
        \includegraphics[width=\linewidth]{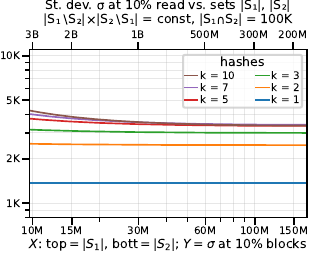}  %
      };    
      \node[
        anchor=north west,
        xshift=8mm,
        yshift=-12mm,
        fill=white,
        inner sep=1pt,
        font=\small\bfseries
      ] at (plot.north west) {(\thesubfigure)};
    \end{tikzpicture}
  \end{subfigure}

  \captionsetup{skip=4pt}
  \caption{Truncated multi-block Bloom filter properties for overlap estimation.}\label{fig:mbf-props}
\end{figure*}

For MBFs, we impose a 5~MiB per-block storage budget to accommodate database cells and parallel processing. With maximum supported expected $\mathtt{density} = \mathtt{MAX\_FPP}^{1/k}$, we choose bits per block $m$ to meet this budget under the compression model, using uncompressed storage when shorter, and set
\begin{equation*}
b = \big\lceil  \big(1 - (1-1/m)^k\big)  \,  /  \,  \big(1 -  (1 - \mathtt{density} )^{1 / \mathtt{MAX\_SIZE}}\big) \big\rceil.
\end{equation*}

Among evaluated configurations, $k = 1$ minimizes both block-pairs and bytes read for target accuracy. Under compression, we therefore favor it over the conventional $k \approx {-}\log_2(\mathtt{MAX\_FPP})$, which optimizes uncompressed BFs at density $1/2$. At $k = 1$, maximum expected density is at most $\mathtt{MAX\_FPP}$. Fig{.}~\ref{fig:mbf-props:accuracy-bfrac} shows accuracy improving with the fraction of blocks read.

At $k = 1$, estimating a 100K overlap with $\sigma = {}$5,000 requires 11 block-pairs (1.9\%) for 1B and 100M sets, 22 (3.7\%) for two 500M sets, 26 (4.4\%) for 3B and 100M, and 42 (7.2\%) for 1B and 500M. These reduce block-pair reads by $14$--$53\times$ relative to a full scan; runtime savings depend on database and decoding costs.

In the plotted regime, larger overlaps tend to increase absolute standard deviation at fixed $b'/b$ (Fig{.}~\ref{fig:mbf-props:stdev-ol}), so target accuracy should reflect the overlaps of interest. A combined tolerance $\sigma = \max(\sigma_{\mathrm{abs}}, \gamma \hat{n}_\cap)$ uses an absolute limit for small overlaps and a relative limit $\gamma$ for large ones.

Larger overlaps also tend to require more blocks at fixed $\sigma$ (Fig{.}~\ref{fig:mbf-props:bfrac-ol}). The disjointness baseline of Sec{.}~\ref{sec:accur:limit} can therefore under-budget Phase~2 for large overlaps, deferring extra work to Phase~3 rather than committing it upfront. Workloads dominated by small overlaps may need little top-up.

The collision-free reference $\sqrt{n_\cap(1 - \theta)/\theta}$, with $\theta = b'/b$ (Remark~\ref{th:overlap:var-overlap-bound}), closely approximates a lower envelope of predicted MBF standard deviation. Unlike the single-block BF bound, it has no $k$ denominator because block sampling retains all or none of an item's $k$ hashes. Our first-order estimator~(\ref{eq:block:var-overlap}) can fall slightly below it, so this bound is approximate.

It gives an approximate minimum block budget $\bprime{\text{min}} = \big\lceil b\hspace{0.5pt}/\hspace{0.5pt}(1 + \sigma^2 / n_\cap)\big\rceil$, independent of exclusive set sizes. For $b \approx 600$, $n_\cap = {}$100K, and $\sigma = {}$5K, this is three block-pairs (Fig{.}~\ref{fig:mbf-props:bpairs-sz}); conversely, 100K overlap at $b'/b = 10\%$ gives a reference ``best'' standard deviation of 948.7. At fixed overlap, larger sets increase I/O by raising both the required block count and the compressed bytes per block-pair (Figs{.}~\ref{fig:mbf-props:bpairs-sz} and~\ref{fig:mbf-props:bytes-sz}).

For fixed overlap, variance estimator~(\ref{eq:block:var-overlap}) depends on set sizes primarily through the product $\big(|S_1| - n_\cap\big)\big(|S_2| - n_\cap\big)$ if both exclusive sets have sparse MBF footprints. As in Remark~\ref{th:overlap:var-overlap-bound}, its $Z$-dependent term evaluated at expectations is proportional to
\begin{equation*}
(r^n - 1) \,+\, r^n(r^c - 1)(r^d - 1) \,\,\approx\,\, (r^n - 1) \,+\, r^n (r - 1)^2 cd,
\end{equation*}
where $n = n_\cap$, $c = |S_1|\,{-}\,n$, $d = |S_2|\,{-}\,n$, $r = 1/p_0$, and $\E Z(S) = b'm\hspace{0.5pt}p_0^{|S|}$ (as in Claim~\ref{th:block:cov-z}). The approximation holds for reasonably sparse MBFs, common at $k = 1$. Fixing $c\hspace{0.5pt}{\cdot}\hspace{0.5pt}d$ gives nearly constant standard deviation (Fig{.}~\ref{fig:mbf-props:stdev-geom-skew}), and an absolute increase in the smaller set changes $\Varh\hat{n}_\cap$ more than an equal increase in the larger set.%

\subsection{Performance on Data}
\label{sec:exper:perform}

We evaluate the cross-overlap method of Sec{.}~\ref{sec:accur:scale} (Alg{.}~\ref{alg:accur:cross}) on two synthetic collections $L$ and $R$ of audiences represented by random keys simulating anonymized customer identifiers. Each collection contains 200 sets: 120 of size 10M, 50 of size 50M, and 10 each of sizes 100M, 500M, and 3B. To model queries testing small overlaps against a minimum commercially significant threshold, every pair in $L\times R$ has overlap 50K. We generate $200 \times 200 \times 50{,}000$ distinct random keys for the pairwise intersections and fill the remaining audience capacity with unique random keys.

Experiments use a Snowflake 2X-Large warehouse with the standard scaling policy, one cluster, and natural table ingestion order. The implementation combines SQL with Java UDFs/UDTFs for BF-block processing, using Google Guava 33.0.0-jre\footnote{\url{https://github.com/google/guava}}, RoaringBitmap 1.3.0\footnote{\url{https://roaringbitmap.org}} for in-memory blocks, and the rANS serialization of Sec{.}~\ref{sec:compress}. One table stores each side, with one row per block and four columns: \texttt{AUDIENCE\_ID} identifies the audience, \texttt{BLOOM\_ID} its MBF UUID, \texttt{BLOOM\_BLOCK\_ID} the block index in $[0,\,b\,{-}\,1]$, and \texttt{BLOOM\_BLOCK} the serialized binary. Each block must fit the 8~MiB cell limit.

Within each experiment, all MBFs share parameters and hash functions. We vary $k$, $b$, and serialization. For $|S| = {}$3B and $n=|S|/b$ expected insertions per block, we choose $m = \big\lceil {-}\hspace{0.5pt}kn\hspace{0.5pt}/\hspace{0.5pt}\log(1 - \mathtt{FPP}^{1/k})\big\rceil$ bits per block, targeting FPP${}=1\%$. With rANS enabled, it serializes blocks at densities outside $(1/3,2/3)$; all other blocks use RoaringBitmap serialization.

Our best observed configurations use $k=1$, approximately 1,500--2,000 blocks, and rANS. As predicted in Sec{.}~\ref{sec:exper:tradeoff}, $k=1$ improves accuracy per byte or block read; we compare it with Guava's default $k=7$, obtained by rounding $\log_2(1/\mathtt{FPP})$. Larger blocks reduce parallelism and coarsen truncation granularity, whereas smaller blocks increase operational overhead. We report $b=2{,}000$ and a smaller block count approaching the 8~MiB cap.

Table~\ref{tab:exper:comp-speed} compares storage and runtime. For $k=1$, the three-phase method is approximately $13$--$18\times$ faster than full-sketch cross-overlaps at $\sigma=5{,}000$ and $34$--$42\times$ faster at $\sigma=10{,}000$. At $k=1$, rANS reduces storage by approximately $1.85\times$ relative to RoaringBitmap serialization, with little apparent runtime penalty.

Tables~\ref{tab:exper:best-overlaps-5K-10K} and~\ref{tab:exper:best-blocks-5K-10K} examine $k=1$, $b=2{,}000$, with rANS. Each row aggregates pairs of the indicated set sizes; Table~\ref{tab:exper:best-overlaps-5K-10K} also reports the pair count.

Table~\ref{tab:exper:best-overlaps-5K-10K} compares sample standard deviations with the requested $\sigma$. For \texttt{3B:3B} at $\sigma=5{,}000$, all 2,000 block-pairs are consumed without reaching the target. Conversely, for \texttt{10M:10M} and \texttt{50M:10M} at $\sigma=10{,}000$, budgets just above one block-pair often round up to two, yielding substantially smaller standard deviations. Excluding these cases, sample standard deviations are within 15\% of their targets. For reference, independent normal samples give approximately 13.9\% and 9.8\% sampling uncertainty at the 95\% level for 100 and 200 observations, respectively; pairs sharing audiences are not independent, so these are only benchmarks. Block granularity and approximate $b'$ estimation also contribute. The ``avg{.}~err$\,$\%'' and ``max{.}~err$\,$\%'' columns report absolute errors relative to the true overlap of 50K. At $\sigma=5{,}000$, maximum errors are below 40\% in most size groups, suggesting utility for minimum-overlap filtering.

Table~\ref{tab:exper:best-blocks-5K-10K} reports mean block-pairs read in Phase~2 (\textsc{Main}) and Phase~3 (\textsc{Top-Up}), together with their serialized size in MiB per set-pair, an I/O proxy that ignores caching. ``Total~I/O$\,$\%'' is each row's share of bytes read across all pairs, counting repeated reads of shared blocks as separate. Pairs involving 3B sets dominate: \texttt{3B:10M} and \texttt{3B:50M} are numerous, while \texttt{3B:500M} and \texttt{3B:3B} require many block-pairs. Phase~2 generally dominates block reads, but Phase~3 dominates for small-set pairs whose overlap determines the budget. Thus, disjointness regularization may be less useful for workloads with many small sets and large overlaps.

\begin{table}[htbp]
  \centering
  \caption{Accuracy-limited MBF cross-overlaps: comparative storage and runtime. ``MiB/blk'' is max{.} (densest) serialized block size, ``GiB/side'' is total MBF storage for $L$ or $R$ side.}
  \label{tab:exper:comp-speed}
  \begingroup
    \setlength{\tabcolsep}{3pt}
    \begin{tabular}{|c|r|r|r|r|r|r|r|}
    \hline
      \multicolumn{3}{|c|}{MBF params} & \multicolumn{2}{|c|}{Storage} & \multicolumn{3}{|c|}{Cross-overlap runtime} \\
    \hline
      $k$ &   $b$   &  rANS  & MiB/blk &  GiB/side  &   Full   & $\sigma = 5$K & $\sigma = 10$K \\
    \hline
       1  &  2,000  &   yes  &  1.458  &   42.7     &  1h 54m  &    8m 22s     &    3m 24s      \\
       1  &  2,000  &    no  &  2.864  &   78.8     &  2h 23m  &    8m 04s     &    3m 41s      \\
       1  &    800  &   yes  &  3.645  &   42.7     &  2h 5m   &    9m 58s     &    3m 36s      \\
       1  &    800  &    no  &  7.160  &   79.2     &  2h 27m  &    8m 54s     &    3m 30s      \\
    \hline
       7  &  2,000  &   yes  &  1.720  &   82.8     &  3h 9m   &   24m 00s     &   13m 28s      \\
       7  &  2,000  &    no  &  1.720  &  113.3     &  3h 36m  &   23m 48s     &   12m 20s      \\
       7  &    600  &   yes  &  5.724  &   82.8     &  3h 41m  &   20m 28s     &   11m 39s      \\
       7  &    600  &    no  &  5.724  &  113.3     &  3h 26m  &   20m 3s      &   10m 15s      \\
    \hline
    \end{tabular}
  \endgroup
\end{table}

\begin{table}[htbp]
  \centering
  \caption{Accuracy-limited MBF cross-overlaps: overlap accuracy ($k = 1$, $b = 2{,}000$, $\sigma = 5{,}000$ and $10{,}000$)}
  \label{tab:exper:best-overlaps-5K-10K}
  \begingroup
    \setlength{\tabcolsep}{2.5pt}
    \begin{tabular}{|c|r|r|r|r|r|r|r|}
    \hline
      \multicolumn{2}{|c|}{} & \multicolumn{3}{|c|}{$\sigma = 5{,}000$} & \multicolumn{3}{|c|}{$\sigma = 10{,}000$} \\
    \hline
          Set    &  \# set   & sample      & avg.        & max.         & sample      & avg.        & max.          \\   
         sizes   &  pairs    & st.dev.     & err$\,$\%   & err$\,$\%    & st.dev.     & err$\,$\%   & err$\,$\%     \\
    \hline
       10M:10M   &  14,400   &    4,861    &  7.7$\,$\%  &  46.7$\,$\%  &    8,277    & 13.4$\,$\%  &  68.7$\,$\%   \\
    \hline
       50M:10M   &  12,000   &    4,862    &  7.7$\,$\%  &  61.3$\,$\%  &    8,465    & 13.9$\,$\%  &  71.3$\,$\%   \\
       50M:50M   &   2,500   &    4,874    &  7.7$\,$\%  &  37.0$\,$\%  &    9,206    & 14.9$\,$\%  &  73.0$\,$\%   \\
    \hline
      100M:10M   &   2,400   &    4,919    &  7.8$\,$\%  &  46.7$\,$\%  &    8,681    & 14.5$\,$\%  &  62.8$\,$\%   \\
      100M:50M   &   1,000   &    4,784    &  7.7$\,$\%  &  33.5$\,$\%  &    9,046    & 14.0$\,$\%  &  61.7$\,$\%   \\
      100M:100M  &     100   &    5,117    &  8.0$\,$\%  &  33.6$\,$\%  &    8,581    & 13.4$\,$\%  &  48.9$\,$\%   \\
    \hline
      500M:10M   &   2,400   &    4,745    &  7.6$\,$\%  &  31.2$\,$\%  &    8,966    & 13.9$\,$\%  &  68.2$\,$\%   \\
      500M:50M   &   1,000   &    4,885    &  7.9$\,$\%  &  32.6$\,$\%  &    9,248    & 14.7$\,$\%  &  78.9$\,$\%   \\
      500M:100M  &     200   &    5,003    &  8.2$\,$\%  &  28.2$\,$\%  &   10,065    & 16.1$\,$\%  &  54.7$\,$\%   \\
      500M:500M  &     100   &    5,440    &  9.0$\,$\%  &  27.2$\,$\%  &    9,562    & 16.0$\,$\%  &  43.2$\,$\%   \\
    \hline
        3B:10M   &   2,400   &    4,833    &  7.7$\,$\%  &  33.4$\,$\%  &    9,225    & 15.0$\,$\%  &  67.8$\,$\%   \\
        3B:50M   &   1,000   &    5,118    &  8.3$\,$\%  &  35.5$\,$\%  &    9,523    & 15.2$\,$\%  &  69.2$\,$\%   \\
        3B:100M  &     200   &    4,770    &  7.8$\,$\%  &  25.4$\,$\%  &    8,905    & 14.4$\,$\%  &  48.6$\,$\%   \\
        3B:500M  &     200   &    5,113    &  8.3$\,$\%  &  30.7$\,$\%  &   11,122    & 17.2$\,$\%  &  92.4$\,$\%   \\
        3B:3B    &     100   &    5,736    &  9.3$\,$\%  &  31.1$\,$\%  &   10,698    & 16.8$\,$\%  &  55.7$\,$\%   \\
    \hline
    \end{tabular}
  \endgroup
\end{table}

\begin{table}[htbp]
  \centering
  \caption{Accuracy-limited MBF cross-overlaps: block-pair usage ($k = 1$, $b = 2{,}000$, $\sigma = 5{,}000$ and $10{,}000$, with rANS)}
  \label{tab:exper:best-blocks-5K-10K}
  \begingroup
    \setlength{\tabcolsep}{2.5pt}
    \begin{tabular}{|c|r|r|r|r|r|r|r|r|}
    \hline
                 &       \multicolumn{4}{|c|}{$\sigma = 5{,}000$}     &      \multicolumn{4}{|c|}{$\sigma = 10{,}000$}     \\
    \hline
         Set     & \multicolumn{3}{|c|}{Blocks$/$set-pair:} & Total   &  \multicolumn{3}{|c|}{Blocks$/$set-pair:} & Total  \\
        sizes    &  Ph.2   &  Ph.3   &  MiB     &         I/O$\,$\%   &  Ph.2   &  Ph.3   &  MiB     &          I/O$\,$\%  \\
    \hline
       10M:10M   &    1    &   3.5   &    0.09  &   0.14$\,$\%        &    1    &   0.5   &    0.03  &   0.16$\,$\%        \\
    \hline
       50M:10M   &    1    &   3.6   &    0.24  &   0.32$\,$\%        &    1    &   0.5   &    0.08  &   0.37$\,$\%        \\
       50M:50M   &    1    &   4.1   &    0.43  &   0.12$\,$\%        &    1    &   0.8   &    0.15  &   0.14$\,$\%        \\
    \hline
      100M:10M   &    1    &   3.7   &    0.41  &   0.11$\,$\%        &    1    &   0.6   &    0.14  &   0.13$\,$\%        \\
      100M:50M   &    2    &   3.8   &    0.70  &   0.08$\,$\%        &    1    &   0.9   &    0.23  &   0.09$\,$\%        \\
      100M:100M  &    3    &   4.2   &    1.13  &   0.01$\,$\%        &    1    &   1.1   &    0.33  &   0.01$\,$\%        \\
    \hline
      500M:10M   &    2    &   3.8   &    1.93  &   0.52$\,$\%        &    1    &   1.0   &    0.65  &   0.60$\,$\%        \\
      500M:50M   &    7    &   4.1   &    4.04  &   0.46$\,$\%        &    2    &   1.1   &    1.11  &   0.43$\,$\%        \\
      500M:100M  &   14    &   3.8   &    7.13  &   0.16$\,$\%        &    4    &   0.9   &    1.97  &   0.15$\,$\%        \\
      500M:500M  &   67.8  &   3.7   &   45.9   &   0.52$\,$\%        &   17    &   1.1   &   11.6   &   0.45$\,$\%        \\
    \hline
        3B:10M   &    8.8  &   3.7   &   18.4   &   4.98$\,$\%        &    2.8  &   0.7   &    5.15  &   4.80$\,$\%        \\
        3B:50M   &   41    &   3.8   &   67.2   &   7.57$\,$\%        &   11    &   0.7   &   17.5   &   6.81$\,$\%        \\
        3B:100M  &   81.3  &   3.8   &  130.7   &   2.95$\,$\%        &   21    &   0.9   &   33.6   &   2.61$\,$\%        \\
        3B:500M  &  405    &   2.9   &  725.9   &  16.4$\,$\%         &  101.7  &   0.8   &  182.4   &  14.2$\,$\%         \\
        3B:3B    & 2000    &   0     &   5832   &  65.7$\,$\%         &  610    &   0.4   & 1780     &  69.1$\,$\%         \\
    \hline
    \end{tabular}
  \endgroup
\end{table}

\section{Conclusion and Future Work}
\label{sec:concl}

We presented a study of single and blocked Bloom filters for set-expression queries, derived variance estimators, adapted rANS compression, and implemented a scalable 3-phase accuracy-limited $L\,{\times}\,R$ cross-overlap method. Future work may extend this method to any $\{\cup\,, \cap\,, \setminus\}$-expressions, do a thorough experimental comparison of blocked BFs to other sketches, and investigate hybrid queries \eg{} overlapping an MBF with a  hashed set sample.

\begin{acks}
We thank Vimalraj Anbarasu, Ayush Tyagi, Kevin Wong, Sekar Krishnamurthy, and Sandeep Nawathe for their valuable feedback and for their broader efforts that made this work possible.
\end{acks}

\clearpage
\bibliographystyle{ACM-Reference-Format}
\bibliography{bloom}

\clearpage
\appendix

\section{Proofs}
\label{sec:app:prf}

\subsection{Basic Properties of Truncated BFs}
\label{app:basic-props}

\paragraph{Membership query.} Truncated BFs allow us to test membership of any item $v\in\Omega$ in the sketched set~$S$ with one-sided guarantee: no false negatives, \ie{} every response ``$v\notin S$'' always implies that $v$ is not in~$S$. For ordinary BFs, where set~$S$ is sketched as $h(S) = \cup_{i=1}^k h_i(S)$, a membership test involves hashing the item~$v$ as $h(v) := \{h_1(v), \ldots, h_k(v)\}$ and then checking whether $h(v)\subseteq h(S)$. If $h(v)\not\subseteq h(S)$, we know for sure that $v\notin S$. This test naturally generalizes to truncated BFs as follows:
\begin{gather*}
\textrm{Let}\;\;h(v)|_{<\theta} \;:=\; \big\{h_1(v), \ldots, h_k(v)\big\} \cap [0, \theta); \;\,\textrm{then,} \\
h(v)|_{<\theta}\,\not\subseteq\, h(S)|_{<\theta} \;\implies\; v\notin S.
\end{gather*}
In other words, after hashing $v\in\Omega$ with the $k$ hash functions, we discard all hashes at or above~$\theta$ and test the remaining hashes for being in~$h(S)|_{<\theta}$. The one-sided guarantee of no false negatives is thus preserved. The false positive probability (FPP) becomes larger for truncated BFs and can be approximately computed using the formula below. Here $v\notin S$ is the item being tested, $n = |S|$, and $m' = |\mathcal{H}\cap[0,\theta)|$ is the hash range under~$\theta$:
\begin{align*}
\mathrm{FPP} \;&=\; \Pr\big[h(v)|_{<\theta}\,\subseteq\, h(S)|_{<\theta}\big] \;={}\\
{}&=\;\sum_{l=0}^k \,\,\Pr\big[\textrm{$l$ out of $k$ hashes${} < \theta$}\big] \cdot \Pr\big[\textrm{the $l$ hashes${} \in h(S)$}\big]\\
{}&\approx\!\!\footnotemark \hspace{0.6pt}\sum_{l=0}^k \,\,\Pr\big[\textrm{$l$ out of $k$ hashes${} < \theta$}\big] \cdot \Pr\big[\textrm{one hash}{} \in h(S)\big]^l\\
{}&\approx\;\sum_{l=0}^k \,\,\binom{k}{l} \cdot \Big(\frac{m'}{m}\Big)^l \Big(1 - \frac{m'}{m}\Big)^{k-l}\cdot\, \bigg[ 1 - \Big(1 - \frac{1}{m}\Big)^{nk}\bigg]^l.
\end{align*}\footnotetext{Ignoring collisions among the $l$ hashes inside $\mathcal{H}|_{<\theta}$ and dependencies in~$h(S)$.}
For Theta-sketches ($k\,{=}\,1$ and $m'\,{=}\,\theta m\,{\to}\,\infty$) we get FPP${} = 1\,{-}\,\theta$.

\paragraph{Size estimation.} The standard approach to derive a size estimator from the BF sketch of set~$S$ is by analyzing how $n = |S|$ affects the distribution of the hashed set size $|h(S)|$, \aka{} the sum of BF's bit-vector bits. The same idea applies to truncated BFs: we derive the expected value for the sum-of-bits statistic $\varsigma := \big|h(S)|_{<\theta}\big|$ given the set size~$n$, and invert it. 

Denote by $\delta_{\mathcal{A}}$ the indicator of event $\mathcal{A}$, which returns~1 if $\mathcal{A}$~occurs and~0 otherwise. If all hashes $\big(h_1(v), ..., h_k(v)\big)_{v\in\Omega}$ are random, independent, and uniformly distributed in~$\mathcal{H}$, we have:
\begin{align*}
\E\varsigma \,=\! \sum_{\updown{\eta\,\in\,\mathcal{H}}{\eta \,<\, \theta}} \hspace{-0.5pt}\E \delta_{\eta\in h(S)} =\, m'\Pr[\eta\,{\in}\,h(S)] \,=\,
    m'\Big[1 - \Big(1\hspace{0.5pt}{-}\hspace{0.5pt}\frac{1}{m}\Big)^{kn}\Big]  \\[-8pt]  %
\implies\quad n\,\,=\,\, \frac{1}{k}\cdot \log\Big(1 - \frac{\E\varsigma}{m'}\Big) \:\Big/\: \log\Big(1 - \frac{1}{m}\Big)\nonumber
\end{align*}
Replacing the expected hash count $\E\varsigma$ with the observed size $\varsigma$ of $h(S)|_{<\theta}$ gives the well-known BF set size estimator~\cite{Papapetrou2010:cardinality}, extended to truncated Bloom filters:
\begin{equation}
\hat{n} \;:=\; \frac{1}{k}\cdot \frac{\log(1 - \varsigma/m')}{\log(1 - 1/m)} \;\approx\; -\frac{m}{k}\cdot \log\Big(1 - \frac{\varsigma}{m'}\Big)
\label{app:basic-props:set-size-predictor}
\end{equation}
For a full filter, the occupancy variance computed below is the exact variance analyzed by Grandi~\cite{Grandi2018:analysis}; we give the calculation for the observed prefix. Variance of the sum-of-bits statistic is computed as follows:
\begin{multline*}
\Var\varsigma \,=\, \sum_{\updown{\eta\,\in\,\mathcal{H}}{\eta \,<\, \theta}} \Var \delta_{\eta\in h(S)}\,+\!\!\!\! 
\sum_{\updown{\eta, \varkappa\,\in\,\mathcal{H}}{\eta, \varkappa \,<\, \theta, \, \eta \neq \varkappa }}
\!\!\!\! \Cov(\delta_{\eta\in h(S)}, \delta_{\varkappa\in h(S)}) \,={}\\[-15pt]
{}=\,m'\Big(1\hspace{0.5pt}{-}\frac{1}{m}\Big)^{kn}\Big[1\,{-}\,\Big(1\hspace{0.5pt}{-}\frac{1}{m}\Big)^{kn}\Big]
\,+\, m'\hspace{-0.5pt}(m'\!{-}1)\Big[
\overbrace{\!\Big(1\hspace{0.5pt}{-}\frac{2}{m}\Big)^{kn} \!\!\!{-}\, \Big(1\hspace{0.5pt}{-}\frac{1}{m}\Big)^{2kn}\!}^{\textrm{covariance}}
\Big]
\end{multline*}
Let's rewrite the covariance term to reveal its asymptotic behavior:
\begin{align*}
&\Big(1\hspace{0.5pt}-\frac{2}{m}\Big)^{kn} \!\! - \, \Big(1\hspace{0.5pt}-\frac{1}{m}\Big)^{2kn} \!\! =\, 
\Big(1\hspace{0.5pt}-\frac{1}{m}\Big)^{2kn} \!\!\! \cdot \Big[\Big(1\hspace{0.5pt}-\frac{1}{(m\,{-}\,1)^2}\Big)^{kn} \!\!\! - 1\Big] \,\approx{}\nonumber\\
&{}\approx\, 
- \,\Big(1\hspace{0.5pt}-\frac{1}{m}\Big)^{2kn} \!\!\! \cdot \frac{kn}{(m\,{-}\,1)^2} \,\approx\, - \,\frac{kn}{m^2}\Big(1 - \frac{\E\varsigma}{m'}\Big)^2\!\!. \quad\textrm{Plug it back in:}\nonumber\\
&\quad\Var\varsigma \,\,\approx\,\, (m' - \E\varsigma) \,-\, \Big(\frac{1}{m'} + \frac{kn}{m^2}\Big) \cdot (m' - \E\varsigma)^2 %
\end{align*}
Variance of the set size estimator $\hat{n} = \hat{n}(\varsigma)$ can be approximated from $\E\varsigma$ and $\Var\varsigma$ using the Delta method applied to~(\ref{app:basic-props:set-size-predictor}):
\begin{equation*}
\Var\hat{n} \,\,\approx\,\, \Big(\frac{m}{k}\Big)^2 \!{\cdot}\, \frac{\Var\varsigma}{(m' - \E\varsigma)^2} \,\,\approx\,\, 
\frac{m^2}{k^2}\Big(\frac{1}{m' - \E\varsigma} - \frac{1}{m'}\Big) \,-\, \frac{n}{k}. \label{app:basic-props:set-size-predictor-variance}
\end{equation*}
Replace $\E\varsigma$ with just $\varsigma$ and replace $n$ with~$\hat{n}$ in~(\ref{app:basic-props:set-size-predictor-variance}) to obtain a practical variance estimator.

If we set $k = 1$, $m' = \theta m$, and take the limit as $m\to\infty$, from $\hat{n}$ we obtain the standard set size estimator for Theta-sketches:
\begin{equation*}
\hat{n}_{\,\mathrm{theta}} \,\,=\,\, \lim_{m\to\infty} \frac{\log\big(1 - \varsigma/(\theta m)\big)}{\log(1 - 1/m)} \,\,=\,\, \lim_{m\to\infty} \frac{- \varsigma/(\theta m)}{- 1/m} \,\,=\,\, \varsigma / \theta
\end{equation*}

\subsection{Inclusion-Exclusion Transform}
\label{app:ie-transform}

Let us restate Claim~\ref{th:overlap:incl-excl} from Sec{.}~\ref{sec:overlap:incl-excl}. Given the sizes $|U_x|$ of the partial unions of sets $S_1, \ldots, S_L$, we can compute the sizes $|P_y|$ of their IE-partitions as follows:
\begin{equation}
|P_y| \;= \sum_{x:\, \updown{x\lor y \,=\, \bar{1}}{x \,>\, \bar{0}}} (-1)^{|x \land y| \,+\, 1} \cdot |U_x|
\label{eq:app:incl-excl}
\end{equation}
\begin{proof}
The inverse linear operation, namely computing the partial union sizes $|U_x|$ from the IE-partition sizes $|P_y|$, is obvious: $U_x$ is the disjoint union of all partitions $P_y$ that contain items from any of the $S_l$'s marked with $x_l=1$, i.e., of all $P_y$ such that $x \land y$ has a~``1'' at any position:
\begin{equation}
U_x \;= \!\!\bigsqcup_{y:\, x \land y \,>\, \bar{0}} P_y
\;\;\implies\;\;
|U_x| \;= \!\!\sum_{y:\, x \land y \,>\, \bar{0}} |P_y| \label{eq:app:inv-incl-excl}
\end{equation}
To prove that~(\ref{eq:app:incl-excl}) gives the linear inverse of~(\ref{eq:app:inv-incl-excl}), let us compose equations~(\ref{eq:app:incl-excl}) and~(\ref{eq:app:inv-incl-excl}) to verify if we obtain the identity:
\begin{eqnarray}
(\ref{eq:app:incl-excl}) \circ (\ref{eq:app:inv-incl-excl}) \;=
\sum_{x:\, \updown{x\lor y \,=\, \bar{1}}{x \,>\, \bar{0}}} (-1)^{|x \land y| \,+\, 1} \!\!\!\! \sum_{y':\, x \land y' \,>\, \bar{0}} |P_{y'}| \; = {} \nonumber \\
{} = \; \sum_{y':\, y' \,>\, \bar{0}}  |P_{y'}|  \cdot \boxed{\sum_{x:\, \updown{x\lor y \,=\, \bar{1}}{x \land y' \,>\, \bar{0}}}  (-1)^{|x \land y| \,+\, 1}} \label{eq:app:incl-excl-id}
\end{eqnarray}
To obtain the identity, we must verify that the boxed expression in~(\ref{eq:app:incl-excl-id}) equals $\delta_{y = y'}$, that is, returns 1 if $y = y'$ and 0 otherwise. Let us prove it as Lemma~\ref{lemma:app:boxed-delta} below.
\end{proof}

\begin{lemma}
The boxed expression in Eq{.}~(\ref{eq:app:incl-excl-id}) equals $\delta_{y = y'}$, that is, returns 1 if $y = y'$ and 0 otherwise:
\begin{equation}
\forall\,y, y' \in \{0, 1\}^L \setminus \{\bar{0}\}:\;
\sum_{x:\, \updown{x\lor y \,=\, \bar{1}}{x \land y' \,>\, \bar{0}}}  (-1)^{|x \land y| \,+\, 1} \;=\; \delta_{y = y'}.\label{eq:app:boxed-delta}
\end{equation}
\vspace{-10pt}
\label{lemma:app:boxed-delta}
\end{lemma}
\begin{proof}
Let us evaluate the sum in~(\ref{eq:app:boxed-delta}) in three possible cases: $y \not\leq y'$, $y < y'$, and $y = y'$.

\paragraph{Case $y \not\leq y'$.} Let $l$ be such that $y_l = 1$ and $y'_l = 0$; then the set $\left\{x\!: \updown{x\,\lor\, y \,=\, \bar{1}}{x \land y' \,>\, \bar{0}}\right\}$ consists of disjoint pairs $\{x|_{x_l=0}\,,\, x|_{x_l=1}\}$ that differ only in the $l^{\mathstrut}$-th bit. Indeed, $x_l$ enters $x\lor y = \bar{1}$ as $x_l \lor 1$ and enters $x \land y' > \bar{0}$ as $x_l \land 0$, so $x_l$ has no effect on either condition. Summing $(-1)^{|x \land y| \,+\, 1}$ over each such pair adds $-1$ and $+1$, hence the entire expression equals~0.

\paragraph{Case $y < y'$.} Let $l$ be such that $y_l = 0$ and $y'_l = 1$. Here the set $\left\{x\!: \updown{x\,\lor\, y \,=\, \bar{1}}{x \land y' \,>\, \bar{0}}\right\}$ equals just $\{{x\!:}\; x\lor y = \bar{1}\}$ because $x_l\lor y_l = 1$ implies $x_l = 1$ and $x_l\land y'^{\mathstrut}_l = 1$. Pick any $l'$ such that $y_{l'} = 1$ (it exists since always $y\neq \bar{0}$) and again the set consists of disjoint pairs because $x_{l'}$ has no effect on condition $x\lor y = \bar{1}$. Summing $(-1)^{|x \land y| \,+\, 1}$ over such pairs, with $y_{l'} = 1$ and $x_{l'}$ flipping, again results in a~0.

\paragraph{Case $y = y'$.} Here our set is $X = \left\{x\!: \updown{x\,\lor\, y \,=\, \bar{1}}{x \land y \,>\, \bar{0}}\right\}$. It differs from set $X' = \{{x\!:}\; x\lor y = \bar{1}^{\mathstrut}\}$ by a single vector: $X'\setminus X = \{\neg y\}$ where $\neg y$ is bitwise NOT-$y$. As in the previous case, the summation in~(\ref{eq:app:boxed-delta}) equals~0 if $x$ ranges over~$X'$, and $(-1)^{|x \land y| \,+\, 1}$ gives a $-1$ for $x = \neg y$, hence the summation over~$X$ evaluates to~$+1$.

Thus, the sum in~(\ref{eq:app:boxed-delta}) equals $\delta_{y = y'}$ for all inputs.
\end{proof}

\subsection{Set Expression Cardinality Estimation}
\label{app:overlap:formula}

Here we prove Claim~\ref{th:overlap:formula} from Sec{.}~\ref{sec:overlap:formula}. Given a $\{\cup\,, \cap\,, \setminus\}$-formula $F(S_1, \ldots, S_L)$, we convert it into a Boolean formula $F^B(x_1, \ldots, x_L)$ where each occurrence of $S_l$ converts to $x_l$ and each instance of\/ $\cup\,, \cap\,, \setminus$ converts to Boolean operation $\lor\,, \land\,, \land\neg$ respectively. Then, the claim states: $F(S_1, \ldots, S_L) \,=\, \bigcup_{x:\,F^B(x)=1} P_x$. Since the union is disjoint, its size estimator is:$\;\hat{n}[F] \,=\, \sum_{x:\,F^B(x)=1} \hat{n}[P_x]$.
\begin{proof}
By induction. For $F\equiv S_l$ we get $S_l = \bigcup_{x:\,x_l=1} P_x$, which is true. Suppose we proved the claim for expressions $F_1$ and $F_2$, let us prove for $F_1 \mathop{\textrm{[op]}} F_2$. Note that each nonempty IE-partition in each disjoint union behaves ``atomically,'' like a single element: it is either all-in or all-out. So, by induction hypothesis,
\begin{align*}
(F_1 \cup F_2)(S) \; & = \!\! \bigcup_{x:\,F_1^B(x)=1} \!\!\!\! P_x \;\,\,\cup \!\! \bigcup_{x:\,F_2^B(x)=1} \!\!\!\! P_x \;\,\, = \!\!\!\!\!\! \bigcup_{x:\,F_1^B(x) \,\lor\, F_2^B(x) \,=\, 1} \!\!\!\!\!\! P_x \\
(F_1 \cap F_2)(S) \; & = \!\! \bigcup_{x:\,F_1^B(x)=1} \!\!\!\! P_x \;\,\,\cap \!\! \bigcup_{x:\,F_2^B(x)=1} \!\!\!\! P_x \;\,\, = \!\!\!\!\!\! \bigcup_{x:\,F_1^B(x) \,\land\, F_2^B(x) \,=\, 1} \!\!\!\!\!\! P_x \\
(F_1 \setminus F_2)(S) \; & = \!\! \bigcup_{x:\,F_1^B(x)=1} \!\!\!\! P_x \;\,\,
\mathbin{\mathmakebox[0pt][l]{\setminus}\hphantom{\cap}}
\!\! \bigcup_{x:\,F_2^B(x)=1} \!\!\!\! P_x \;\,\, = \!\!\!\!\!\!\! \bigcup_{x:\,F_1^B(x) \,\land\, \neg F_2^B(x) \,=\, 1} \!\!\!\!\!\!\! P_x \qedhere
\end{align*}
\end{proof}

\subsection{Covariance of Zero-Bit Counts}
\label{app:overlap:cov-z}

Below is the proof of Claim~\ref{th:overlap:cov-z} about the covariance between zero-bit counts of sets $S$ and $S'$ sketched using $(m, k, \theta)$-Bloom filters. Let us repeat formula~(\ref{eq:overlap:cov-z}) for convenience:
\begin{align*}
&\Cov\big(Z(S), Z(S')\big) \,\,=\,\, m' \Big[\Big(1\,{-}\,\frac{1}{m}\Big)^{k|S\hspace{0.5pt}\cup\hspace{0.5pt}S'|}  \!\!{-}\,\Big(1\,{-}\,\frac{1}{m}\Big)^{k(|S|\hspace{0.5pt}+\hspace{0.5pt}|S'|)} \Big] \,\,{+}\\
&{+}\,\,m'(m'\,{-}\,1) \Big(1\,{-}\,\frac{1}{m}\Big)^{k(|S|\hspace{0.5pt}+\hspace{0.5pt}|S'|)}   \Big[ \Big(1\,{-}\,\frac{1}{(m\,{-}\,1)^2}\Big)^{k|S\hspace{0.5pt}\cap\hspace{0.5pt}S'|} \!\!{-}\, 1\Big].
\end{align*}
\begin{proof}
Use the bilinearity of covariance:
\begin{align}
&\Cov\big(Z(S), Z(S')\big) \,=\,
\Cov\bigg(\sum_{\updown{\eta\,\in\,\mathcal{H}}{\eta \,<\, \theta}} \hspace{-0.5pt}\delta_{\eta\,\notin\, h(S)}\,,
\!\sum_{\updown{\varkappa\,\in\,\mathcal{H}}{\varkappa \,<\, \theta}} \hspace{-0.5pt}\delta_{\varkappa\,\notin\, h(S')}\bigg)\,={}\nonumber\\
&\!{=}\,\sum_{\updown{\eta\,\in\,\mathcal{H}}{\eta \,<\, \theta}} \!\Cov \big(\delta_{\eta\hspace{0.5pt}\notin\hspace{0.5pt}h(S)},\,  \delta_{\eta\hspace{0.5pt}\notin\hspace{0.5pt}h(S')}\big) \,+\!\!
\sum_{\updown{\eta \,\neq\,\varkappa}{\eta,\varkappa \,<\, \theta}} \!\!\Cov \big(\delta_{\eta\hspace{0.5pt}\notin\hspace{0.5pt}h(S)},  \delta_{\varkappa\hspace{0.5pt}\notin\hspace{0.5pt}h(S')}\big) \label{eq:app:overlap:cov-z-1}
\end{align}
Let us call the left term \emph{same-hash}\/ and the right term \emph{cross-hash}. The left term consists of $m'$ identical same-hash covariances:
\begin{align}
& \Cov \big(\delta_{\eta\hspace{0.5pt}\notin\hspace{0.5pt}h(S)},\, \delta_{\eta\hspace{0.5pt}\notin\hspace{0.5pt}h(S')}\big) \,=\,
\E \delta_{\eta\hspace{0.5pt}\notin\hspace{0.5pt}h(S)\hspace{0.5pt}\cup\hspace{0.5pt}h(S')} \,-\,
\E \delta_{\eta\hspace{0.5pt}\notin\hspace{0.5pt}h(S)} \E \delta_{\eta\hspace{0.5pt}\notin\hspace{0.5pt}h(S')}\nonumber\\
& \hspace{10pt}{}=\, \Big(1\,{-}\,\frac{1}{m}\Big)^{k|S\hspace{0.5pt}\cup\hspace{0.5pt}S'|}  -\, \Big(1\,{-}\,\frac{1}{m}\Big)^{k|S|}\Big(1\,{-}\,\frac{1}{m}\Big)^{k|S'|} \label{eq:app:overlap:cov-z-2}
\end{align}
The cross-hash sum consists of $m'(m'-1)$ identical cross-hash covariances with $\eta$ and $\varkappa$ as different hash values from $\mathcal{H}|_{<\theta}$:
\begin{align}
& \Cov \big(\delta_{\eta\hspace{0.5pt}\notin\hspace{0.5pt}h(S)},\, \delta_{\varkappa\hspace{0.5pt}\notin\hspace{0.5pt}h(S')}\big) \,=\,
\E \delta_{\eta\hspace{0.5pt}\notin\hspace{0.5pt}h(S)}    \delta_{\varkappa\hspace{0.5pt}\notin\hspace{0.5pt}h(S')}\,-\,
\E \delta_{\eta\hspace{0.5pt}\notin\hspace{0.5pt}h(S)} \E \delta_{\varkappa\hspace{0.5pt}\notin\hspace{0.5pt}h(S')}\nonumber\\
& \hspace{10pt}{}=\, \Pr\big[\eta \,{\notin}\, h(S) \land \varkappa \,{\notin}\, h(S')\big] \,-\, \Big(1\,{-}\,\frac{1}{m}\Big)^{k|S|}\Big(1\,{-}\,\frac{1}{m}\Big)^{k|S'|} \label{eq:app:overlap:cov-z-3}
\end{align}
To evaluate the left probability, split $S$ and $S'$ into 3 disjoint parts: $A = S\cap S'$, $B = S\setminus S'$, and $C = S'\setminus S$, so that $S = A \cup B$ and $S' = A \cup C$. Disjoint sets $A$, $B$, and $C$ generate hashes independently, hence
\begin{align}
& \Pr \big[\eta \,{\notin}\, h(S) \land \varkappa \,{\notin}\, h(S')\big] \,=\,
\Pr\big[\eta \,{\notin}\, h(A)\hspace{0.5pt}{\cup}\hspace{0.5pt}h(B) \land \varkappa \,{\notin}\, h(A)\hspace{0.5pt}{\cup}\hspace{0.5pt}h(C)\big]\nonumber\\
& \hspace{10pt}{}=\, \Pr\big[\{\eta, \varkappa\} \,{\cap}\, h(A) \,{=}\,\emptyset\big] \,\cdot\, \Pr\big[\eta \,{\notin}\, h(B)\big] \,\cdot\, \Pr\big[\varkappa \,{\notin}\, h(C)\big] \,={}\nonumber\\
& \hspace{10pt}{}=\, \Big(1\,{-}\,\frac{2}{m}\Big)^{k|A|} \Big(1\,{-}\,\frac{1}{m}\Big)^{k|B|} \Big(1\,{-}\,\frac{1}{m}\Big)^{k|C|} \!={}\nonumber\\
& \hspace{10pt}{}=\, \Big[\hspace{0.5pt}\Big(1\,{-}\,\frac{1}{(m\,{-}\,1)^2}\Big) \Big(1\,{-}\,\frac{1}{m}\Big)^{2\,}\Big]^{k|A|} \,\Big(1\,{-}\,\frac{1}{m}\Big)^{k(|B|\hspace{0.5pt}+\hspace{0.5pt}|C|)} \!={}\nonumber\\
& \hspace{10pt}{}=\, \Big(1\,{-}\,\frac{1}{(m\,{-}\,1)^2}\Big)^{k|S\hspace{0.5pt}\cap\hspace{0.5pt}S'|} \Big(1\,{-}\,\frac{1}{m}\Big)^{k(|S|\hspace{0.5pt}+\hspace{0.5pt}|S'|)} \label{eq:app:overlap:cov-z-4}
\end{align}
Combining (\ref{eq:app:overlap:cov-z-1}), (\ref{eq:app:overlap:cov-z-2}), (\ref{eq:app:overlap:cov-z-3}), and (\ref{eq:app:overlap:cov-z-4}) completes the proof of (\ref{eq:overlap:cov-z}).
\end{proof}

\subsection{Variance of Pairwise Overlap}
\label{app:var-overlap}

Below is the proof of Corollary~\ref{th:overlap:var-overlap}. We will use shorthand $\hat{n}_\cap$ for the pairwise overlap estimator:
\begin{equation}
\hat{n}_\cap \,:=\, - \frac{m}{k}\left(\log\frac{Z_1}{m'} \,+\, \log\frac{Z_2}{m'} \,-\, \log\frac{Z_\cup}{m'}\right) \label{eq:app:simple-overlap}
\end{equation}
where $Z_1$, $Z_2$, and $Z_\cup$ are zero-bit counts observed in truncated BFs for $S_1$, $S_2$, and $S_1\,{\cup}\,S_2$ respectively. Let us repeat the proposed formula (\ref{eq:overlap:var-overlap}) for convenience:
\begin{equation*}
\Var \hat{n}_\cap \,\approx\, \frac{m^2}{k^2} \left(\frac{2\hspace{0.5pt}Z_\cup}{Z_1 Z_2} + \frac{1}{Z_\cup} - \frac{1}{Z_1} - \frac{1}{Z_2} - \frac{1}{m'}\right) \,-\, \frac{\hat{n}_\cap}{k}
\end{equation*}

\begin{proof}
Instantiate the Delta method (\ref{eq:overlap:delta}) for (\ref{eq:app:simple-overlap}):
\begin{align*}
&\setlength{\arraycolsep}{2pt}
\hat{n}_\cap \,:=\, - \frac{m}{k} \begin{bmatrix} +1 & +1 & -1 \end{bmatrix} \begin{bmatrix} \log Z_1 \\ \log Z_2 \\ \log Z_\cup \end{bmatrix} \,+\, \frac{m}{k} \log m';
\;\;\;\Var \hat{n}_\cap \,\approx\,{}\frac{m^2}{k^2} \,\times{}\\
&\setlength{\arraycolsep}{2pt}
\hspace{-3pt}\begin{bmatrix}\displaystyle \frac{1}{Z_1} & \displaystyle \frac{1}{Z_2} & \displaystyle \frac{-1}{Z_\cup} \end{bmatrix} \,
\begin{bmatrix}
\Var Z_1          & \Cov(Z_1, Z_2)    & \Cov(Z_1, Z_\cup) \\
\Cov(Z_1, Z_2)    & \Var Z_2          & \Cov(Z_2, Z_\cup) \\
\Cov(Z_1, Z_\cup) & \Cov(Z_2, Z_\cup) & \Var Z_\cup
\end{bmatrix}\,
\begin{bmatrix} 1/Z_1 \\ 1/Z_2 \\ -1/Z_\cup \end{bmatrix}
\end{align*}
Use Eq{.}~(\ref{eq:overlap:cov-z-est}) to approximate the entries in the covariance matrix:
\begin{align*}
\Var Z_1          \,&\approx\, \mathrlap{Z_1}\hphantom{Z_\cup} \,-\, \big(1/m' \,-\, 1/m   \,\log(Z_1/m')     \big) \cdot Z_1^2       \\
\Var Z_2          \,&\approx\, \mathrlap{Z_2}\hphantom{Z_\cup} \,-\, \big(1/m' \,-\, 1/m   \,\log(Z_2/m')     \big) \cdot Z_2^2       \\
\Var Z_\cup       \,&\approx\,                         Z_\cup  \,-\, \big(1/m' \,-\, 1/m   \,\log(Z_\cup/m')  \big) \cdot Z_\cup^2    \\
\Cov(Z_1, Z_\cup) \,&\approx\,                         Z_\cup  \,-\, \big(1/m' \,-\, 1/m   \,\log(Z_1/m')     \big) \cdot Z_1 Z_\cup  \\
\Cov(Z_2, Z_\cup) \,&\approx\,                         Z_\cup  \,-\, \big(1/m' \,-\, 1/m   \,\log(Z_2/m')     \big) \cdot Z_2 Z_\cup  \\
\Cov(Z_1, Z_2)    \,&\approx\,                         Z_\cup  \,-\, \big(1/m' \,+\, k/m^2 \cdot \hat{n}_\cap \big) \cdot Z_1 Z_2
\end{align*}
Substitute these into the covariance matrix and perform vector multiplications:
\begin{align*}
\Var \hat{n}_\cap \,\approx\, \frac{m^2}{k^2} \,\times\, \big[\, 
         1 / \mathrlap{Z_1}\hphantom{Z_\cup} \,&-\, \hphantom{2}\big(1/m' \,-\, 1/m   \,\log(Z_1/m')     \big)  \\[-2.1pt]
  {} +\, 1 / \mathrlap{Z_2}\hphantom{Z_\cup} \,&-\, \hphantom{2}\big(1/m' \,-\, 1/m   \,\log(Z_2/m')     \big)  \\
  {} +\, 1 /                         Z_\cup  \,&-\, \hphantom{2}\big(1/m' \,-\, 1/m   \,\log(Z_\cup/m')  \big)  \\
  {} -\, 2 / \mathrlap{Z_1}\hphantom{Z_\cup} \,&+\,           2 \big(1/m' \,-\, 1/m   \,\log(Z_1/m')     \big)  \\
  {} -\, 2 / \mathrlap{Z_2}\hphantom{Z_\cup} \,&+\,           2 \big(1/m' \,-\, 1/m   \,\log(Z_2/m')     \big)  \\
  {} +\, 2\,Z_\cup / (Z_1 Z_2)               \,&-\,           2 \big(1/m' \,+\, k/m^2 \cdot \hat{n}_\cap \big) 
\,\big]\;\;=\;\; \text{(\ref{eq:overlap:var-overlap})}. \qedhere
\end{align*}
\end{proof}

\subsection{Theta Sketch Lower Bound}
\label{app:overlap:var-overlap-bound}

Let us prove the Theta sketch based lower bound on the BF overlap estimator~(\ref{eq:overlap:var-overlap}) written with expectations of zero-bit counts (rather than the observed counts). Denote $n = |S_1\,{\cap}\,S_2|$. Here is the inequality we want to prove, repeated for convenience:
\begin{equation*}
\frac{m^2}{k^2} \left(\frac{2\,{\cdot}\hspace{0.5pt}\E Z_\cup}{\E Z_1 \E Z_2} + \frac{1}{\E Z_\cup} - \frac{1}{\E Z_1} - \frac{1}{\E Z_2} - \frac{1}{\theta m}\right) \,-\, \frac{n}{k} \,\geq\,
\frac{n\hspace{0.5pt}(1\,{-}\,\theta)}{k\hspace{0.5pt}{\cdot}\hspace{0.5pt}\theta}.
\end{equation*}
\begin{proof}
From the basic BF properties~(\ref{eq:basics:set-size-predictor}) we have $\E Z(S) = m'\cdot (1\,{-}\,1/m)^{k|S|}$. If we denote $r = 1/(1\,{-}\,1/m)^k$ and keep in mind that $m' = \theta m$, we have $1\hspace{0.5pt}/\hspace{0.5pt}\E Z(S) = r^{|S|}\hspace{0.5pt}/\hspace{0.5pt}(\theta m)$. Substituting this into the variance estimator, denoting $c = |S_1\,{\setminus}\,S_2|$ and $d = |S_2\,{\setminus}\,S_1|$, we obtain:
\begin{align*}
&\frac{m^2}{k^2} \left(\frac{2\,{\cdot}\hspace{0.5pt}\E Z_\cup}{\E Z_1 \E Z_2} + \frac{1}{\E Z_\cup} - \frac{1}{\E Z_1} - \frac{1}{\E Z_2} - \frac{1}{\theta m}\right) \,-\, \frac{n}{k} \,\,={}\\
{}&=\,\, m\hspace{0.5pt}/\hspace{0.5pt}(\theta k^2) \cdot \big(2 r^n + r^{n+c+d} - r^{n+c} - r^{n+d} - 1\big) \,-\, n/k \,\,={}\\
{}&=\,\, m\hspace{0.5pt}/\hspace{0.5pt}(\theta k^2) \cdot \big((r^n - 1) \,+\, r^n(r^c - 1)(r^d - 1)\big) \,-\, n/k \,\,\geq{}\\
{}&\geq\,\, m\hspace{0.5pt}/\hspace{0.5pt}(\theta k^2) \cdot (r^n - 1) \,-\, n/k  \qquad \textrm{because of $r > 1$.}
\end{align*}
By $e^x - 1\geq x$ and $-\log(1\,{-}\,1/m)\geq 1/m$ we have: $r^n\,{-}\,1 \geq n \log r \geq nk/m$. Substitute this into the above:
\begin{equation*}
m\hspace{0.5pt}/\hspace{0.5pt}(\theta k^2) \cdot (r^n - 1) \,-\, n/k \,\,\geq\,\, n/(\theta k) - n/k \,=\, \frac{n\hspace{0.5pt}(1\,{-}\,\theta)}{k\hspace{0.5pt}{\cdot}\hspace{0.5pt}\theta}.\qedhere
\end{equation*}
\end{proof}

\subsection{rANS Compression for BFs}
\label{app:rans-bf}

A typical rANS algorithm compresses strings over a finite alphabet $\mathcal{A}$ where the frequency of each character $a\in\mathcal{A}$ is specified as $f_a/M$, a ratio of integers with the same denominator~$M = \sum_a f_a$. The algorithm's basic idea is to maintain a state $\xi\in\mathbb{N}$ and encode each next character~$a$ by updating the state as follows:
\begin{equation*}
\xi \,\,\gets\,\, \big\lfloor\xi \,{/}\, f_a\big\rfloor \cdot M \,+\, F_a  \,+\,  (\xi \bmod f_a),  \;\; \textrm{where} \;\; F_a \,=\, {\textstyle\sum_{a'<a} f_{a'}}
\end{equation*}
Here we assume that the alphabet $\mathcal{A}$ is listed in order, and $F_a / M$ is the cumulative frequency of the letters in $\mathcal{A}$ listed before~$a$. In base-$M$ (``$M$-ary'') representation of~$\xi$, this update
\begin{itemize}
\item Breaks $\xi$ into two parts: $\big\lfloor\xi \,{/}\, f_a\big\rfloor$ and $(\xi \bmod f_a)$;
\item Shifts $\big\lfloor\xi \,{/}\, f_a\big\rfloor$ to make space for a new $M$-ary ``digit''~$d$;
\item Fits $d$ in interval $d\in[F_a, F_a\,{+}\,f_a)$ to encode $(\xi \bmod f_a)$.
\end{itemize}
Decoding characters proceeds in reverse; to decode a character, compute $d = (\xi \bmod M)$, find its interval $d\in[F_a, F_a\,{+}\,f_a)$ to extract the character~$a$, and remove the ``digit'' from~$\xi$ by updating
\begin{equation*}
\xi \,\,\gets\,\, \big\lfloor\xi \,/\, M\big\rfloor \cdot f_a \,+\, (\xi \bmod M) \,-\, F_a.
\end{equation*}
To keep $\xi$ bounded, a rANS algorithm picks a segment $[L, H]$ where $L = Mt$ and $H = bMt - 1$ for $b, t \in \mathbb{N}$, and writes/reads bits to/from storage once $\xi$ moves outside $[L, H]$. Since each encoded character~$a$ expands $\xi$ to $\lfloor\xi / f_a\rfloor \,{\cdot}\,M$, adding roughly $\log_2(M/f_a)$ bits, encoding a string with $m\,{\cdot}\,f_a/M$ instances of each character results in adding roughly the Shannon bound number of bits.

For BF compression, our alphabet is $\mathcal{A} = \{0, 1\}$ ordered as $0 < 1$, with $M = m$, $f_1 = C$, and the state $\xi$ is bounded within segment $[L, H] = [m, 2m\,{-}\,1]$. In this setting, the rANS encoder is equivalent to Algorithm~\ref{alg:compress:rans-encode} and the decoder to Algorithm~\ref{alg:compress:rans-decode-plain}. Since decoding is time-critical for overlap and membership quer\-ies, Algorithm~\ref{alg:compress:rans-decode-plain} warrants further optimization. First, in the top while-loop (line~\ref{alg:compress:rans-decode-plain:decoding-loop}) we can recompute quotient $q := {}$\hbox{$\big\lfloor(\xi - m) \,{/}\, C\big\rfloor$} without performing division if we observe that line~\ref{alg:compress:rans-decode-plain:xi-update} implies
\begin{align*}
&\psi \,\,\gets\,\, (m \bmod C) \,+\, 2 \!\cdot \! \big( (\xi - m) \bmod C \big) \,+\, \textproc{Read-Bit}(v)\\
&q := \big\lfloor(\xi - m) \,{/}\, C\big\rfloor  \,\,\gets\,\,  \lfloor m/C \rfloor  \,-\, 2 \,+\, \lfloor\psi/C\rfloor
\end{align*}
where $0\leq\psi\leq 3C-2$ and hence $\lfloor\psi/C\rfloor \in \{0, 1, 2\}$ computable without the division. Second, in the bottom while-loop (line~\ref{alg:compress:rans-decode-plain:reading-loop}) we can read $\nu := \lfloor\log_2(m\,{+}\,1)\rfloor - \lfloor\log_2(\xi\,{+}\,1)\rfloor$ bits ``upfront,'' because after reading $\nu\,{-}\,1$ bits into $\xi$ we still have $\xi < m$ even if the bits are all ones:
\begin{align*}
&\nu  \,\,<\,\,  \log_2(m + 1) \,-\, \big(\log_2(\xi + 1) - 1\big) \;\;\Rightarrow{}\\
&{}\Rightarrow\;\;  2^{\nu - 1}  \,<\,  (m + 1) \,/\, (\xi + 1) \;\;\Rightarrow\;\; 2^{\nu - 1}\xi \,+\, (2^{\nu - 1} - 1)  \,<\,  m.
\end{align*}
Incorporating these two optimizations results in Algorithm~\ref{alg:compress:rans-decode-faster}.

\begin{algorithm}
\caption{Decodes a sparse bit vector from rANS-compressed encoding to run-length representation.}
\label{alg:compress:rans-decode-plain}
\begin{algorithmic}[1]%
\Require $v = {}$the rANS-compressed form of sparse $m$-bit vector
\Ensure $\langle r_1, r_2, \ldots, r_C, r_{C+1} \rangle = {}$the bit vector in run-length format
\Procedure{RANS-Decode}{$v, m$}
    \State $C \gets {}$\Call{Read-Integer}{$v$}; $\;\xi \gets m + {}$\Call{Read-Integer}{$v$}
    \ForAll{$i = C+1,\, C, \ldots, 2, 1$}
        \State $r_i \gets 0$ \Comment{Decode the $i$-th run of zeros}
        \While{$\xi \leq 2m - 1 - C$} \label{alg:compress:rans-decode-plain:decoding-loop}
            \State $r_i \gets r_i + \big\lfloor(\xi - m) \,{/}\, C\big\rfloor + 1$
            \State $\xi \gets 2\big(m - C + (\xi - m) \bmod C\big) + {}$\Call{Read-Bit}{$v$} \label{alg:compress:rans-decode-plain:xi-update}
        \EndWhile
        \State $\xi \gets \xi - 2(m - C)$
        \While{$\xi < m$} \label{alg:compress:rans-decode-plain:reading-loop}
            \State $\xi \gets 2\xi + {}$\Call{Read-Bit}{$v$}
        \EndWhile            
    \EndFor
    \State \textbf{return} $r_1, r_2, \ldots, r_C, r_{C+1}$
\EndProcedure
\end{algorithmic}
\end{algorithm}

\subsection{Cardinality Estimation for MBFs}
\label{app:block:cardinal}

As with ordinary BFs, here set size estimation is the key operation when estimating overlaps and the sizes of arbitrary $\{\cup\,, \cap\,, \setminus\}$-expressions evaluated over multiple sets. For multi-block BFs, each block represents a disjoint part of the sketched set, so we can estimate the set size by adding together the size estimates for individual blocks. However, this turns out to be suboptimal for very large sets. To see why, consider an extreme example where one of the blocks $B_j$ has been ``maxed out'' so that $H|_j = \mathcal{H}$, \ie{} its bit-vector has only~1s. Then $B_j$ has no upper limit on its set size estimate, essentially returning $+\infty$ set size. If we add the estimates from all blocks, a single $+\infty$ results in the overall estimate of~$+\infty$. Clearly, we should do better if the other blocks return finite estimates.

An alternative approach is to add up the hashed set sizes (bit-vector bits) across all blocks and use this statistic to estimate the overall set size. Then each block's contribution is limited to a $1/b$-th fraction of the total bits, reducing the bias and improving the statistical robustness of our predictor. The formula for the estimator follows the same idea as for the ordinary BF: derive the expected value for the sum-of-bits statistic given the true set size, then invert it. Let the true size $|S|$ be~$n$ and assume that all hashes $\big(h_0(v), h_1(v), ..., h_k(v)\big)_{v\in\Omega}$ are random, independent, uniformly distributed within their hash ranges. Denote by $\varsigma$ the sum-of-bits statistic; then we have:
\begingroup
\allowdisplaybreaks
\begin{align}
\E\varsigma\,&=\, \E \,\sum_{j=1}^b \sum_{\eta\in\mathcal{H}} \delta_{\eta\in H|_j} \,=\,  \sum_{j=1}^b \sum_{\eta\in\mathcal{H}} \E \delta_{\eta\in H|_j} =\, bm\,\Pr[\eta\in H|_j] \,={}\nonumber\\
{}&=\, bm\cdot\Big(1 \;-\; \Pr\big[\forall v \in S: \textrm{$v$ does not hash to $(j, \eta)$}\big]\Big) \,={}\nonumber\\
{}&=\, bm\cdot\Big(1 \;-\; \Pr\big[\textrm{$v$ does not hash to $(j, \eta)$}\big]^n\Big) \,={}\nonumber\\
{}&=\, bm\cdot\bigg(1 \;-\; \Big(\Pr[h_0(v) \neq j] \,\, + {}\nonumber\\[-5pt]
{}& \qquad +\; \Pr\big[h_0(v) = j \;\;\textrm{and}\;\;   \forall i = 1,\ldots, k\!:\, h_i(v) \neq \eta\big] \Big)^n \;\bigg) \,={}\nonumber\\
{}&=\, bm\cdot\bigg(1 \;-\; \Big[\,\frac{b-1}{b} \;+ \; \frac{1}{b}\cdot \Big(1 - \frac{1}{m}\Big)^k\,\Big]^n\bigg).
\label{eq:app:block:expected-sum-of-bits}
\end{align}%
We will also use the shorthand
\begin{equation}
Q \,=\, \frac{b-1}{b} \, + \, \frac{1}{b}\Big(1 - \frac{1}{m}\Big)^k.
\label{eq:app:block:q}
\end{equation}
\endgroup
Inverting the above formula, we obtain the following estimator:
\begin{equation}
\label{eq:app:block:mbf-size}
\hat{n} \;=\; \log\Big(1 - \frac{\varsigma}{bm}\Big) \,\,\Big/\,\, \log\Big[\,\frac{b-1}{b} \;+ \; \frac{1}{b}\cdot \Big(1 - \frac{1}{m}\Big)^k\,\Big]
\end{equation}
It is insightful to compare it to the naive estimator that adds up the per-block size estimates. Denote by $\hat{n}|_j$ the ordinary BF size estimator and by $\varsigma|_j$ the hashed set size (sum-of-bits) for block~$B_j$; then we have:
\begin{equation}
\label{eq:app:block:bf-size}
\hat{n}|_j \;=\; \log\Big(1 - \frac{\varsigma|_j}{m}\Big) \,\,\Big/\,\, \log\Big[\,\Big(1 - \frac{1}{m}\Big)^k\,\Big]
\end{equation}
Combining (\ref{eq:app:block:mbf-size}) with (\ref{eq:app:block:bf-size}) and using $\varsigma = \sum_{j=1}^b \varsigma|_j$ we get the formula:
\begin{equation*}
\begin{gathered}
\hat{n} \;=\; \alpha^{-1} \!\cdot \log\Big[\,\frac{1}{b} \,\sum_{j=1}^b \,\exp\big(\beta\cdot\hat{n}|_j\big)\,\Big], \quad \textrm{where} \\
\alpha \,=\, \log \Big[\,\frac{b-1}{b} \;+ \; \frac{1}{b} \,\Big(1 - \frac{1}{m}\Big)^k\,\Big], \quad \beta \,=\, \log\Big[\,\Big(1 - \frac{1}{m}\Big)^k\,\Big]
\end{gathered}
\end{equation*}
So, the more robust estimator aggregates per-block size estimates using \emph{log-average-exp} rather than the sum. If a few of the block estimates $\hat{n}|_j$ are too large or infinite, their $\exp(\beta\!\cdot\!\hat{n}|_j)$-terms will be nearly~0 (as $\beta < 0$) with limited effect on the overall average.

We can use the set size estimator to predict the size of overlaps and $\{\cup\,, \cap\,, \setminus\}$-expressions over multiple sets given their MBF sketches. First, we arrange their BF-blocks so that same-ID blocks from all sketches appear together. Then, we read each group of same-ID blocks, compute all partial hashed unions within the group, and determine the number of hashes (\ie{} sum-of-bits) in each union. Finally, we aggregate the number of hashes in each union across all block IDs and apply estimator~(\ref{eq:app:block:mbf-size}) to obtain estimated partial union sizes. Now we can use the inclusion-exclusion principle described in Section~\ref{sec:overlap:incl-excl} to evaluate any $\{\cup\,, \cap\,, \setminus\}$-expression.

To validate why we can compute the hashed unions for each block ID independently, let $S_1, \ldots, S_L$ be the unsketched sets and let $U_x := \bigcup_{l:\,x_l=1} S_l$ be their partial unions. We have\footnote{Reminder: $f(A\cup B) \equiv f(A)\cup f(B)$, but $f(A\cap B) \not\equiv f(A)\cap f(B)$.}:
\begin{multline*}
\!\!\!U_x|_j \,=\, \Big(\!\bigcup_{l:\,x_l=1} \! S_l\Big) \cap\, \Omega|_j \,= \bigcup_{l:\,x_l=1} \big(S_l \cap \Omega|_j\big) \,= \bigcup_{l:\,x_l=1} \! S_l|_j \,;\! \;\; H_x|_j ={}\\
{}=\, h(U_x|_j) \,=\,\, \bigcup_{i = 1}^k \,h_i \Big(\!\bigcup_{l:\,x_l=1} \! S_l|_j\Big) \,\,= \bigcup_{l:\,x_l=1} \bigcup_{i = 1}^k \,h_i(S_l|_j) \,= \bigcup_{l:\,x_l=1}  H_l|_j.
\end{multline*}
Hence, hashing the full union $U_x$ into the blocks is the same as computing the hashed union within each block and aggregating their sum-of-bits counts:
\begin{equation*}
\varsigma_x \,=\, \sum_{j=1}^b \varsigma_x|_j \quad\textrm{where}\quad \varsigma_x|_j \,\,:=\,\, \big|h(U_x|_j)\big| \,\,=\,\, \bigg|\bigcup_{l:\,x_l=1} \!H_l|_{j\,}\bigg|.
\end{equation*}

We do not have to read and aggregate all $b$ blocks before computing a set size prediction. Suppose we have only seen the first $b'<b$ blocks $B_1, \ldots, B_{b'}$ of the sketch for set~$S$. Let $\varsigma|_{\leq b'}$ denote the sum-of-bits statistic aggregated over blocks $B_1, \ldots, B_{b'}$. The same reasoning as in~(\ref{eq:app:block:expected-sum-of-bits}) produces
\begin{equation*}
\E\varsigma|_{\leq b'}\,=\,\sum_{j=1}^{b'} \sum_{\eta\in\mathcal{H}} \E \delta_{\eta\in H|_j} =\, b'm\,\bigg(1 - \Big[\,\frac{b\,{-}\,1}{b} + \frac{1}{b}\, \Big(1\,{-}\,\frac{1}{m}\Big)^k\,\Big]^n\bigg)
\end{equation*}
Inverting it gives us a set size predictor from the first $b'$ blocks:
\begin{multline*}
\hat{n}[S] \,\,=\,\, \log\Big(1 - \frac{\varsigma|_{\leq b'}}{b'm}\Big) \,\,\Big/\,\, \log\Big[\,\frac{b-1}{b} \;+ \; \frac{1}{b}\cdot \Big(1 - \frac{1}{m}\Big)^k\,\Big]\,\,={}\\
{}=\,\, \logonep\Big({-}\,\frac{\varsigma|_{\leq b'}}{b'm}\Big)\,\,\Big/\,\, \logonep \Big[\,\frac{1}{b} \,\expmone\Big(k\, \logonep\Big({-}\,\frac{1}{m}\Big)\Big)\Big]
\end{multline*}
where functions $\expmone(x) := e^x - 1$ and $\logonep(x) := \ln(1 + x)$ help reduce roundoff errors. This formula is the same as~(\ref{eq:block:mbf-size-prime}) in Sec{.}~\ref{sec:block}.

\subsection{Covariance of Zero-Bit Counts for MBFs}
\label{app:block:cov-z}

Below is the proof of Claim~\ref{th:block:cov-z} that gives formula~(\ref{eq:block:cov-z}) for the exact covariance between $Z(S)$ and $Z(S')$, zero-bit counts observed within the first $b'\leq b$ blocks of $(b, m, k)$-MBFs representing sets $S$ and~$S'$. Let us repeat the formula for convenience:
\begin{align*}
&\Cov\big(Z(S), Z(S')\big) \,\,=\,\,\, b'm \, p_0^{|S \,\cup\, S'|} \,+\, \big[(b'\,{-}\,1)m\,(p_d/p_0^2)^{|S\,\cap\,S'|} \,+{}\nonumber\\
&\qquad{} +\, (m - 1)\, (p_s/p_0^2)^{|S\,\cap\,S'|} \,-\, b'm \big] \cdot b'm \,p_0^{|S| + |S'|}. \,
\end{align*}
Here $p_0$, $p_d$, and $p_s$ denote the following expressions:
\begin{align*}
p_0 \,\,&:=\,\, (b-1) / b \,+\, 1/b \cdot (1-1/m)^k \\[-3pt]
p_d \,\,&:=\,\, (b-2) / b \,+\, 2/b \cdot (1-1/m)^k \\[-3pt]
p_s \,\,&:=\,\, (b-1) / b \,+\, 1/b \cdot (1-2/m)^k
\end{align*}

\begin{proof}
By definition, $Z(S)$ is the sum of Bernoulli random indicators $\delta_{\eta\,\notin\,H|_j}$ where $j$ ranges over observed BF-blocks ($1\,{\leq}\,j\,{\leq}\,b'$) and $\eta\in\mathcal{H}$ ranges over all hashes in a BF-block. Here  $H|_j \subseteq \mathcal{H}$ represents all hashes in block~$B_j$ ``hit'' by the items of set~$S$ that are hashed into~$B_j$: $H|_j = \cup_{i=1}^k h_i(S|_j)$ where $S|_j = \{v \in S: h_0(v)=j\}$. Let's write out the sums for $Z(S)$ and $Z(S')$ and use the bilinearity of their covariance:
\begin{align*}
\Cov\big(Z(S), Z(S')\big) \,\,&=\,\, \Cov\,\Big(\,\sum_{j=1}^{b'} \sum_{\eta\in\mathcal{H}} \delta_{\eta\,\notin\, H|_j}\, ,\, \sum_{j'=1}^{b'} \sum_{\varkappa\in\mathcal{H}} \delta_{\varkappa\,\notin\, H'|_{j'}}\Big) \,\,={}\\
{}&=\,\,\sum_{j, j'=1}^{b'} \,\sum_{\eta, \varkappa\in\mathcal{H}} \, \Cov\big(\delta_{\eta\,\notin\, H|_j}\, , \delta_{\varkappa\,\notin\, H'|_{j'}}\big).
\end{align*}
Use the definition of covariance and the fact about random indicators that $\E \delta_Q$ $=$ $\Pr[Q]$:
\begin{align*}
&\Cov \,\,=\,\sum_{j, j'=1}^{b'} \,\sum_{\eta, \varkappa\in\mathcal{H}} \, \big(\E \delta_{\eta\,\notin\, H|_j} \hspace{0.5pt} \delta_{\varkappa\,\notin\, H'|_{j'}} \,-\, \E \delta_{\eta\,\notin\, H|_j} \E \delta_{\varkappa\,\notin\, H'|_{j'}}\big)\,\,={}\\
&\sum_{j, j'=1}^{b'} \,\sum_{\eta, \varkappa\in\mathcal{H}} \big(\Pr[\eta\,{\notin}\,H|_j \land \varkappa\,{\notin}\,H'|_{j'}] \,-\, \Pr[\eta\,{\notin}\,H|_j]\,\Pr[\varkappa\,{\notin}\,H'|_{j'}]\,\big)
\end{align*}
For a hash $\eta\in\mathcal{H}$ in BF-block $B_j$, and one item $v\in S$, the probability that $h(v)$ does not ``hit''~$\eta$ equals $p_0$ $=$ $(b-1) / b$ $+$ $(1-1/m)^k/b$, because $v$ hashes to some other block with probability $(b-1) / b$ and, if it hashes to~$B_j$, none of its $k$ hashes ``hit''~$\eta$ with probability $(1-1/m)^k$. For all items in~$S$, the probability that they all ``miss''~$\eta$ equals $p_0^{|S|}$ due to hash independence:
\begin{equation}
\Cov \,\,=\,\sum_{j, j'=1}^{b'} \,\sum_{\eta, \varkappa\in\mathcal{H}} \big(\Pr[\eta\,{\notin}\,H|_j \land \varkappa\,{\notin}\,H'|_{j'}\hspace{-0.5pt}] \,-\, p_0^{|S|} \, p_0^{|S'|}\,\big)
\label{eq:app:block:cov-z-1}
\end{equation}
The probability $\Pr[\eta\,{\notin}\,H|_j \land \varkappa\,{\notin}\,H'|_{j'}\hspace{-0.5pt}]$ depends only on the following three cases; let's denote it as $P_1$, $P_2$, and $P_3$ for each of these three cases:
\begin{description}
\item[\hspace{13pt}$P_1$:] Different BF-blocks ($j \neq j'$);
\item[\hspace{13pt}$P_2$:] Same block, different hashes ($j = j'$ $\land$ $\eta \neq \varkappa$);
\item[\hspace{13pt}$P_3$:] Same block and same hash ($j = j'$ $\land$ $\eta = \varkappa$).
\end{description}
In the sum (\ref{eq:app:block:cov-z-1}) there are $b'(b'-1)m^2$ cases with different blocks, $b'm(m-1)$ cases with the same block and different hashes, and $b'm$ cases with the same block and hash:
\begin{equation*}
\Cov \,=\, b'(b'\hspace{0.5pt}{-}\,1)m^2 P_1 \,+\, b'm(m\,{-}\,1) P_2 \,+\, b'm\hspace{0.5pt}P_3 \,-\, (b'm)^2 p_0^{|S|+|S'|}
\end{equation*}

For $P_3$, since $j = j'$ and $\eta = \varkappa$, we have $P_3 = \Pr[\eta\,{\notin}\,H|_j\,{\cup}\,H'|_j]$ where $H|_j\,{\cup}\,H'|_j$ is the hashed image of the union $S\,{\cup}\,S'$ into block~$B_j$. Each item $v\in S\,{\cup}\,S'$ does not ``hit''~$\eta$ with probability~$p_0$, independently, hence $P_3 = p_0^{|S\,{\cup}\,S'|}$.

However, for $P_1$ and $P_2$ we must address the dependence between items of~$S$ ``hitting'' hash~$\eta$ and items of~$S'$ ``hitting'' hash~$\varkappa$ because the two sets share some items. Let us partition $S$ and~$S'$ as in their Venn diagram: $S = A\,{\cup}\,B$ and $S' = A\,{\cup}\,C$ where $A = S\,{\cap}\,S'$, $B = S\,{\setminus}\,S'$, and $C = S'\,{\setminus}\,S$.  Then,
\begin{align*}
&\Pr[\eta\,{\notin}\,H|_j \land \varkappa\,{\notin}\,H'|_{j'}\hspace{-0.5pt}] \,\,={}\\
&{}=\,\,\Pr\!\big[\eta \notin h(A|_j)\,{\cup}\,h(B|_j) \,\land\, \varkappa \notin h(A|_{j'})\,{\cup}\,h(C|_{j'})\big]\,\,={}\\
&{}=\,\,\Pr\!\big[\eta \notin h(A|_j) \,\land\, \varkappa \notin h(A|_{j'}) \,\land\, \eta \notin h(B|_j) \,\land\, \varkappa \notin h(C|_{j'})\big]\,\,={}\\
&\,\Pr\!\big[\eta \notin h(A|_j) \land \varkappa \notin h(A|_{j'})\big] \,\cdot\, \Pr\!\big[\eta \notin h(B|_j)\big] \,\cdot\, \Pr\!\big[\varkappa \notin h(C|_{j'})\big]
\end{align*}
The last step of factoring the probability was possible because sets $A$, $B$, and $C$ are disjoint and hash independently. As before, the single-hash $\Pr[\eta\,{\notin}\,h(B|_j)]$ equals $p_0^{|B|}$ and $\Pr[\varkappa\,{\notin}\,h(C|_{j'})]$ equals $p_0^{|C|}$. The leftmost double-hash factor, however, is different in $P_1$ versus~$P_2$:
\begin{description}[leftmargin=*]
\item[\hspace{13pt}$P_1$:] For one item $v\in A$, ``hitting'' neither hash~$\eta$ in block~$j$ nor hash~$\varkappa$ in block~$j'$ occurs either if $h_0(v)\notin\{j, j'\}$ or if none of its $k$ in-block hashes ``hit''~$\eta$ (or~$\varkappa$). The probability of such event is $p_d$ $=$ $(b-2)/b$ $+$ $2/b\,(1-1/m)^k$, where ``$d$'' stands for ``different blocks.'' For all items in~$A$, the probability that they all ``miss'' both targets equals $p_d^{|A|}$ due to hash independence:
\begingroup
\allowdisplaybreaks
\begin{align*}
\hspace{13.6pt}&\hspace{11.8pt}\Pr\genfrac{[}{]}{0pt}{0}{\,\eta \notin h(A|_j) \,\land\, \varkappa \notin h(A|_{j'})}{\,\textrm{different blocks:}\,\,j \neq j'} \,\,={}\\
{}&=\,\, \Pr\big[\forall v \in A: \textrm{$v$ hashes to neither $(j, \eta)$ nor $(j', \varkappa)$}\big] \,\,={}\\
{}&=\,\, \Pr\big[\textrm{$v$ hashes to neither $(j, \eta)$ nor $(j', \varkappa)$}\big]^{|A|} \,\,={}\\
{}&=\,\, \bigg(\,\Pr\big[h_0(v) \notin \{j, j'\}\big] \,\,+\,\, \Pr\genfrac{[}{]}{0pt}{0}{h_0(v) = j \;\;\textrm{and}}{\,\forall i = 1,\ldots, k\!:\, h_i(v) \neq \eta} \,\,+{}\\
{}&\qquad +\,\,\Pr\genfrac{[}{]}{0pt}{0}{h_0(v) = j' \;\;\textrm{and}}{\,\forall i = 1,\ldots, k\!:\, h_i(v) \neq \varkappa}\,\,\bigg)^{|A|} \,={}\\
{}&=\,\,\big((b-2)/b\,+\,2/b\cdot (1-1/m)^k\big)^{|A|} \,\,=\,\, p_d^{|A|}.
\end{align*}%
\endgroup
\item[\hspace{13pt}$P_2$:] For one item $v\in A$, ``hitting'' neither hash~$\eta$ nor~$\varkappa$ in the same block~$j$ occurs either if $h_0(v)\neq j$ or if each of its $k$ in-block hashes ``miss'' both $\eta$ and~$\varkappa$. The probability of such event is $p_s$ $=$ $(b-1)/b$ $+$ $1/b\,(1-2/m)^k$, where ``$s$'' stands for ``same block.'' For all items in~$A$, the probability that they all ``miss'' both targets equals $p_s^{|A|}$ due to hash independence:
\begingroup
\allowdisplaybreaks
\begin{align*}
\hspace{15.6pt}&\hspace{11.8pt}\Pr\genfrac{[}{]}{0pt}{0}{\,\eta \notin h(A|_j) \,\land\, \varkappa \notin h(A|_{j'})}{\textrm{same block:}\,\,j = j'} \,\,={}\\
{}&=\,\, \Pr\big[\forall v \in A: \textrm{$v$ hashes to neither $(j, \eta)$ nor $(j, \varkappa)$}\big] \,\,={}\\
{}&=\,\, \Pr\big[\textrm{$v$ hashes to neither $(j, \eta)$ nor $(j, \varkappa)$}\big]^{|A|} \,\,={}\\
{}&=\,\, \left(\,\Pr[h_0(v) \neq j] \,\,+\,\, \Pr\genfrac{[}{]}{0pt}{0}{h_0(v) = j \;\;\textrm{and}}{\,\forall i = 1,\ldots, k\!:\, h_i(v) \notin \{\eta, \varkappa\}} \,\right)^{|A|} \!\!\!={}\\
{}&=\,\,\big((b-1)/b\,+\,1/b\cdot (1-2/m)^k\big)^{|A|} \,\,=\,\, p_s^{|A|}.
\end{align*}%
\endgroup
\end{description}
To summarize, we have: $P_1 = p_d^{|A|}p_0^{|B|+|C|}$ and $P_2 = p_s^{|A|}p_0^{|B|+|C|}$. Substitute these back into our covariance formula:
\begin{align*}
&\Cov\big(Z(S), Z(S')\big) \,\,=\,\,\, b'(b'\hspace{0.5pt}{-}\,1)m^2 p_d^{|A|}p_0^{|B|+|C|} \,+{}\\
&\quad{}+\, b'm(m\,{-}\,1) p_s^{|A|}p_0^{|B|+|C|} \,+\, b'm\hspace{0.5pt}p_0^{|S\,{\cup}\,S'|} \,-\, (b'm)^2 p_0^{|S|+|S'|}
\end{align*}
To finish, it remains to observe that $p_0^{|B|+|C|}$ $=$ $p_0^{|S|+|S'|}/p_0^{2|A|}$ and that $|A|$ $=$ $|S\,{\cap}\,S'|$:
\begin{align*}
&\Cov\big(Z(S), Z(S')\big) \,\,=\,\,\, b'(b'\hspace{0.5pt}{-}\,1)m^2   p_d^{|A|}  p_0^{|S|+|S'|} / p_0^{2|A|} \,+{}\\
{}&+\, b'm(m\,{-}\,1) p_s^{|A|} p_0^{|S|+|S'|} / p_0^{2|A|}  \,-\, (b'm)^2 p_0^{|S|+|S'|}   \,+\, b'm\hspace{0.5pt}p_0^{|S\,{\cup}\,S'|}
\end{align*}
Now a simple rearrangement of terms gives us formula~(\ref{eq:block:cov-z}).
\end{proof}

\subsection{Estimating Zero-Bit Covariance for MBFs}
\label{app:block:cov-z-est}

Below is the proof of Corollary~\ref{th:block:cov-z-est}, specifically of the approximation formula (\ref{eq:block:cov-z-appr}) restated below for convenience:
\begin{align*}
&\Cov\hspace{-0.5pt}\big(Z(S), Z(S')\big) \,\,\approx\,\, \E Z(S\,{\cup}\,S') \,+{}\\
&\quad{}+\, \frac{1}{b'm}\Big(\frac{(b - b')k^2 - bk}{b^2 m}\,|S\,{\cap}\,S'| \,-\, 1\Big) \cdot \E Z(S) \,\E Z(S').
\end{align*}
Here $Z(S)$, $Z(S')$, and $Z(S\,{\cup}\,S')$ are zero-bit counts observed within the first $b'\leq b$ blocks of the MBFs for sets $S$, $S'$, and $S\,{\cup}\,S'$ respectively, and we assume $m \gg k^2$.

\begin{proof}
Let us start by rewriting the exact covariance formula~(\ref{eq:block:cov-z}) with a couple of helper functions, $f_d(x)$ and $f_s(x)$:
\begin{align}
&\Cov\big(Z(S), Z(S')\big) \,\,=\,\,\, b'm \, p_0^{|S \,\cup\, S'|} \,+\, \big[(b'\,{-}\,1)m\,f_d(1/m)^{|S\,\cap\,S'|} \,+{}\nonumber\\
&\qquad{} +\, (m - 1)\, f_s(1/m)^{|S\,\cap\,S'|} \,-\, b'm \big] \cdot b'm \,p_0^{|S| + |S'|} \,
\label{eq:app:cov-z}
\end{align}
where $p_0$ $:=$ $(b-1) / b \,+\, (1-1/m)^k / b$ and
\begin{align*}
f_d(x) \,&:=\, \frac{(b-2) / b \,+\, 2(1-x)^k / b}{\big((b-1) / b \,+\, (1-x)^k / b\big)^2},\\
f_s(x) \,&:=\, \frac{(b-1) / b \,+\, (1-2x)^k / b}{\big((b-1) / b \,+\, (1-x)^k / b\big)^2}.
\end{align*}
Rational functions $f_d(x)$ and $f_s(x)$ are analytical to at least $1/k$ distance from $x=0$ and have the following derivatives at $x=0$:
\begin{align*}
f'_d(0) &= 0 & f''_d(0) &= - 2k^2/b^2         & f'''_d(0) &\sim O(k^3/b^2) \\
f'_s(0) &= 0 & f''_s(0) &= 2k(bk - b - k)/b^2 & f'''_s(0) &\sim O(k^3/b)
\end{align*}
Hence, we can use Taylor series around $x=0$ and evaluate them at $x=1/m$:
\begin{align*}
f_d(1/m) \,&=\, 1 \,-\, k^2\,/\,(b^2 m^2) \,+\, O\big(k^3/(b^2 m^3)\big)\\
f_s(1/m) \,&=\, 1 \,+\, k(bk - b - k)\,/\,(b^2 m^2) \,+\, O\big(k^3/(b m^3)\big)
\end{align*}
Then we can raise them to the power of $n = |S\,{\cap}\,S'| \sim O(bm)$ using the fact that $(1 + \eps)^n$ $=$ $1 + n\eps + o(n\eps)$ when $\eps, n\eps \to 0$:
\begin{align*}
f_d(1/m)^n \,&=\, 1 \,-\, nk^2\,/\,(b^2 m^2) \,+\, o\big(nk^2\,/\,(b^2 m^2)\big)\\
f_s(1/m)^n \,&=\, 1 \,+\, nk(bk - b - k)\,/\,(b^2 m^2) \,+\, o\big(nk^2\,/\,(b m^2)\big)
\end{align*}
Finally, substitute these approximations back into Eq{.}~(\ref{eq:app:cov-z}), while also replacing $b'm p_0^{|S \,\cup\, S'|}$ with $\E Z(S\,{\cup}\,S')$ and $(b'm)^2 p_0^{|S| + |S'|}$ with $\E Z(S) \,\E Z(S')$:
\begin{align*}
&\Cov\big(Z(S), Z(S')\big) \,\,\approx\,\,\, \E Z(S\,{\cup}\,S') \,+\, \Big[(b'\,{-}\,1)m\,\Big(1 \,-\, \frac{nk^2}{b^2 m^2}\Big) \,+{}\\
&{}\;\; +\, (m - 1)\, \Big(1 \,+\, \frac{nk(bk - b - k)}{b^2 m^2}\Big) \,-\, b'm \Big] \cdot \frac{\E Z(S) \,\E Z(S')}{b'm} \,\,\approx{}\\
&{}\approx\,\, \E Z(S\,{\cup}\,S') \,+\, \frac{1}{b'm} \,\Big[\frac{(b - b')k^2 - bk}{b^2 m} \, n \,-\, 1\Big] \, \E Z(S) \,\E Z(S'),
\end{align*}
which is the same formula as in (\ref{eq:block:cov-z-appr}).
\end{proof}

\subsection{Variance of Pairwise Overlap for MBFs}
\label{app:block:var-overlap}

Below is the proof of Corollary~\ref{th:block:var-overlap}. For convenience, let us repeat the formula for the pairwise overlap estimator $\hat{n}_\cap$ $=$ $\hat{n}[S_1\,{\cap}\,S_2]$ and the proposed formula (\ref{eq:block:var-overlap}) for its estimated variance:
\begin{align*}
&\hat{n}_\cap \,=\, (\log p_0)^{-1} \big(\log Z_1 + \log Z_2 - \log Z_\cup - \log(b'm)\big); \;\;\Var \hat{n}_\cap \,\approx{}\\
&{}\approx \,\, \left(\frac{bm}{k}\right)^2 \left(\frac{2\hspace{0.5pt}Z_\cup}{Z_1 Z_2} \,{+}\, \frac{1}{Z_\cup} \,{-}\, \frac{1}{Z_1} \,{-}\, \frac{1}{Z_2} \,{-}\, \frac{1}{b'm}\right) \,+\, \left(\frac{b(k\,{-}\,1)}{b'k} - 1\right)\hat{n}_\cap
\end{align*}
Here $Z_1$, $Z_2$, and $Z_\cup$ are zero-bit counts observed within the first $b'\leq b$ blocks of the MBFs for sets $S_1$, $S_2$, and $S_1\,{\cup}\,S_2$ respectively.

\begin{proof}
Instantiate the Delta method (\ref{eq:overlap:delta}) for the MBF overlap estimator, denoting $\bm{c}$ $=$ $- \log(b'm)\,/\,\log p_0$:
\begin{align*}
&\setlength{\arraycolsep}{2pt}
\hat{n}_\cap \,:=\, \frac{1}{\log p_0} \begin{bmatrix} +1 & +1 & -1 \end{bmatrix} \begin{bmatrix} \log Z_1 \\ \log Z_2 \\ \log Z_\cup \end{bmatrix} \,+\, \bm{c}\hspace{0.5pt};  
\;\;\;\Var \hat{n}_\cap \,\approx\,{}\frac{1}{(\log p_0)^2} \,\times{}\\
&\setlength{\arraycolsep}{2pt}
\hspace{-3pt}\begin{bmatrix}\displaystyle \frac{1}{Z_1} & \displaystyle \frac{1}{Z_2} & \displaystyle \frac{-1}{Z_\cup} \end{bmatrix} \,
\begin{bmatrix}
\Var Z_1          & \Cov(Z_1, Z_2)    & \Cov(Z_1, Z_\cup) \\
\Cov(Z_1, Z_2)    & \Var Z_2          & \Cov(Z_2, Z_\cup) \\
\Cov(Z_1, Z_\cup) & \Cov(Z_2, Z_\cup) & \Var Z_\cup
\end{bmatrix}\,
\begin{bmatrix} 1/Z_1 \\ 1/Z_2 \\ -1/Z_\cup \end{bmatrix}
\end{align*}
Denote the MBF set-size estimators for $S_1$, $S_2$, and $S_1\,{\cup}\,S_2$ by $\hat{n}_1$, $\hat{n}_2$, and $\hat{n}_\cup$ correspondingly. Also denote $\alpha := \big((b - b')k^2 - bk\big)\,/\,(b^2 m)$. Use Eq{.}~(\ref{eq:block:cov-z-est}) to approximate the entries in the covariance matrix:
\begin{align*}
\Var Z_1          \,&\approx\, \mathrlap{Z_1}\hphantom{Z_\cup} \,+\, 1/(b'm) \cdot (\alpha\mathrlap{\hat{n}_1}\hphantom{\hat{n}_\cup} - 1) \cdot Z_1^2       \\
\Var Z_2          \,&\approx\, \mathrlap{Z_2}\hphantom{Z_\cup} \,+\, 1/(b'm) \cdot (\alpha\mathrlap{\hat{n}_2}\hphantom{\hat{n}_\cup} - 1) \cdot Z_2^2       \\
\Var Z_\cup       \,&\approx\,                         Z_\cup  \,+\, 1/(b'm) \cdot (\alpha                              \hat{n}_\cup  - 1) \cdot Z_\cup^2    \\
\Cov(Z_1, Z_\cup) \,&\approx\,                         Z_\cup  \,+\, 1/(b'm) \cdot (\alpha\mathrlap{\hat{n}_1}\hphantom{\hat{n}_\cup} - 1) \cdot Z_1 Z_\cup  \\
\Cov(Z_2, Z_\cup) \,&\approx\,                         Z_\cup  \,+\, 1/(b'm) \cdot (\alpha\mathrlap{\hat{n}_2}\hphantom{\hat{n}_\cup} - 1) \cdot Z_2 Z_\cup  \\
\Cov(Z_1, Z_2)    \,&\approx\,                         Z_\cup  \,+\, 1/(b'm) \cdot (\alpha                              \hat{n}_\cap  - 1) \cdot Z_1 Z_2
\end{align*}
Substitute these into the covariance matrix and perform vector multiplications, observing that $\hat{n}_\cap = \hat{n}_1 + \hat{n}_2 - \hat{n}_\cup$ by inclusion-exclusion and that $\log p_0 \approx -k/(bm)$ by first-order approximations given $m \gg k^2$:
\begin{align*}
\Var \hat{n}_\cap \,\approx\, (bm/k)^2 \,\times\, \big[\, 
         1 / \mathrlap{Z_1}\hphantom{Z_\cup} \,&+\, 1/(b'm) \cdot (\alpha\mathrlap{\hat{n}_1}\hphantom{\hat{n}_\cup} - 1)  \\
  {} +\, 1 / \mathrlap{Z_2}\hphantom{Z_\cup} \,&+\, 1/(b'm) \cdot (\alpha\mathrlap{\hat{n}_2}\hphantom{\hat{n}_\cup} - 1)  \\
  {} +\, 1 /                         Z_\cup  \,&+\, 1/(b'm) \cdot (\alpha                              \hat{n}_\cup  - 1)  \\
  {} -\, 2 / \mathrlap{Z_1}\hphantom{Z_\cup} \,&-\, 2/(b'm) \cdot (\alpha\mathrlap{\hat{n}_1}\hphantom{\hat{n}_\cup} - 1)  \\   
  {} -\, 2 / \mathrlap{Z_2}\hphantom{Z_\cup} \,&-\, 2/(b'm) \cdot (\alpha\mathrlap{\hat{n}_2}\hphantom{\hat{n}_\cup} - 1)  \\  
  {} +\, 2\,Z_\cup / (Z_1 Z_2)               \,&+\, 2/(b'm) \cdot (\alpha                              \hat{n}_\cap  - 1)  \,\big]\;\;{=}\\
&\hspace{-110pt}
{=}\;\;\left(\frac{bm}{k}\right)^2 \left(\frac{2\hspace{0.5pt}Z_\cup}{Z_1 Z_2} \,{+}\, \frac{1}{Z_\cup} \,{-}\, \frac{1}{Z_1} \,{-}\, \frac{1}{Z_2} \,{-}\, \frac{1}{b'm}\right)
\,+\, \left(\frac{bm}{k}\right)^2 \!\! \frac{\alpha}{b'm} \, \hat{n}_\cap
\end{align*}
It remains to notice that $(bm/k)^2\hspace{0.5pt}{\cdot}\,\alpha/(b'm)$ $=$ $b(k\,{-}\,1)/(b'k) - 1$, and formula~(\ref{eq:block:var-overlap}) is proven.
\end{proof}

\section{Efficiency, Intervals, and Bias}
\label{sec:app:stt}

This appendix records the i.i.d.-bit (Poissonized) view of the occupancy estimators, the Cram\'er--Rao bounds they attain to first order, Wald intervals from the main-text variance formulas, a second-order bias calculation, and a remark on the three-phase block budget.
Notation follows the main text: $\varsigma$ is a sum of bits, $Z=m'-\varsigma$ (or $Z=b'm-\varsigma|_{\le b'}$ for MBFs) a zero count, $\theta$ the truncation threshold, and $p_0$ the per-item miss probability of Appendix~\ref{app:block:cardinal}.

\subsection{Poissonization and negative association}
\label{app:stt:poisson}

Replace the fixed number of hash placements $K=kn$ by $\tilde K\sim\mathrm{Poisson}(kn)$ (equivalently, a Poisson item count).
Per-position hit counts are then i.i.d.\ Poisson, and the occupancy indicators are i.i.d.\ Bernoulli with success $p=1-e^{-kn/m}$.
For a truncated filter the visible sum of bits is $\varsigma\sim\mathrm{Binomial}(m',p)$ with $m'=\theta m$; for an MBF reading $b'$ blocks it is $\varsigma|_{\le b'}\sim\mathrm{Binomial}(b'm,\,1-p_0^n)$, with $p_0$ from Eq.~(\ref{eq:app:block:q}).

The true fixed-$K$ occupancy indicators are \emph{negatively associated}: distinct-position covariances are nonpositive, so the i.i.d.\ variance slightly overstates the truth.
Keeping that $O(m^{-1})$ correction recovers the first-order variance of Eq.~(\ref{eq:basics:set-size-predictor-variance}), including the $-n/k$ term, and the analogous $b(k-1)/(b'k)-1$ term of Eq.~(\ref{eq:block:var-overlap}).

Poisson splitting also gives independence across blocks: with a Poisson item count, items routed by $h_0$ to different blocks form independent processes, so the per-block count vectors are independent.
The de-Poissonized cross-block covariance is the $O(1/b)$ term already in Claim~\ref{th:block:cov-z}.

\subsection{Cram\'er--Rao bounds}
\label{app:lower-bounds}

Under the i.i.d.-bit model, $N$ visible bits are Bernoulli$(p(n))$ and the Fisher information is $\mathcal I(n)=N\,[p'(n)]^2/[p(n)(1-p(n))]$, so $\Var(\hat n)\ge 1/\mathcal I(n)$ for any unbiased estimator.

\begin{proposition}[CRLB for a single/truncated BF]
\label{prop:crlb:single-bf}
Let a Bloom filter with parameters $(m,k,\theta)$ sketch a set of size $n$, and let $m'=\theta m$ be the number of visible positions. Under the i.i.d.-bit model,
\[
\Var(\hat n)\ \ge\ \frac{1}{m'}\cdot \frac{1-q^{\,n}}{(\ln q)^2 q^{\,n}},
\qquad q=\Bigl(1-\frac{1}{m}\Bigr)^k .
\]
As $q^{\,n}\downarrow 0$ (saturation), the bound diverges.
\end{proposition}

With $\ln q\approx -k/m$ and $q^n=\E Z/m'$, the bound equals $(m/k)^2 \cdot \bigl(1/\E Z-1/m'\bigr)$, which is the first term of Eq.~(\ref{eq:basics:set-size-predictor-variance}).
The occupancy-inversion estimator of Eq.~(\ref{eq:basics:set-size-predictor}) is therefore asymptotically efficient under this model; the remaining $-n/k$ term is the negative-association correction of Section~\ref{app:stt:poisson}.

\begin{proposition}[CRLB for an MBF with partial reads]
\label{prop:crlb:mbf}
Let an MBF with parameters $(b,m,k)$ sketch a set of size $n$. If only $b'\le b$ blocks are read, then under the same model,
\[
\Var(\hat n)\ \ge\ \frac{1}{b'm}\cdot \frac{1-p_0^{\,n}}{(\ln p_0)^2 p_0^{\,n}},
\]
with $p_0$ from Eq.~(\ref{eq:app:block:q}).
\end{proposition}

The bound scales as $1/b'$ and diverges as blocks saturate.
No unbiased estimator under this model can meet a target $\sigma^2$ with materially fewer blocks than Eq.~(\ref{eq:accur:b-prime-overlap}) prescribes.
Evaluate $\ln q=k\,\logonep(-1/m)$ and $\ln p_0=\logonep\bigl(\tfrac1b\,\expmone(k\logonep(-1/m))\bigr)$.

\subsection{Wald intervals}
\label{app:bf-ci-two}

The pairwise overlap estimators of Eqs.~(\ref{eq:overlap:overlap-from-Z_x}) and~(\ref{eq:block:affine}) are smooth functions of the observed zero counts.
A $(1-\delta)$ Wald interval is
\[
\hat n_\cap \;\pm\; z_{1-\delta/2}\,\widehat{\sd}(\hat n_\cap),
\]
with $\widehat{\sd}$ the square root of Eq.~(\ref{eq:overlap:var-overlap}) or Eq.~(\ref{eq:block:var-overlap}).
A multinomial $2\times 2$ table of bit pairs yields the same first-order variance as the main-text delta method, minus the negative-association correction; we do not treat it as a separate estimator.

For finite-sample hygiene, replace a zero count $Z$ by $Z+\epsilon$ with a tiny $\epsilon$ (e.g.\ $1/2$) before taking logarithms, and clamp a negative $\hat n_\cap$ at $0$ as in Section~\ref{sec:accur:scale}.
Clamping bias is quantified in Section~\ref{app:stt:bias}.
These intervals are first-order approximations under the i.i.d.-bit model; they are not finite-sample coverage guarantees.

\subsection{Bias of the overlap estimators}
\label{app:stt:bias}

The estimators are affine in $\langle\log Z_x\rangle$ (Eqs.~(\ref{eq:overlap:affine}) and~(\ref{eq:block:affine})), so the Hessian is diagonal and covariances do not enter the second-order bias.
With $\E\log Z\approx\log\E Z-\Var Z/(2(\E Z)^2)$,
\[
\E\hat n[F]-|F| \;\approx\; -\tfrac12\sum_x a_x\,\frac{\Var Z_x}{(\E Z_x)^2}.
\]
For MBF pairwise overlap, $a=(1/\log p_0)\,(+1,+1,-1)$ and $\log p_0\approx -k/(bm)$.
Using the leading term $\Var Z\approx \E Z-(\E Z)^2/(b'm)$ from Eq.~(\ref{eq:block:cov-z-appr}),
\begin{equation}
\label{eq:stt:bias-overlap}
\E\hat n_\cap-n_\cap \;\approx\; \frac{bm}{2k}\left(\frac{1}{\E Z_1}+\frac{1}{\E Z_2}-\frac{1}{\E Z_\cup}-\frac{1}{b'm}\right).
\end{equation}
The bracket is $O(1/(b'm))$ times a function of the bit densities, so the bias is $O\bigl(b/(k\,b')\bigr)$ items.
In the regime of Sections~\ref{sec:accur:limit} and~\ref{sec:exper} this is a fraction of an item, against $\sigma$ in the thousands (e.g.\ $b=600$, $m=5{\cdot}10^7$, $k=7$, $|S_1|=10^8$, $|S_2|=10^7$, $n_\cap=5{\cdot}10^4$, $b'=1$: $|\mathrm{bias}|<0.01$ and $\sigma\approx 7{\cdot}10^3$).
The $O(k^2/m)$ correction of Eq.~(\ref{eq:block:cov-z-est}) does not change the order.
The single-filter case is analogous, recovering the commented second-order expansion in Section~\ref{sec:basics}.

\paragraph{Clamping.}
If $\hat n_\cap\approx\mathcal N(n_\cap,\sigma^2)$ and negatives are set to $0$, the added bias is $\sigma\,\varphi(n_\cap/\sigma)-n_\cap\,\Phi(-n_\cap/\sigma)$.
It equals $\sigma/\sqrt{2\pi}\approx 0.40\sigma$ at $n_\cap=0$ and drops below $0.01\sigma$ once $n_\cap\gtrsim 2.4\sigma$.
For the borderline-threshold use case the clamped estimator is therefore effectively unbiased; for near-zero overlaps, report the unclamped estimate together with the Wald interval.

\subsection{Validity under the three-phase budget}
\label{app:stt:three-phase}

The number of blocks read, $b''=b'+\Delta b$, is chosen from data that are then reused in $\hat n_\cap$, so the fixed-$b'$ variance is not literally applicable.
The following remark justifies using Eq.~(\ref{eq:block:var-overlap}) with the realized $b''$ asymptotically; it is not a finite-sample coverage guarantee.

\begin{remark}
\label{rmk:stt:three-phase}
Assume the Poissonized model of Section~\ref{app:stt:poisson}, so per-block count vectors are independent.
\begin{enumerate}[leftmargin=*]
\item Algorithm~\ref{alg:accur:cross} computes the Seed budget $b'$ from block~$1$ only (Eq.~(\ref{eq:accur:b-prime-disjoint})).
Blocks $2,\ldots,b'$ used in Main are independent of that decision.
\item Top-up reads blocks $b,b-1,\ldots$ in reverse.
The clamp $b''\le b$ keeps this range disjoint from $1,\ldots,b'$, so Top-up blocks are independent of $\Delta b$.
\item The estimator is a smooth function of the mean of $b''$ i.i.d.\ block vectors, where $b''$ is a continuous function of zero-fractions that concentrate at relative error $O(1/\sqrt{b'm})$.
Thus $b''/b''_{\mathrm{oracle}}\to 1$ in probability, with $b''_{\mathrm{oracle}}$ computed from true expectations.
A standard random-sample-size CLT then yields the same first-order variance, with $b''$ substituted, as in the fixed-budget case.
\end{enumerate}
Hence the Wald interval of Section~\ref{app:bf-ci-two} is asymptotically valid for the three-phase scheme.
The remark is per pair: pairs that share an audience are dependent, and we do not claim a joint statement over $L\times R$.
It also does not claim that the target $\sigma$ is always met (Phase~1 can under-budget; Phase~3 corrects the budget) nor validity under per-block adaptive stopping with many looks, which would need a confidence-sequence construction~\cite{Howard2021:confidence}.
\end{remark}

\subsection{Proofs of the Cram\'er--Rao bounds}
\label{app:stt:proofs}

For i.i.d.\ Bernoulli observations with success $p(n)$,
\[
\mathcal I(n)=N\,\frac{(p'(n))^2}{p(n)\bigl(1-p(n)\bigr)}, \qquad \Var(\hat n)\;\ge\;\mathcal I(n)^{-1}.
\]

\begin{proof}[Proof of Proposition~\ref{prop:crlb:single-bf}]
Here $N=m'$, $p(n)=1-q^n$ with $q=(1-1/m)^k$, and $p'(n)=-q^n\ln q$. Hence
\[
\mathcal I(n)=m'\cdot\frac{(\ln q)^2 q^{n}}{1-q^n},
\quad\Rightarrow\quad
\Var(\hat n)\ \ge\ \frac{1}{m'}\cdot \frac{1-q^n}{(\ln q)^2 q^{n}}.
\]
\end{proof}

\begin{proof}[Proof of Proposition~\ref{prop:crlb:mbf}]
Here $N=b'm$ and $p(n)=1-p_0^n$ with $p_0$ from Eq.~(\ref{eq:app:block:q}). Then $p'(n)=-p_0^n\ln p_0$ and
\[
\mathcal I(n)=b'm\cdot \frac{(\ln p_0)^2 p_0^{n}}{1-p_0^n},
\quad\Rightarrow\quad
\Var(\hat n)\ \ge\ \frac{1}{b'm}\cdot \frac{1-p_0^n}{(\ln p_0)^2 p_0^{n}}.
\]
\end{proof}

\end{document}